\documentclass[%
 reprint,
 amsmath,amssymb,
 aps,
 pre,
 longbibliography,
]{revtex4-2} 

\usepackage[pdftex]{graphicx} 
\usepackage{dcolumn}
\usepackage{bm}
\usepackage[T1]{fontenc}

\newtheorem{lem}{Lemma}
\newtheorem{prop}{Proposition}

\begin{document}


\title{Theoretical analysis of quantum turbulence using the Onsager ``ideal turbulence'' theory}

\author{Tomohiro Tanogami}
 \email{tanogami.tomohiro.84c@st.kyoto-u.ac.jp}
\affiliation{%
Department of Physics, Kyoto University, Kyoto 606-8502, Japan
}%




\date{\today}
\begin{abstract}
We investigate three-dimensional quantum turbulence as described by the Gross-Pitaevskii model using the analytical method exploited in the Onsager ``ideal turbulence'' theory.
We derive the scale-independence of the scale-to-scale kinetic energy flux and establish a double-cascade scenario: at scales much larger than the mean intervortex $\ell_i$, the Richardson cascade becomes dominant, whereas at scales much smaller than $\ell_i$, another type of cascade is induced by quantum stress.
We then evaluate the corresponding velocity power spectrum using a phenomenological argument.
The relation between the novel cascade, which we call quantum stress cascade, and the Kelvin-wave cascade is also discussed.

\end{abstract}

\pacs{Valid PACS appear here}

\maketitle
\section{Introduction}
The phenomenon of fully developed turbulence is well described by Onsager's theory of ``ideal turbulence'' \cite{Onsager_1949,Eyink_Review,Eyink_Sreenivasan}.
According to this theory, in classical incompressible turbulence, kinetic energy is continuously transferred from large to small scales in the intermediate-length scales---inertial range---without dissipation.
This energy transfer is called the energy cascade, or Richardson cascade, because the mechanism of this phenomenon is intuitively explained by Richardson's depiction of a large vortex splitting into smaller vortices \cite{Richardson_1922}.
For the Richardson cascade, the kinetic energy spectrum $E(k)$ follows the Kolmogorov spectrum $E(k)\propto k^{-5/3}$ \cite{Onsager_1949,K41_a,K41_b,K41_c}.
Onsager's theory is based on the empirical fact, first observed by Taylor \cite{Taylor_1917}, that the energy dissipation remains significant even in the inviscid limit.
The existence of this \textit{anomalous dissipation} means that as the viscosity decreases, the inertial range extends to infinitely small scales, i.e., the energy cascade continues endlessly. 
Onsager's theory provides a rigorous method that can describe such a remarkable phenomenon in the absence of viscosity.
This theory has been applied to two-dimensional enstrophy cascade \cite{Eyink_1995_2D,Eyink_2001_2D}, three-dimensional helicity cascade \cite{Chae_2003,Chen_etal_2003}, and magnetohydrodynamic turbulence \cite{Caflisch_etal_1997,Aluie_Eyink_2010_MHD,Aluie_2017} and has recently been extended to the case of compressible turbulence \cite{Feireisl_2017,Eyink_Drivas,Drivas_Eyink,Aluie_PRL,Aluie_scale_locality,Aluie_scale_decomposition}, collisionless plasma turbulence \cite{Eyink_2018}, and relativistic fluid turbulence \cite{Eyink_Drivas_2018_relativistic}.

Note that although Onsager's ``ideal turbulence'' cannot be achieved exactly in a classical system, in which the effect of viscosity cannot be ignored, completely inviscid flow is realized in quantum fluids, such as superfluid helium or atomic Bose-Einstein condensates (BECs).
Such quantum fluids differ from classical fluids in that (i) they exhibit two-fluid behavior at nonzero temperature, (ii) they flow without viscosity, and (iii) their circulation is quantized, as proposed independently by Onsager \cite{Onsager_1949} and Feynman \cite{Feynman}.
Turbulence in quantum fluids is called quantum turbulence and has been intensively investigated both theoretically and experimentally in recent years \cite{Tsubota_etal_2013,Madeira_etal_2020,Tsubota_2008,QT_review,Tsatsos_2016}.

Despite these differences between quantum and classical fluids, recent numerical and experimental results have revealed that there are quantitative similarities between quantum and classical turbulence.
For example, quantum turbulence exhibits both the Richardson cascade and the Kolmogorov spectrum, as in classical turbulence \cite{Nore_etal_1997_PhysFluid,Nore_etal_1997_PRL,Kobayashi_Tsubota,*Kobayashi_Tsubota_PRL,QT_experiments}.
On the basis of the fact that pure quantum turbulence, which consists of quantized vortices and has zero viscosity, behaves like classical turbulence, quantum turbulence has been described as a prototype or ``skeleton'' of turbulence \cite{Kobayashi_Tsubota,*Kobayashi_Tsubota_PRL,Hanninen_2014}.

Although there are similarities between quantum and classical turbulence, there are, of course, striking differences. 
First, quantum turbulence comprises three main characteristic length scales: the characteristic length scale of injection of kinetic energy by external stirring $L$, the mean intervortex distance $\ell_i$, and the vortex core radius $\xi$. 
Second, it is believed that, at scales smaller than $\ell_i$, the Richardson cascade is no longer dominant and the Kelvin-wave cascade excited by vortex reconnections becomes dominant.
The Kelvin-wave cascade is an energy cascade believed to result from the interaction of Kelvin waves \cite{Thomson_1880} of different wave numbers on a single quantum vortex.
It is conjectured that the energy spectrum corresponding to the Kelvin-wave cascade also shows a power-law behavior, and several candidates for the exponent have been theoretically predicted. 
For example, Kozik and Svistunov, who provided the first theoretical analysis of the Kelvin-wave cascade, obtained a value of $-7/5$ using the wave turbulence theory of a six-wave scattering process under the assumption of local interaction \cite{KS}. 
Later, L'vov and Nazarenko criticized this assumption of locality and concluded that the exponent becomes $-5/3$ \cite{LN}.
In addition, several works predict the spectral exponent $-1$ \cite{Vinen_2000,Vinen_Niemela_2002,Nemirovskii_2013}.

The spectral index values obtained from these previous studies have formed the basis of many subsequent theoretical analyses. 
However, because these studies are based on the vortex filament model \cite{Schwarz_1985}, the effects of compressibility, particularly the contribution of quantum stress, are not considered \cite{Hanninen_2014}.
This compressibility effect is particularly important in quantum turbulence in trapped atomic BECs, for which various experimental techniques have been developed in recent years.
Because superfluid density changes significantly in the vicinity of quantum vortices, quantum stress may contribute to the energy transfer across scales.
Therefore, we conjecture that a novel energy cascade induced by quantum stress will occur at scales sufficiently smaller than the mean intervortex distance.
That is, we expect the following double-cascade scenario:
\begin{enumerate}
	\item The energy injected from the large-scale $L$ is transferred to small-scale $\gtrsim\ell_i$ by the Richardson cascade, induced by a tangled structure of quantum vortices behaving like a classical vortex.
	\item At scales smaller than the mean intervortex distance $\ell_i$, the Richardson cascade is no longer dominant, and the effect of the quantum stress becomes significant. 
	As a result, another type of cascade, which we call quantum stress cascade, is induced by the quantum stress.
\end{enumerate}
Because both the quantum stress cascade and Kelvin-wave cascade occur at scales sufficiently smaller than the mean intervortex distance and Kelvin waves are accompanied by rapid density changes, we expect that the quantum stress cascade is related to the Kelvin-wave cascade.
It is important, however, to note that the existence of Kelvin waves may not be a necessary condition for the quantum stress cascade to exist.

We validate our conjecture by studying the Gross-Pitaevskii (GP) model \cite{Gross_1961,Pitaevskii_1961}, which can describe quantum turbulence in atomic BECs.
Because there are no viscous effects in quantum turbulence as described by the GP model, the cascade can extend to infinitely small scales although it may be cut off at the vortex core radius $\xi$.
Such a situation may be appropriately described using the Onsager ``ideal turbulence'' theory.
Therefore, to establish the validity of the above double-cascade scenario, we take a phenomenological approach based on the Onsager theory.
In other words, we investigate the dynamics of quantum vortices as described by the \textit{quantum Euler equations}, which is a hydrodynamic form of the GP model, using the Onsager theory.
Because the quantum Euler equations have a form similar to that of the ordinary compressible Euler equations, we can exploit recent developments that extend Onsager's theory to classical compressible turbulence \cite{Feireisl_2017,Eyink_Drivas,Drivas_Eyink,Aluie_PRL,Aluie_scale_locality,Aluie_scale_decomposition}.

This paper is organized as follows.
In the next section, we introduce the quantum Euler equations and derive the local energy balance equations.
Subsequently, we introduce a coarse-graining approach that can resolve the turbulent fields both in scale and in space.
Next, in Section \ref{Main assumptions}, we explain the main assumptions used in our analysis.
The main claims of this paper are presented in Section \ref{Main claims}.
In Section \ref{Derivation and explanation}, we derive and explain the main claims.
Concluding remarks are provided in Section \ref{Concluding remarks}.

\section{Preliminaries\label{Preliminaries}}
\subsection{Setup\label{Setup}}
We consider quantum turbulence as described by the GP model:
\begin{equation}
i\hbar\dfrac{\partial}{\partial t}\Psi({\bf x},t)=\left[-\dfrac{\hbar^2}{2m}\nabla^2+V_{\mathrm{ext}}+g|\Psi|^2-\mu\right]\Psi({\bf x},t).
\label{eq:GP}
\end{equation}
Here, $\Psi({\bf x},t)$ denotes the condensate's complex wave function, $m$ the boson mass, $\mu$ the chemical potential, $g$ is a positive constant that represents the strength of the interaction between bosons, and $V_{\mathrm{ext}}({\bf x},t)$ expresses both the trapping potential and external stirring, e.g., rotation along distinct axes \cite{Kobayashi_Tsubota_2007}.
Note that our analysis can be applied for both decaying and forced turbulence.
To avoid questions regarding boundaries, we assume periodic boundary conditions with periodic box $\Omega=[0,\mathcal{L}]^d$.
Although we focus on the case in which the spatial dimension $d=3$, the calculations in Sections \ref{Proof that three types of scale-to-scale kinetic energy flux exist} and \ref{Scale dependence of the energy fluxes} are formally valid in arbitrary space dimensions.
By introducing the Madelung transformation $\Psi=\sqrt{n}\exp(i\theta)$ \cite{Madelung_1926,Madelung_1927}, which relates $\Psi$ to the superfluid mass density $\rho=mn$ and velocity ${\bf v}=(\hbar/m)\nabla\theta$, we can obtain a hydrodynamic description of the system:
\begin{equation}
\partial_t\rho+\nabla\cdot(\rho{\bf v})=0,
\label{eq:QHD-1}
\end{equation}
\begin{equation}
\partial_t(\rho{\bf v})+\nabla\cdot\left(\rho{\bf v}{\bf v}+p{\bf I}-{\bf \Sigma}\right)={\bf f},
\label{eq:QHD-2}
\end{equation}
where $p:=g\rho^2/(2m^2)$, ${\bf f}$ denotes the external force due to $V_{\mathrm{ext}}$, and ${\bf \Sigma}$ is the \textit{quantum stress}:  
\begin{equation}
{\bf \Sigma}:=\dfrac{\hbar^2}{4m^2}\Delta\rho{\bf I}-\dfrac{\hbar^2}{m^2}\nabla\sqrt{\rho}\nabla\sqrt{\rho},
\end{equation}
where ${\bf I}$ denotes the unit tensor.
The \textit{quantum Euler equations} (\ref{eq:QHD-1}) and (\ref{eq:QHD-2}) become identical to the ordinary compressible Euler equations by taking the classical limit $\hbar\rightarrow0$.
The important point here is that the quantum stress $\nabla\cdot{\bf \Sigma}$ becomes relevant in the vicinity of quantum vortices because it contains higher-order spatial derivatives.
Note that because the density changes are significant even outside the vortex cores \cite{Leoni,Fujimoto_2015,Villois_2017}, we cannot ignore the quantum stress even in that region in general.

The crossover length scale at which the quantum stress becomes comparable to the momentum flux, $\rho{\bf v}{\bf v}$, can be estimated as follows.
Let $v_0$ be a typical velocity, such as the critical velocity \cite{Hanninen_Schoepe_2010}, and $\rho_0$ be the volume averaged density.
Then, using an estimation that $\rho{\bf v}{\bf v}\sim\rho_0v^2_0$ and ${\bf \Sigma}\sim\kappa^2\rho_0/\ell^2$, where $\ell$ is a length scale and $\kappa:=h/m$ is the quantum circulation, one obtains
\begin{eqnarray}
\rho{\bf v}{\bf v}&\sim&{\bf \Sigma}\notag\\
\rho_0v^2_0&\sim&\kappa^2\rho_0/\ell^2\notag\\
\ell&\sim&\kappa/v_0.
\end{eqnarray}
This is on the order of the mean intervortex distance $\ell_i$ \cite{Hanninen_Schoepe_2010}.
This fact already suggests the existence of a double-cascade process consisting of the Richardson cascade, induced by the momentum flux, and the quantum stress cascade, induced by the quantum stress.

The total energy of the system can be written as comprising three components, the kinetic energy $\int d^d{\bf x}\rho|{\bf v}|^2/2$, interaction energy $\int d^d{\bf x}p$, and quantum energy $\int d^d{\bf x}\hbar^2|\nabla\sqrt{\rho}|^2/2m^2$ \cite{Nore_etal_1997_PRL,Nore_etal_1997_PhysFluid}:
\begin{equation}
E=\int_{\Omega}d^d{\bf x}\left[\dfrac{1}{2}\rho|{\bf v}|^2+p+\dfrac{\hbar^2}{2m^2}|\nabla\sqrt{\rho}|^2\right].
\label{total energy}
\end{equation}
For smooth solutions of (\ref{eq:QHD-1}) and (\ref{eq:QHD-2}), one can immediately obtain the balance equations for the energy densities. 
The local energy balance equation is given by
\begin{widetext}
\begin{eqnarray}
\partial_t\left(\dfrac{1}{2}\rho|{\bf v}|^2+p+\dfrac{\hbar^2}{2m^2}|\nabla\sqrt{\rho}|^2\right)&+&\nabla\cdot\left\{\left[\left(\dfrac{1}{2}\rho|{\bf v}|^2+p+\dfrac{\hbar^2}{2m^2}|\nabla\sqrt{\rho}|^2\right){\bf I} \right.\right.\notag\\
&&\left.+p{\bf I}-{\bf \Sigma}\Bigr]\cdot{\bf v}+\dfrac{\hbar^2}{4m^2}(\nabla\rho)\nabla\cdot{\bf v}\right\}={\bf v}\cdot{\bf f}.
\label{eq:local total energy balance}
\end{eqnarray}
\end{widetext}
The evolution equation for the kinetic energy density is given by
\begin{eqnarray}
&&\partial_t\left(\dfrac{1}{2}\rho|{\bf v}|^2\right)+\nabla\cdot\left[\left(\dfrac{1}{2}\rho|{\bf v}|^2+p\right){\bf v}-{\bf \Sigma}\cdot{\bf v}\right]\notag\\
&&=p\nabla\cdot{\bf v}-{\bf \Sigma}:\nabla{\bf v}+{\bf v}\cdot{\bf f}.
\label{eq:local kinetic energy balance}
\end{eqnarray}
Using (\ref{eq:local total energy balance}) and (\ref{eq:local kinetic energy balance}), one can also derive the evolution equation for the sum of the interaction and quantum energy densities as follows:
\begin{widetext}
\begin{equation}
\partial_t\left(p+\dfrac{\hbar^2}{2m^2}|\nabla\sqrt{\rho}|^2\right)+\nabla\cdot\left[\left(p+\dfrac{\hbar^2}{2m^2}|\nabla\sqrt{\rho}|^2\right){\bf v}+\dfrac{\hbar^2}{4m^2}(\nabla\rho)\nabla\cdot{\bf v}\right]=-p\nabla\cdot{\bf v}+{\bf \Sigma}:\nabla{\bf v}.
\label{eq:local interaction energy balance}
\end{equation}
\end{widetext}
Note that, even with no external force, the total kinetic energy is not conserved because of the effect of the first two terms on the right-hand side of (\ref{eq:local kinetic energy balance}).
The first term $-p\nabla\cdot{\bf v}$ is known as the pressure-dilatation, which represents the conversion of kinetic energy into interaction or quantum energy and vice versa.
As shown through numerical calculation \cite{Wang-Yang-Shi}, the pressure-dilatation is considered to convert kinetic energy into interaction or quantum energy on average.
The existence of the second term $\Sigma:\nabla{\bf v}$ is specific to quantum turbulence.
As the form of this term is similar to that of pressure-dilatation, we refer to it as \textit{quantum-stress--strain}.
It also represents the conversion of kinetic energy into interaction or quantum energy and vice versa.
We note that the additional energy flux in (\ref{eq:local total energy balance}) and (\ref{eq:local interaction energy balance}), $\hbar^2(\nabla\rho)\nabla\cdot{\bf v}/4m^2$, does not exist in ordinary hydrodynamics.
This term corresponds to \textit{interstitial working} in the Navier-Stokes-Korteweg equations \cite{Dunn_Serrin,Gavage}.

\subsection{Coarse-graining}
Hereafter, we use the theoretical framework and notation given in \cite{Aluie_PRL,Aluie_scale_locality}.
Additionally, to simplify the notation, we often suppress argument $t$.
We study  properties of kinetic energy transfer across scales by a coarse-graining approach that can resolve turbulent fields both in scale and in space \cite{Leonard_1974,Germano_1992}.
For any locally integrable function ${\bf a}({\bf x})$, we define a coarse-grained field at length scale $\ell$ as
\begin{equation}
\bar{{\bf a}}_\ell({\bf x}):=\int_{\Omega}d^d{\bf r}G_\ell({\bf r}){\bf a}({\bf x}+{\bf r}),
\label{coarse-graining}
\end{equation}
where $G({\bf r})$ is the Friedrichs mollifier, i.e., $G:\Omega\rightarrow [0,\infty)$ is a smooth symmetric function supported in the open unit ball and with $\int_{\Omega}G=1$, and $G_\ell({\bf r}):=\ell^{-d}G({\bf r}/\ell)$ is the rescaling defined for each $\ell>0$.
We also define a residual field as
\begin{equation}
{\bf a}'_\ell({\bf x}):={\bf a}({\bf x})-\bar{{\bf a}}_\ell({\bf x}).
\end{equation}
As a direct consequence of the Riemann-Lebesgue lemma, the coarse-graining procedure (\ref{coarse-graining}) substantially reduces the amplitude of high-frequency Fourier components in space.
Hence, it is reasonable to describe $\bar{{\bf a}}_\ell$ as the large-scale component of ${\bf a}$ and ${\bf a}'_\ell$ as the small-scale field.
Here, we note that 
\begin{eqnarray}
\overline{(\bar{{\bf a}}_\ell)}_\ell&\neq&\bar{{\bf a}}_\ell\\
\overline{({\bf a}'_\ell)}_\ell&\neq&{\bf 0},
\end{eqnarray}
in general.

Because the coarse-graining operation commutes with space and time derivatives, coarse-graining of (\ref{eq:QHD-1}) and (\ref{eq:QHD-2}) gives
\begin{equation}
\partial_t\bar{\rho}_\ell+\nabla\cdot\overline{(\rho{\bf v})}_\ell=0,
\label{eq:coarse-grained QHD-1}
\end{equation}
\begin{equation}
\partial_t\overline{(\rho{\bf v})}_\ell+\nabla\cdot\overline{(\rho{\bf v}{\bf v})}_\ell=-\nabla\bar{p}_\ell+\nabla\cdot\bar{{\bf \Sigma}}_\ell+\bar{{\bf f}}_\ell.
\label{eq:coarse-grained QHD-2}
\end{equation}
Note that to express (\ref{eq:coarse-grained QHD-1}) and (\ref{eq:coarse-grained QHD-2}) in terms of large-scale quantities $\bar{\rho}_\ell$ and $\bar{{\bf v}}_\ell$, we must introduce several cumulants, such as $\bar{\tau}_\ell(\rho,{\bf v}):=\overline{(\rho{\bf v})}_\ell-\bar{\rho}_\ell\bar{{\bf v}}_\ell$.
We introduce the density-weighted Favre average to reduce the number of additional cumulant terms and obtain a simple physical interpretation \cite{Favre}:
\begin{equation}
\tilde{{\bf a}}_\ell:=\dfrac{\overline{(\rho{\bf a})}_\ell}{\bar{\rho}_\ell}.
\end{equation}
Subsequently, we rewrite (\ref{eq:coarse-grained QHD-1}) and (\ref{eq:coarse-grained QHD-2}) using the Favre average:
\begin{equation}
\partial_t\bar{\rho}_\ell+\nabla\cdot(\bar{\rho}_\ell\tilde{{\bf v}}_\ell)=0,
\label{eq:Favre coarse-grained QHD-1}
\end{equation}
\begin{equation}
\bar{\rho}_\ell(\partial_t+\tilde{{\bf v}}_\ell\cdot\nabla)\tilde{{\bf v}}_\ell+\nabla\cdot(\bar{\rho}_\ell\tilde{\tau}_\ell({\bf v},{\bf v}))=-\nabla\bar{p}_\ell+\nabla\cdot\bar{{\bf \Sigma}}_\ell+\bar{{\bf f}}_\ell,
\label{eq:Favre coarse-grained QHD-2}
\end{equation}
where $\tilde{\tau}_\ell({\bf v},{\bf v}):=\widetilde{({\bf v}{\bf v})}_\ell-\tilde{{\bf v}}_\ell\tilde{{\bf v}}_\ell$.

Note that the coarse-grained equations (\ref{eq:Favre coarse-grained QHD-1}) and (\ref{eq:Favre coarse-grained QHD-2}) are not closed in terms of large-scale fields $\bar{\rho}_\ell$, $\tilde{{\bf v}}_\ell$, $\bar{p}_\ell$, and $\bar{{\bf \Sigma}}_\ell$ because of the appearance of the additional cumulant term, i.e., $\tilde{\tau}_\ell({\bf v},{\bf v})$.
This cumulant term depends on the small-scale ($<\ell$) velocity field and can be regarded as the source of the energy cascade, as demonstrated below.

\section{Main assumptions\label{Main assumptions}}
\subsection{Assumption 1: Regularity of velocity and density fields\label{Assumption 1}}
We assume that the velocity field is H\"older continuous, i.e., 
\begin{equation}
\delta{\bf v}({\bf r};{\bf x})=O(|{\bf r}|^h)\quad\text{as}\quad|{\bf r}|/L\rightarrow0,
\label{assumption_v_naive}
\end{equation}
with $h\in(0,1]$, where $L$ is the characteristic energy injection scale and $\delta{\bf a}({\bf r};{\bf x}):={\bf a}({\bf x}+{\bf r})-{\bf a}({\bf x})$ for any field ${\bf a}({\bf x})$.
Note that the local H\"older condition (\ref{assumption_v_naive}) is not in contradiction with the blow-up of the velocity field as $1/r$ with distance $r$ from a quantum vortex.
From the perspective of the multifractal hypothesis \cite{Parisi-Frisch,Frisch}, there is a spectrum of H\"older singularities $[h_{\mathrm{min}},h_{\mathrm{max}}]$.
Imposing the assumption (\ref{assumption_v_naive}) implies that we are considering a point ${\bf x}\in\Omega$ with $h>0$, just to permit a simplified analysis.
This does not contradict the fact that there might be other points with $h<0$.
To correctly examine negative local H\"older singularities, we must perform a more sophisticated space-global analysis using Besov spaces (see Appendix \ref{Appendix: Assumption 1: Regularity of velocity and density fields}).
We remark that the H\"older-type condition for the velocity field was first conjectured by Onsager \cite{Onsager_1949,Eyink_Sreenivasan} and is well established empirically in classical turbulence.

In addition, we assume that the density field satisfies
\begin{eqnarray}
\delta\rho({\bf r};{\bf x})&=&O(|{\bf r}|)\quad\text{as}\quad|{\bf r}|/L\rightarrow0,\label{assumption_rho_naive}\\
1/\bar{\rho}_\ell\le M&<&\infty \quad\text{for}\quad\ell\ge\xi,\label{eq:inverse_rho_naive}
\end{eqnarray}
where $\xi:=\hbar/\sqrt{2mg\rho_0}$ is the vortex core radius.
Requirement (\ref{assumption_rho_naive}) is reasonable because the energy density of the system (\ref{total energy}) contains the density gradient term $\propto|\nabla\rho|^2$.
In the last condition (\ref{eq:inverse_rho_naive}), $M$ can be defined, for instance, as $M:=\sup_{\ell\ge\xi,{\bf x}\in\Omega}1/\bar{\rho}_\ell$.
Note that this requirement is not strong enough to prohibit the existence of vacuum regions, $\{{\bf x}\in\Omega|\rho({\bf x})=0\}$, where the quantized vortices are located.
In the following, we always assume that $\ell\ge\xi$ because we are interested only in the intermediate scales $\gg\xi$.

\subsection{Assumption 2: Existence of steady state\label{Assumption 2}}
We assume that the system eventually reaches a steady state in which the mean kinetic energy is constant.
The following two facts support the validity of this assumption.
First, although there are no viscous effects in pure quantum turbulence, in which no normal fluid component remains, experimental and numerical calculations imply that at least the incompressible component of kinetic energy can be dissipated \cite{Barenghi_2008,Bradley_2006,Nore_etal_1997_PRL}. 
This is attributed to the emission of compressible excitations from vortex reconnections \cite{Leadbeater_etal_2001,Zuccher_etal_2012} and Kelvin waves \cite{Vinen_2000,Vinen_2001}.
Second, the Onsager-type regularity condition for weak solutions (solutions in the sense of distribution) of the Euler-Korteweg equations, including the quantum Euler equations as its special case, was recently formulated \cite{Debiec_etal_2018}.
This result implies that weak solutions that satisfy the condition may dissipate energy because of their singularity even in the absence of viscosity. 

There are two remarks regarding this assumption.
First, to make this assumption more plausible, it might be beneficial to introduce an energy dissipation term that eliminates the small-scale kinetic energy as in previous studies \cite{Kobayashi_Tsubota_PRL,Kobayashi-Ueda,Bradley_Anderson_2012}.
Such a phenomenological dissipation term is considered to model the dissipation caused by the presence of thermal particles \cite{Tsatsos_2016}. 
Even in that case, if the dissipation works on a scale sufficiently smaller than the mean intervortex distance $\ell_i$, the inertial range can be properly defined, and the result of the present analysis does not change.
Second, instead of a steady state, it is also possible to assume a quasi-steady-state with the total energy of the system increasing (forced turbulence) or conserved (decaying turbulence).
Such quasi-steady states have been well investigated through numerical simulations \cite{Leoni,Shukla_etal_2019}.
We note that our analysis is also applicable to this case.

\subsection{Assumption 3: Scale-independence of pressure-dilatation at small scales\label{Assumption 3}}
We impose an additional assumption concerning the co-spectrum of pressure-dilatation.
The pressure-dilatation co-spectrum is defined as
\begin{equation}
E^{(p)}(k):=-\dfrac{1}{\Delta k}\sum_{k-\Delta k/2<|{\bf k}|<k+\Delta k/2}\hat{p}({\bf k})\widehat{\nabla\cdot{\bf v}}(-{\bf k}),
\end{equation}
where $\Delta k:=2\pi/\mathcal{L}$, and the symbol $\hat{f}$ denotes the Fourier coefficient of $f$.
The requirement of the pressure-dilatation co-spectrum is that it decays sufficiently quickly at large $k$:
\begin{equation}
E^{(p)}(k)=O(k^{-\alpha}),\quad \alpha>1.
\label{eq:E^PD condition}
\end{equation}
Note that the condition (\ref{eq:E^PD condition}) does not require a power-law behavior.
This assumption is based on the decorrelation effects, namely cancellations between $\bar{p}_\ell$, which acts at large scales $\approx L$, and $\nabla\cdot\bar{{\bf v}}_\ell$, which varies rapidly in space \cite{Aluie_PRL,Aluie_scale_locality}.
This condition is the same as that imposed by Aluie \cite{Aluie_PRL,Aluie_scale_locality}, and its validity is confirmed numerically for the classical case \cite{Aluie_2012,Wang-Yang-Shi}.

\section{Main claims\label{Main claims}}

\subsection{Three types of energy flux\label{Three types of energy flux}}
In quantum turbulence, three types of scale-to-scale kinetic energy flux are capable of directly transferring kinetic energy across scales. 
Two of these fluxes contribute to the energy cascade.

\subsubsection{Deformation work}
The first type is deformation work \cite{A_first_course_in_turbulence},
\begin{equation}
\Pi_\ell:=-\bar{\rho}_\ell\nabla\tilde{{\bf v}}_\ell:\tilde{\tau}_\ell({\bf v},{\bf v}),
\end{equation}
which corresponds to the energy flux of the Richardson cascade in classical turbulence.
In particular, in classical homogeneous ($\rho=\mathrm{const}$) incompressible turbulence, deformation work is the only scale-to-scale energy flux that transfers kinetic energy across scales.
Deformation work represents work done by the large-scale ($>\ell$) strain $\nabla\tilde{{\bf v}}_\ell$ against the small-scale ($<\ell$) stress $\bar{\rho}_\ell\tilde{\tau}_\ell({\bf v},{\bf v})$.
Here, we refer to $\bar{\rho}_\ell\tilde{\tau}_\ell({\bf v},{\bf v})$ as ``small-scale stress'' because it is the residual field obtained by subtracting the contribution of large-scale stress $\bar{\rho}_\ell\tilde{{\bf v}}_\ell\tilde{{\bf v}}_\ell$ from $\bar{\rho}_\ell\widetilde{({\bf v}{\bf v})}_\ell$.

\subsubsection{Baropycnal work}
The second type is baropycnal work \cite{Aluie_scale_decomposition,Aluie_PRL,Lees_Aluie}, which arises owing to compressibility:
\begin{equation}
\Lambda^{(p)}_\ell:=\dfrac{1}{\bar{\rho}_\ell}\nabla\bar{p}_\ell\cdot\bar{\tau}_\ell(\rho,{\bf v}).
\end{equation}
Baropycnal work represents work done by the large-scale ($>\ell$) pressure gradient force $-(1/\bar{\rho}_\ell)\nabla\bar{p}_\ell$ against the small-scale ($<\ell$) mass flux $\bar{\tau}_\ell(\rho,{\bf v})$.
Here, we refer to $\bar{\tau}_\ell(\rho,{\bf v})$ as ``small-scale mass flux'' because it is the residual field obtained by subtracting the contribution of large-scale mass flux $\bar{\rho}_\ell\bar{{\bf v}}_\ell$ from $\overline{(\rho{\bf v})}_\ell$.
The presence of baropycnal work is peculiar to compressible fluids and it does not exist in homogeneous incompressible fluids.
In classical compressible turbulence, not only deformation work but also baropycnal work contributes to the transfer of kinetic energy across scales.
The physical mechanism of baropycnal work is studied in detail in \cite{Aluie_scale_decomposition,Lees_Aluie}.

Under Assumption 1, we can prove that
\begin{eqnarray}
\Lambda^{(p)}_\ell
&=&O\left(\ell^{h+1}\right).
\end{eqnarray}
Therefore, the baropycnal work vanishes at least at $O(\ell^{h+1})$ for $\ell\rightarrow0$.
Thus, in quantum turbulence, baropycnal work does not contribute to the transfer of kinetic energy across scales.

\subsubsection{Quantum baropycnal work}
The third type is the energy flux that is specific to quantum turbulence,
\begin{equation}
\Lambda^{(\Sigma)}_\ell:=-\dfrac{1}{\bar{\rho}_\ell}\nabla\cdot\bar{{\bf \Sigma}}_\ell\cdot\bar{\tau}_\ell(\rho,{\bf v}),
\label{eq:Quantum baropycnal work}
\end{equation}
which represents the energy transfer due to quantum stress.
We call $\Lambda^{(\Sigma)}_\ell$ \textit{quantum baropycnal work} because its form is similar to that of baropycnal work.
Quantum baropycnal work represents work done by the large-scale ($>\ell$) force due to quantum stress $(1/\bar{\rho}_\ell)\nabla\cdot\bar{{\bf \Sigma}}_\ell$ against the small-scale ($<\ell$) mass flux $\bar{\tau}_\ell(\rho,{\bf v})$.
Note that, at small scales near a vortex core, this energy flux is expected to be greater than deformation work because it contains higher-order spatial derivatives.
This implies the existence of a double-cascade process, as stated in Section \ref{Existence of a double-cascade process_claim}.

\subsection{``Kolmogorov's $4/5$-law'' for quantum turbulence}
Under Assumptions 2 and 3, we can derive the relation that is analogous to the Kolmogorov $4/5$-law \cite{K41_c,Frisch}:
\begin{eqnarray}
\langle Q^{\mathrm{flux}}_\ell\rangle\approx\epsilon_{\mathrm{eff}}\quad\text{for}\quad\ell_{\mathrm{small}}\ll\ell\ll\ell_{\mathrm{large}},
\label{4/5-law}
\end{eqnarray}
where the symbol $\langle\cdot\rangle$ denotes a volume average $\int_{\Omega}\cdot d^d{\bf x}/\mathcal{L}^d$.
Here, $\langle Q^{\mathrm{flux}}_\ell\rangle:=\langle\Pi_\ell\rangle+\langle\Lambda^{(p)}_\ell\rangle+\langle\Lambda^{(\Sigma)}_\ell\rangle$ denotes the total mean scale-to-scale kinetic energy flux and $\epsilon_{\mathrm{eff}}:=\langle p\nabla\cdot{\bf v}\rangle+\langle\epsilon^{\mathrm{in}}_L\rangle$, where $\epsilon^{\mathrm{in}}_L\approx\epsilon^{\mathrm{in}}_\ell:=\tilde{{\bf v}}_\ell\cdot\bar{{\bf f}}_\ell$, denotes the effective mean energy injection rate.
The intermediate asymptotic limit $\ell_{\mathrm{small}}\ll\ell\ll\ell_{\mathrm{large}}$ can be interpreted as the ``inertial range'' for quantum turbulence.
The $\ell_{\mathrm{large}}$ is defined using pressure-dilatation co-spectrum $E^{(p)}(k)$:
\begin{eqnarray}
\ell_{\mathrm{large}}:=\dfrac{\sum_kk^{-1}E^{(p)}(k)}{\sum_kE^{(p)}(k)},
\end{eqnarray}
and $\ell_{\mathrm{small}}$ is defined using quantum-stress--strain co-spectrum $E^{(\Sigma)}(k)$:
\begin{equation}
\ell_{\mathrm{small}}:=\dfrac{\sum_kk^{-1}E^{(\Sigma)}(k)}{\sum_kE^{(\Sigma)}(k)},
\end{equation}
where $E^{(\Sigma)}(k)$ is defined as
\begin{equation}
E^{(\Sigma)}(k):=\dfrac{1}{\Delta k}\sum_{k-\Delta k/2<|{\bf k}|<k+\Delta k/2}\hat{{\bf \Sigma}}({\bf k}):\widehat{\nabla{\bf v}}(-{\bf k}).
\end{equation}
Because $\langle Q^{\mathrm{flux}}_\ell\rangle$ can be expressed in terms of field increments $\delta{\bf v}({\bf r};{\bf x})$ and $\delta\rho({\bf r};{\bf x})$, the relation (\ref{4/5-law}) plays the same role as the $4/5$-law.

\subsection{Existence of a double-cascade process\label{Existence of a double-cascade process_claim}}
Under Assumptions 1--3, we can show that
\begin{eqnarray}
\langle\Lambda^{(\Sigma)}_\ell\rangle\ll\langle\Pi_\ell\rangle&=&O(1)\quad\text{for}\quad\ell_i\ll\ell\ll\ell_{\mathrm{large}},\\
\langle\Pi_\ell\rangle\ll\langle\Lambda^{(\Sigma)}_\ell\rangle&=&O(1)\quad\text{for}\quad\ell_{\mathrm{small}}\ll\ell\ll\ell_i.
\end{eqnarray}
Thus, a double-cascade occurs in quantum turbulence: one in the scale range much larger than the mean intervortex distance $\ell_i$, $\ell_i\ll\ell\ll\ell_{\mathrm{large}}$, and the other in $\ell_{\mathrm{small}}\ll\ell\ll\ell_i$.
The former is induced by deformation work and is thus the same as in classical turbulence, i.e., the Richardson cascade. 
The latter is induced by quantum baropycnal work and is referred to as the \textit{quantum stress cascade} instead of the Kelvin-wave cascade because the existence of Kelvin waves may not be a prerequisite for the existence of this cascade.

\begin{figure}[t]
\includegraphics[width=8.6cm]{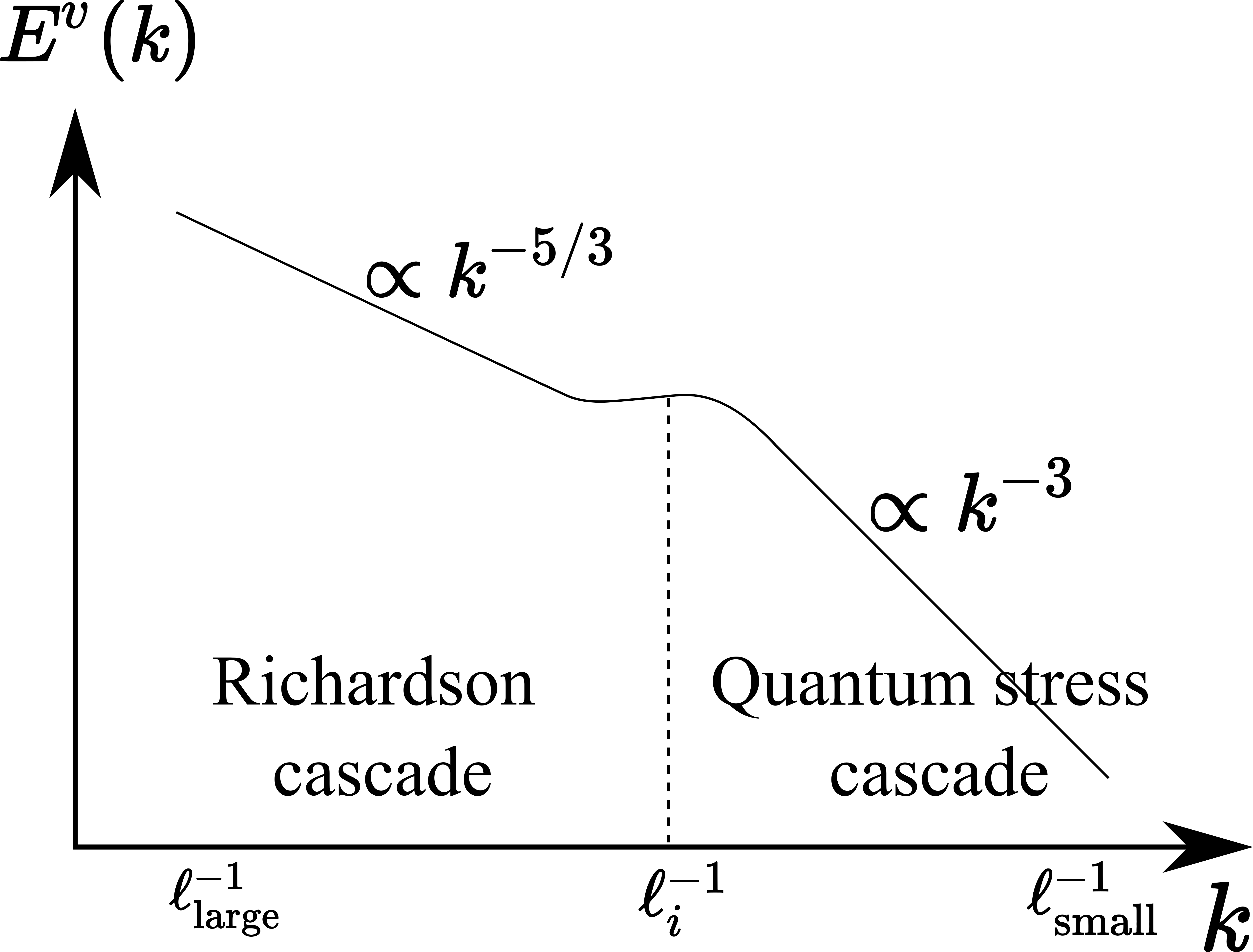}
\caption{Velocity power spectrum $E^v(k)$ in the inertial range $\ell^{-1}_{\mathrm{large}}\ll k\ll\ell^{-1}_{\mathrm{small}}$.}
\label{fig:energy_spectrum}
\end{figure}

As a consequence of this double-cascade process, we expect that the velocity power spectrum $E^v(k)$ exhibits the following asymptotic behavior (see Fig.~\ref{fig:energy_spectrum}):
\begin{equation}
E^v(k)\sim 
\begin{cases}
C_{\mathrm{large}} k^{-5/3}\quad\text{for}\quad \ell^{-1}_{\mathrm{large}}\ll k\ll\ell^{-1}_i,\\
C_{\mathrm{small}}k^{-3}\quad\text{for}\quad \ell^{-1}_i\ll k\ll\ell^{-1}_{\mathrm{small}},
\end{cases}
\end{equation}
where $C_{\mathrm{large}}$ and $C_{\mathrm{small}}$ are positive constants.
For the power spectrum of the density-weighted velocity field $\sqrt{\rho}{\bf v}$, the same asymptotic behavior is expected (see Appendix \ref{Estimation of the spectral index of the velocity power spectrum}).
Here, we note that there is a possibility that the spectrum $k^{-3}$ in $\ell^{-1}_i\ll k\ll\ell^{-1}_{\mathrm{small}}$ becomes shallower because of the \textit{depletion of nonlinearity} or regularity of the density gradient field (see Section \ref{Note on the spectral index of quantum stress cascade}).

The detailed scenario of the energy transport in quantum turbulence implied by our results is as follows:
\begin{enumerate}
	\item Kinetic energy is injected from the large scale ($\sim L$) through external stirring.
	\item In the scale range larger than $\ell_{\mathrm{large}}$, the injected kinetic energy is gradually transferred to a smaller scale because of the effect of deformation work, and part of the energy is converted to other forms through pressure-dilatation.
	\item In the inertial range $\ell_{\mathrm{small}}\ll\ell\ll\ell_{\mathrm{large}}$, the following double-cascade process occurs:
	\begin{enumerate}
		\item In the scale range $\ell_i\ll\ell\ll\ell_{\mathrm{large}}$, the Richardson cascade, induced by deformation work, is dominant. Intuitively, this is because quantum vortices form a tangled structure that effectively behaves like a classical vortex.
		\item At scales smaller than the mean intervortex distance $\ell_i$, the effect of the quantum stress due to quantum vortices becomes significant. Therefore, the Richardson cascade is no longer dominant, and the quantum stress cascade, induced by quantum baropycnal work, develops.
	\end{enumerate}
	\item In the scale range smaller than $\ell_{\mathrm{small}}$, the kinetic energy transferred by the double-cascade is further transferred to a smaller scale through quantum baropycnal work, and part of the energy is converted to other forms through quantum-stress--strain.
 \end{enumerate}

\section{Derivation and explanation\label{Derivation and explanation}}
In this section, we present the derivation and explanation of the assertion discussed in the previous section.
First, we prove that three types of scale-to-scale kinetic energy flux exist that can contribute to energy transfer across scales and describe the scale dependence of these fluxes.
Next, we define the inertial range for quantum turbulence and derive ``Kolmogorov's $4/5$-law'' for quantum turbulence.
Finally, we prove the existence of the double-cascade process using the scale dependence of the energy fluxes and the definition of the inertial range.
For more sophisticated derivation and explanation using the $L^p$-norm, see Appendix \ref{More sophisticated analysis using Besov spaces}.

\subsection{Proof that three types of scale-to-scale kinetic energy flux exist\label{Proof that three types of scale-to-scale kinetic energy flux exist}}
We show that three types of scale-to-scale kinetic energy flux exist in quantum turbulence by deriving the large-scale kinetic energy budget equation.
Here, the large-scale kinetic energy density is defined as $\bar{\rho}_\ell|\tilde{{\bf v}}_\ell|^2/2$ and satisfies the inequality \cite{Eyink_Drivas}
\begin{eqnarray}
\int_\Omega d^d{\bf x}\dfrac{1}{2}\bar{\rho}_\ell|\tilde{{\bf v}}_\ell|^2\le\int_\Omega d^d{\bf x}\dfrac{1}{2}\rho|{\bf v}|^2.
\end{eqnarray}

Using the coarse-grained equations (\ref{eq:Favre coarse-grained QHD-1}) and (\ref{eq:Favre coarse-grained QHD-2}), we can obtain the large-scale kinetic energy budget equation,
\begin{equation}
\partial_t\left(\dfrac{1}{2}\bar{\rho}_\ell|\tilde{{\bf v}}_\ell|^2\right)+\nabla\cdot{\bf J}_\ell=\bar{p}_\ell\nabla\cdot\bar{{\bf v}}_\ell-\bar{{\bf \Sigma}}_\ell:\nabla\bar{{\bf v}}_\ell-Q^{\mathrm{flux}}_\ell+\epsilon^{\mathrm{in}}_\ell,
\label{eq:coarse-grained kinetic energy balance}
\end{equation}
where the various terms are defined as follows:
\begin{eqnarray}
{\bf J}_\ell&:=&\left(\dfrac{1}{2}\bar{\rho}_\ell|\tilde{{\bf v}}_\ell|^2+\bar{p}_\ell\right)\tilde{{\bf v}}_\ell+\bar{\rho}_\ell\tilde{{\bf v}}_\ell\cdot\tilde{\tau}_\ell({\bf v},{\bf v})-\dfrac{\bar{p}_\ell}{\bar{\rho}_\ell}\bar{\tau}_\ell(\rho,{\bf v})\notag\\
&&-\bar{{\bf \Sigma}}_\ell\cdot\tilde{{\bf v}}_\ell+\dfrac{\bar{{\bf \Sigma}}_\ell}{\bar{\rho}_\ell}\cdot\bar{\tau}_\ell(\rho,{\bf v}),
\end{eqnarray}
\begin{equation}
\epsilon^{\mathrm{in}}_\ell:=\tilde{{\bf v}}_\ell\cdot\bar{{\bf f}}_\ell,
\end{equation}
\begin{equation}
Q^{\mathrm{flux}}_\ell:=\Pi_\ell+\Lambda^{(p)}_\ell+\Lambda^{(\Sigma)}_\ell,
\label{eq:energy flux}
\end{equation}
\begin{equation}
\Pi_\ell:=-\bar{\rho}_\ell\nabla\tilde{{\bf v}}_\ell:\tilde{\tau}_\ell({\bf v},{\bf v}),
\label{eq:Deformation work}
\end{equation}
\begin{equation}
\Lambda^{(p)}_\ell:=\dfrac{1}{\bar{\rho}_\ell}\nabla\bar{p}_\ell\cdot\bar{\tau}_\ell(\rho,{\bf v}),
\end{equation}
\begin{equation}
\Lambda^{(\Sigma)}_\ell:=-\dfrac{1}{\bar{\rho}_\ell}\nabla\cdot\bar{{\bf \Sigma}}_\ell\cdot\bar{\tau}_\ell(\rho,{\bf v}).
\label{eq:Quantum baropycnal work}
\end{equation}
Here, ${\bf J}_\ell$ represents the spatial transport of large-scale kinetic energy, which does not contribute to the transfer of kinetic energy across scales, and $\epsilon^{\mathrm{in}}_\ell$ denotes the energy injection rate due to external stirring at scale $\ell$.
The first two terms on the right-hand side of (\ref{eq:coarse-grained kinetic energy balance}), $-\bar{p}_\ell\nabla\cdot\bar{{\bf v}}_\ell$ and $\bar{{\bf \Sigma}}_\ell:\nabla\bar{{\bf v}}_\ell$, are the large-scale pressure-dilatation and quantum-stress--strain, respectively. 
Note that these two terms contain no modes at small scales $<\ell$.
Therefore, they only contribute to the conversion of the large-scale kinetic energy into other forms of energy, i.e., interaction and quantum energies, and vice versa.

The third term on the right-hand side of (\ref{eq:coarse-grained kinetic energy balance}), $Q^{\mathrm{flux}}_\ell$, consists of three components: deformation work $\Pi_\ell$ \cite{A_first_course_in_turbulence}, baropycnal work $\Lambda^{(p)}_\ell$ \cite{Aluie_scale_decomposition,Aluie_PRL,Lees_Aluie}, and quantum baropycnal work $\Lambda^{(\Sigma)}_\ell$.
Note that each of the three components has the following form: ``large-scale ($>\ell$) quantity $\times$ small-scale ($<\ell$) quantity,'' while other terms on the right-hand side of (\ref{eq:coarse-grained kinetic energy balance}) do not.
Therefore, only these three terms are capable of directly transferring kinetic energy across scales, as stated in Section \ref{Three types of energy flux}, and $Q^{\mathrm{flux}}_\ell$ represents the total scale-to-scale kinetic energy flux.

We note that these three terms are all Galilean invariant.
Another possible definition is $\tilde{v}_j\partial_i(\bar{\rho}\tilde{\tau}_\ell(v_i,v_j))$, where $i\in\{1,2,\cdots,d\}$ and a summation convention for equal indices is adopted.
This definition differs from ours by the total gradient $\partial_j(\bar{\rho}_\ell\tilde{v}_i\tilde{\tau}_\ell(v_i,v_j))$.
However, this quantity is not Galilean invariant, so the amount of energy transferred from large to small scales at point ${\bf x}$ in the flow would depend on the velocity of the observer.
As observed by Eyink \cite{Eyink_2005}, Galilean invariance is necessary for the scale locality of the energy cascade.
Note that non-Galilean-invariant terms in (\ref{eq:coarse-grained kinetic energy balance}) do not contribute to the energy transfer across scales.

\subsection{Scale dependence of the energy fluxes\label{Scale dependence of the energy fluxes}}
In this subsection, we explain the scale dependence of the three energy fluxes, deformation work, baropycnal work, and quantum baropycnal work, using arguments similar to those used in the Onsager ``ideal turbulence'' theory.

\subsubsection{Deformation work}
We now examine the scale $\ell$ dependence of the deformation work $\Pi_\ell=-\bar{\rho}_\ell\nabla\tilde{{\bf v}}_\ell:\tilde{\tau}_\ell({\bf v},{\bf v})$.
Expressing the energy flux in terms of increments, $\delta{\bf a}({\bf r};{\bf x})$, is a crucial aspect of Onsager's theory.
Using the Cauchy-Schwarz inequality and the fact that $\bar{\rho}_\ell\le\sup_{{\bf x}\in\Omega}\rho({\bf x})$, we obtain
\begin{eqnarray}
|\Pi_\ell|&=&|\bar{\rho}_\ell\nabla\tilde{{\bf v}}_\ell:\tilde{\tau}_\ell({\bf v},{\bf v})|\notag\\
&\le&\left(\sup_{{\bf x}\in\Omega}\rho({\bf x})\right)|\nabla\tilde{{\bf v}}_\ell||\tilde{\tau}_\ell({\bf v},{\bf v})|,
\label{Pi_cauchy-Schwarz}
\end{eqnarray}
where $|A|$ for matrix $A=(a_{ij})$ denotes the Frobenius norm, i.e., $|\nabla\tilde{{\bf v}}_\ell|:=\sqrt{\sum^d_{i=1}\sum^d_{j=1}|\partial_i\widetilde{(v_j)}_\ell({\bf x})|^2}$ and $|\tilde{\tau}_\ell({\bf v},{\bf v})|:=\sqrt{\sum^d_{i=1}\sum^d_{j=1}|\tilde{\tau}_\ell(v_i,v_j)|^2}$.

For the large-scale strain $\nabla\tilde{{\bf v}}_\ell$, using the relation
\begin{equation}
\tilde{{\bf v}}_\ell=\bar{{\bf v}}_\ell+\dfrac{\bar{\tau}_\ell(\rho,{\bf v})}{\bar{\rho}_\ell},
\label{eq:tilde_u decompose}
\end{equation}
one obtains
\begin{eqnarray}
\nabla\tilde{{\bf v}}_\ell=\nabla\bar{{\bf v}}_\ell+\dfrac{1}{\bar{\rho}_\ell}\nabla\bar{\tau}_\ell(\rho,{\bf v})+\dfrac{\bar{\tau}_\ell(\rho,{\bf v})}{\bar{\rho}^2_\ell}\nabla\bar{\rho}_\ell.
\label{eq:nabla tilde u_naive}
\end{eqnarray}
For the first term of (\ref{eq:nabla tilde u_naive}), as $\int d^d{\bf r}\nabla G({\bf r})={\bf 0}$, for any locally integrable function ${\bf a}({\bf x})$,
\begin{equation}
\nabla\bar{{\bf a}}_\ell({\bf x})=-\dfrac{1}{\ell}\int_{\Omega}d^{d}{\bf r}(\nabla G)_\ell({\bf r})\delta{\bf a}({\bf r};{\bf x}).
\end{equation}
Therefore,
\begin{eqnarray}
|\nabla\bar{{\bf a}}_\ell|&=&\left|\dfrac{1}{\ell}\int_{\Omega}d^{d}{\bf r}(\nabla G)_\ell({\bf r})\delta{\bf a}({\bf r};{\bf x})\right| \notag\\
&\le&\dfrac{1}{\ell}\int_{\Omega}d^{d}{\bf r}|(\nabla G)_\ell({\bf r})||\delta{\bf a}({\bf r};{\bf x})| \notag\\
&\le&\dfrac{\mathrm{(const)}}{\ell}\sup_{|{\bf r}|<\ell}|\delta{\bf a}({\bf r};{\bf x})|,
\label{eq:estimate_nabla_a_naive}
\end{eqnarray}
and hence
\begin{equation}
\nabla\bar{{\bf v}}_\ell=O\left(\dfrac{\delta v(\ell;{\bf x})}{\ell}\right),
\label{eq:estimate_nabla_v_tilde_1_naive}
\end{equation}
where $\delta a(\ell;{\bf x}):=\sup_{|{\bf r}|<\ell}|\delta{\bf a}({\bf r};{\bf x})|$.
For the second and third terms of (\ref{eq:nabla tilde u_naive}), using the assumption (\ref{eq:inverse_rho_naive}) and Propositions \ref{prop:1} and \ref{prop:2} in Appendix \ref{Commutator estimates}, we obtain
\begin{eqnarray}
\dfrac{1}{\bar{\rho}_\ell}\nabla\bar{\tau}_\ell(\rho,{\bf v})=O\left(\dfrac{M\delta\rho(\ell;{\bf x})\delta v(\ell;{\bf x})}{\ell}\right),
\label{eq:estimate_nabla_v_tilde_2_naive}
\end{eqnarray}
\begin{eqnarray}
\dfrac{\bar{\tau}_\ell(\rho,{\bf v})}{\bar{\rho}^2_\ell}\nabla\bar{\rho}_\ell=O\left(\dfrac{M^2(\delta\rho(\ell;{\bf x}))^2\delta v(\ell;{\bf x})}{\ell}\right).
\label{eq:estimate_nabla_v_tilde_3_naive}
\end{eqnarray}
Thus, combining the results (\ref{eq:estimate_nabla_v_tilde_1_naive}), (\ref{eq:estimate_nabla_v_tilde_2_naive}), and (\ref{eq:estimate_nabla_v_tilde_3_naive}), we obtain
\begin{eqnarray}
\nabla\tilde{{\bf v}}_\ell&=&\dfrac{O(\delta v(\ell;{\bf x}))}{\ell}\notag\\
&&\times\Bigl[1+O(M\delta\rho(\ell;{\bf x}))+O(M^2(\delta\rho(\ell;{\bf x}))^2)\Bigr] \notag\\
&=&O\left(\dfrac{\delta v(\ell;{\bf x})}{\ell}\right).
\label{eq:estimate nabla_tilde_u_naive}
\end{eqnarray}

For the small-scale stress $\tilde{\tau}_\ell({\bf v},{\bf v})$, using the relation
\begin{equation}
\tilde{\tau}_\ell({\bf v},{\bf v})=\bar{\tau}_\ell({\bf v},{\bf v})+\dfrac{1}{\bar{\rho}_\ell}\bar{\tau}_\ell(\rho,{\bf v},{\bf v})-\dfrac{1}{\bar{\rho}^2_\ell}\bar{\tau}_\ell(\rho,{\bf v})\bar{\tau}_\ell(\rho,{\bf v}),
\label{eq:tilde_u_u decompose}
\end{equation}
where $\bar{\tau}_\ell(\rho,{\bf v},{\bf v})$ is defined by
\begin{equation}
\bar{\tau}_\ell(\rho,{\bf v},{\bf v}):=\overline{(\rho{\bf v}{\bf v})}_\ell-\bar{\rho}_\ell\overline{({\bf v}{\bf v})}_\ell-2\bar{{\bf v}}_\ell\overline{(\rho{\bf v})}_\ell-\bar{\rho}_\ell\bar{{\bf v}}_\ell\bar{{\bf v}}_\ell,
\end{equation}
and the assumption (\ref{eq:inverse_rho_naive}) and Proposition \ref{prop:1} in Appendix \ref{Commutator estimates}, one obtains
\begin{eqnarray}
\tilde{\tau}_\ell({\bf v},{\bf v})&=&O((\delta v(\ell;{\bf x}))^2)\notag\\
&&\times\Bigl[1+O(M\delta\rho(\ell;{\bf x}))+O(M^2(\delta\rho(\ell;{\bf x}))^2)\Bigr] \notag\\
&=&O((\delta v(\ell;{\bf x}))^2).
\label{eq:estimate tilde_tau_u_u_naive}
\end{eqnarray}

Thus, from (\ref{Pi_cauchy-Schwarz}), (\ref{eq:estimate nabla_tilde_u_naive}), (\ref{eq:estimate tilde_tau_u_u_naive}), and the assumption (\ref{assumption_v_naive}), we finally obtain
\begin{eqnarray}
\Pi_\ell&=&-\bar{\rho}_\ell\nabla\tilde{{\bf v}}_\ell:\tilde{\tau}_\ell({\bf v},{\bf v})\notag\\
&=&O\left(\dfrac{(\delta v(\ell;{\bf x}))^3}{\ell}\right) \notag\\
&=&O\left(\ell^{3h-1}\right),
\label{eq:Estimation of the second term_naive}
\end{eqnarray}
as a rigorous upper bound.
This implies that the deformation work vanishes at least at $O(\ell^{3h-1})$ for $\ell\rightarrow0$ if $h>1/3$.
The scale-independent upper bound is obtained in the case of $h=1/3$.

\subsubsection{Baropycnal work}
We now explain the scale $\ell$ dependence of the baropycnal work $\Lambda^{(p)}_\ell=(1/\bar{\rho}_\ell)\nabla\bar{p}_\ell\cdot\bar{\tau}_\ell(\rho,{\bf v})$.
Using the Cauchy-Schwarz inequality, we obtain
\begin{eqnarray}
|\Lambda^{(p)}_\ell|&=&|(1/\bar{\rho}_\ell)\nabla\bar{p}_\ell\cdot\bar{\tau}_\ell(\rho,{\bf v})|\notag\\
&\le&|(1/\bar{\rho}_\ell)\nabla\bar{p}_\ell||\bar{\tau}_\ell(\rho,{\bf v})|.
\label{Lambda^p_Cauchy-Schwarz}
\end{eqnarray}

For the large-scale pressure gradient force $-(1/\bar{\rho}_\ell)\nabla\bar{p}_\ell$, from the assumption (\ref{eq:inverse_rho_naive}) and the inequality (\ref{eq:estimate_nabla_a_naive}), we obtain
\begin{eqnarray}
\left|\dfrac{1}{\bar{\rho}_\ell}\nabla\bar{p}_\ell\right|&\le&M\dfrac{\mathrm{(const)}}{\ell}\delta p(\ell;{\bf x})\notag\\
&=&O\left(\dfrac{\delta\rho(\ell;{\bf x})}{\ell}\right),
\label{Pressure gradient force}
\end{eqnarray}
where we have used the fact that
\begin{equation}
\delta p(\ell;{\bf x})=O(\delta\rho(\ell;{\bf x})),
\end{equation}
which follows from the definition of $p\propto\rho^2$ and the mean value theorem.

For the small-scale mass flux $\bar{\tau}_\ell(\rho,{\bf v})$, using Proposition \ref{prop:1} in Appendix \ref{Commutator estimates}, we obtain
\begin{equation}
\bar{\tau}_\ell(\rho,{\bf v})=O(\delta\rho(\ell;{\bf x})\delta v(\ell;{\bf x})).
\label{eq:estimate tau_rho_u_naive}
\end{equation}

Thus, from (\ref{Lambda^p_Cauchy-Schwarz}), (\ref{Pressure gradient force}), (\ref{eq:estimate tau_rho_u_naive}), and the assumptions (\ref{assumption_v_naive}) and (\ref{assumption_rho_naive}), we obtain
\begin{eqnarray}
\Lambda^{(p)}_\ell&=&(1/\bar{\rho}_\ell)\nabla\bar{p}_\ell\cdot\bar{\tau}_\ell(\rho,{\bf v}) \notag\\
&=&O\left(\dfrac{(\delta\rho(\ell;{\bf x}))^2\delta v(\ell;{\bf x})}{\ell}\right) \notag\\
&=&O\left(\ell^{h+1}\right).
\label{eq:Estimation of baropycnal work_naive}
\end{eqnarray}
This implies that the baropycnal work vanishes at least at $O(\ell^{h+1})$ for $\ell\rightarrow0$.
Therefore, unlike in the case of classical compressible turbulence, baropycnal work does not contribute to the transfer of kinetic energy across scales.
This result is a consequence of the functional form of $p=g\rho^2/(2m^2)$ and the assumption (\ref{assumption_rho_naive}).

\subsubsection{Quantum baropycnal work\label{Quantum baropycnal work}}
Next, we investigate the scale $\ell$ dependence of the quantum baropycnal work $\Lambda^{(\Sigma)}_\ell=-(1/\bar{\rho}_\ell)\nabla\cdot\bar{{\bf \Sigma}}_\ell\cdot\bar{\tau}_\ell(\rho,{\bf v})$.
Using the Cauchy-Schwarz inequality, we obtain
\begin{eqnarray}
|\Lambda^{(\Sigma)}_\ell|&=&|(1/\bar{\rho}_\ell)\nabla\cdot\bar{{\bf \Sigma}}_\ell\cdot\bar{\tau}_\ell(\rho,{\bf v})|\notag\\
&\le&|(1/\bar{\rho}_\ell)\nabla\cdot\bar{{\bf \Sigma}}_\ell||\bar{\tau}_\ell(\rho,{\bf v})|.
\label{Lambda^Sigma_Cauchy-Schwarz}
\end{eqnarray}

For the large-scale force due to quantum stress $(1/\bar{\rho}_\ell)\nabla\cdot\bar{{\bf \Sigma}}_\ell$, using the assumption (\ref{eq:inverse_rho_naive}) and the inequality (\ref{eq:estimate_nabla_a_naive}), one obtains
\begin{widetext}
\begin{eqnarray}
\left|\dfrac{1}{\bar{\rho}_\ell}\nabla\cdot\bar{{\bf \Sigma}}_\ell\right|
&\le&\left|M\dfrac{\hbar^2}{4m^2}\nabla\Delta\bar{\rho}_\ell\right|+\left|M\dfrac{\hbar^2}{m^2}\nabla\cdot\overline{(\nabla\sqrt{\rho}\nabla\sqrt{\rho})}_\ell\right| \notag\\
&\le&M\dfrac{\hbar^2}{4m^2\ell^3}\int_\Omega d^d{\bf r}|(\nabla\Delta G)_\ell({\bf r})||\delta\rho({\bf r};{\bf x})|+M\dfrac{\hbar^2}{4m^2\ell}\int_\Omega d^d{\bf r}|(\nabla G)_\ell({\bf r})|\left|\delta\left(\dfrac{1}{\rho}\right)({\bf r};{\bf x})\right|\left(\sup_{{\bf x}\in\Omega}|\nabla\rho|\right)^2\notag\\
&\le&M\dfrac{\mathrm{(const)}}{\ell^3}\delta\rho(\ell;{\bf x})+M\dfrac{\mathrm{(const)}}{\ell}\delta\rho(\ell;{\bf x}).
\label{eq:estimate nabla_sigma_naive}
\end{eqnarray}
\end{widetext}

Therefore, from (\ref{Lambda^Sigma_Cauchy-Schwarz}), (\ref{eq:estimate tau_rho_u_naive}), (\ref{eq:estimate nabla_sigma_naive}), and the assumptions (\ref{assumption_v_naive}) and (\ref{assumption_rho_naive}), we obtain
\begin{eqnarray}
\Lambda^{(\Sigma)}_\ell&=&-(1/\bar{\rho}_\ell)\nabla\cdot\bar{{\bf \Sigma}}_\ell\cdot\bar{\tau}_\ell(\rho,{\bf v})\notag\\
&=&O\left(\dfrac{1}{\ell^3}(\delta\rho(\ell;{\bf x}))^2\delta v(\ell;{\bf x})\right)\notag\\
&&+O\left(\dfrac{1}{\ell}(\delta\rho(\ell;{\bf x}))^2\delta v(\ell;{\bf x})\right) \notag\\
&=&O\left(\ell^{h-1}\right).
\label{eq:Estimation of the last term_naive}
\end{eqnarray}
Note that, for any $h\in(0,1]$, we cannot conclude that the quantum baropycnal work vanishes for $\ell\rightarrow0$.
In other words, unlike deformation work, quantum baropycnal work can contribute to the transfer of kinetic energy across scales regardless of the regularity of the velocity field.
The scale-independent upper bound is obtained in the case of $h=1$.

\subsection{Derivation of ``Kolmogorov's $4/5$-law'' for quantum turbulence\label{Derivation of ``Kolmogorov's $4/5$-law'' for quantum turbulence}}
Suppose that the external stirring force ${\bf f}$ varies at scales ($\sim L$) much larger than the mean intervortex distance $\ell_i$ and that the steady state in which the total mean kinetic energy is constant is realized.
Then, in the steady state, spatial averaging of (\ref{eq:coarse-grained kinetic energy balance}) gives
\begin{eqnarray}
\langle Q^{\mathrm{flux}}_\ell\rangle=\langle\bar{p}_\ell\nabla\cdot\bar{{\bf v}}_\ell\rangle-\langle\bar{{\bf \Sigma}}_\ell:\nabla\bar{{\bf v}}_\ell\rangle+\langle\epsilon^{\mathrm{in}}_L\rangle,
\label{eq:quantum_Q}
\end{eqnarray}
where $\langle\epsilon^{\mathrm{in}}_L\rangle$ denotes the large-scale energy injection rate due to external stirring.
Here, we have used the approximation $\langle\epsilon^{\mathrm{in}}_\ell\rangle\approx\langle\epsilon^{\mathrm{in}}_L\rangle$ \cite{Aluie_scale_decomposition}.
Below, following the definition of the inertial range for classical compressible turbulence suggested by Aluie \cite{Aluie_PRL,Aluie_scale_locality}, we aim to define the inertial range for quantum turbulence $\ell_{\mathrm{small}}\ll\ell\ll\ell_{\mathrm{large}}$, in which the total mean scale-to-scale kinetic energy flux becomes scale-independent, i.e., $\langle Q^{\mathrm{flux}}_\ell\rangle=O(1)$.

To this end, we determine the scale range in which the pressure-dilatation and quantum-stress--strain, which appear on the right-hand side of (\ref{eq:quantum_Q}), become scale-independent.
To ensure that the mean pressure-dilatation $\langle\bar{p}_\ell\nabla\cdot\bar{{\bf v}}_\ell\rangle$ becomes scale-independent at small scales, we impose the assumption on the pressure-dilatation stated in Section \ref{Assumption 3}.
From this assumption, it follows that the mean pressure-dilatation $\langle\bar{p}_\ell\nabla\cdot\bar{{\bf v}}_\ell\rangle$ converges to the finite constant $\langle p\nabla\cdot{\bf v}\rangle$ and becomes independent of $\ell$ at sufficiently small scales:
\begin{equation}
\lim_{\ell\rightarrow0}\langle\bar{p}_\ell\nabla\cdot\bar{{\bf v}}_\ell\rangle=-\lim_{K\rightarrow\infty}\sum_{0\le k<K}E^{(p)}(k)=\langle p\nabla\cdot{\bf v}\rangle.
\label{eq:saturation PD}
\end{equation}
Let $\ell_{\mathrm{large}}$ be a characteristic length scale of the pressure-dilatation.
It may be defined, for instance, as
\begin{equation}
\ell_{\mathrm{large}}:=\dfrac{\sum_kk^{-1}E^{(p)}(k)}{\sum_kE^{(p)}(k)}.
\end{equation}
Then, it follows that
\begin{eqnarray}
\langle\bar{p}_\ell\nabla\cdot\bar{{\bf v}}_\ell\rangle\approx\langle p\nabla\cdot{\bf v}\rangle\quad\text{for}\quad\ell\ll\ell_{\mathrm{large}}.
\end{eqnarray}

For the mean quantum-stress--strain $\langle\bar{{\bf \Sigma}}_\ell:\nabla\bar{{\bf v}}_\ell\rangle$, through evaluations similar to (\ref{eq:estimate nabla_sigma_naive}), one obtains
\begin{eqnarray}
&&|\bar{{\bf \Sigma}}_\ell:\nabla\bar{{\bf v}}_\ell|\notag\\
&\le&\left(\dfrac{\hbar^2}{4m^2}|\Delta\bar{\rho}_\ell|+\dfrac{\hbar^2}{m^2}\left|\overline{(\nabla\sqrt{\rho}\nabla\sqrt{\rho})}_\ell\right|\right)|\nabla\bar{{\bf v}}_\ell|\notag\\
&\le&\left(\dfrac{(\mathrm{const})}{\ell^2}\delta\rho(\ell;{\bf x})+(\mathrm{const})\delta\rho(\ell;{\bf x})\right)\notag\\
&&\times\dfrac{(\mathrm{const})}{\ell}\delta v(\ell;{\bf x})\notag\\
&=&O(\ell^{h-2}).
\end{eqnarray}
Therefore, there is a characteristic length scale of the quantum-stress--strain, which may be defined, for instance, as
\begin{equation}
\ell_{\mathrm{small}}:=\dfrac{\sum_kk^{-1}E^{(\Sigma)}(k)}{\sum_kE^{(\Sigma)}(k)},
\end{equation}
where $E^{(\Sigma)}(k)$ is the quantum-stress--strain co-spectrum defined by
\begin{equation}
E^{(\Sigma)}(k):=\sum_{k-\Delta k/2<|{\bf k}|<k+\Delta k/2}\hat{{\bf \Sigma}}({\bf k}):\widehat{\nabla{\bf v}}(-{\bf k}).
\end{equation}
Then, it follows that
\begin{eqnarray}
\langle\bar{{\bf \Sigma}}_\ell:\nabla\bar{{\bf v}}_\ell\rangle\approx0\quad\text{for}\quad\ell\gg\ell_{\mathrm{small}}.
\end{eqnarray}
Note that, unlike the pressure-dilatation, there are no expected decorrelation effects between $\bar{{\bf \Sigma}}_\ell$ and $\nabla\cdot{\bf v}_\ell$ because quantum stress ${\bf \Sigma}$ changes rapidly in space.

We now consider the intermediate asymptotic limit $\ell_{\mathrm{small}}\ll\ell\ll\ell_{\mathrm{large}}$.
In this scale range, the steady-state mean kinetic energy budget (\ref{eq:quantum_Q}) becomes
\begin{eqnarray}
\langle Q^{\mathrm{flux}}_\ell\rangle&\approx&\langle p\nabla\cdot{\bf v}\rangle+\langle\epsilon^{\mathrm{in}}_L\rangle\notag\\
&=:&\epsilon_{\mathrm{eff}},
\label{eq:quantum_Q_inertial_range}
\end{eqnarray} 
where $\epsilon_{\mathrm{eff}}$ denotes the effective mean energy injection rate.
Note that the right-hand side of (\ref{eq:quantum_Q_inertial_range}) does not include length scale $\ell$.
The scale independence of the mean scale-to-scale kinetic energy flux implies that kinetic energy is transferred conservatively to smaller scales on average.
Therefore, it is reasonable to call scale range $\ell_{\mathrm{small}}\ll\ell\ll\ell_{\mathrm{large}}$ the inertial range of quantum turbulence.

We provide some remarks related to the stationarity assumption.
If we introduce a phenomenological dissipation term to eliminate the small-scale kinetic energy, the artificial dissipation term $-\langle D_\ell\rangle$ acting at scale $\ell_d$ is added on the right-hand side of (\ref{eq:quantum_Q}).
In that case, the inertial range is modified to $\max\{\ell_{\mathrm{small}},\ell_d\}\ll\ell\ll\ell_{\mathrm{large}}$.
It is also possible to assume a quasi-steady-state instead of a steady state, as mentioned in Section \ref{Assumption 2}.
In that case, if the total energy is conserved (decaying turbulence), $-\langle\partial_t(\bar{\rho}_\ell|\tilde{{\bf v}}_\ell|^2/2)\rangle$ plays the role of $\langle\epsilon^{\mathrm{in}}_L\rangle$, while if the total energy is increasing (forced turbulence), $\langle\epsilon^{\mathrm{in}}_L\rangle$ in (\ref{eq:quantum_Q}) is replaced by $\langle\epsilon^{\mathrm{in}}_L\rangle-\langle\partial_t(\bar{\rho}_\ell|\tilde{{\bf v}}_\ell|^2/2)\rangle$.

\subsection{Proof of the existence of a double-cascade process\label{Proof of the existence of a double-cascade process}}
\subsubsection{Existence of a double-cascade process}
We first note that the contribution to the energy transfer from the baropycnal work $\Lambda^{(p)}_\ell$ can be ignored because it converges to zero as $\ell/L\rightarrow0$, as shown in (\ref{eq:Estimation of baropycnal work_naive}).
Therefore, ``Kolmogorov's $4/5$-law'' (\ref{eq:quantum_Q_inertial_range}) can be further approximated as
\begin{eqnarray}
\langle\Pi_\ell\rangle+\langle\Lambda^{(\Sigma)}_\ell\rangle\approx\epsilon_{\mathrm{eff}}.
\end{eqnarray}

From (\ref{eq:Estimation of the second term_naive}) and (\ref{eq:Estimation of the last term_naive}), it immediately follows that the upper bounds of the mean deformation work $\langle\Pi_\ell\rangle$ and mean quantum baropycnal work $\langle\Lambda^{(\Sigma)}_\ell\rangle$ have different $\ell$ dependences.
In the case of $h=1/3$,
\begin{eqnarray}
\langle\Pi_\ell\rangle&=&O(1),\\
\langle\Lambda^{(\Sigma)}_\ell\rangle&=&O(\ell^{-2/3}),
\end{eqnarray}
whereas in the case of $h=1$,
\begin{eqnarray}
\langle\Pi_\ell\rangle&=&O(\ell^2),\\
\langle\Lambda^{(\Sigma)}_\ell\rangle&=&O(1).
\end{eqnarray}

\begin{figure}[t]
\includegraphics[width=8.6cm]{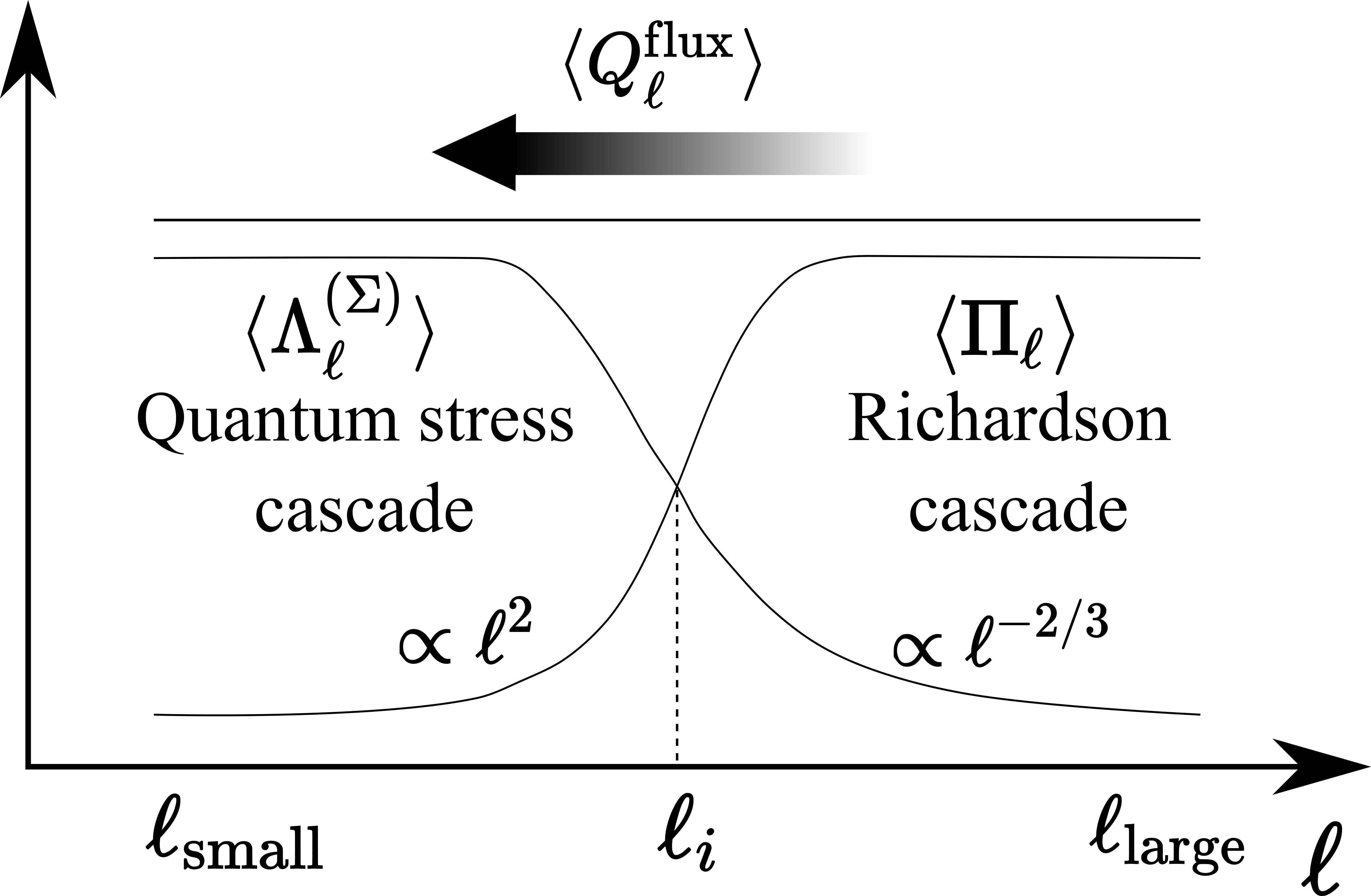}
\caption{Scale dependence of the scale-to-scale kinetic energy fluxes. The solid lines indicate the upper bounds of the energy fluxes, and the arrow indicates the direction of energy transfer.}
\label{fig:crossover}
\end{figure}

From the above observation and the fact that the sum of the mean deformation work and mean quantum baropycnal work $\langle\Pi_\ell\rangle+\langle\Lambda^{(\Sigma)}_\ell\rangle$ becomes scale-independent in the inertial range $\ell_{\mathrm{small}}\ll\ell\ll\ell_{\mathrm{large}}$, it follows that a characteristic length scale $\lambda$ exists such that the energy cascade due to deformation work is dominant in $\lambda\ll\ell\ll\ell_{\mathrm{large}}$, whereas that due to quantum baropycnal work is dominant in $\ell_{\mathrm{small}}\ll\ell\ll\lambda$ (see Fig.~\ref{fig:crossover}):
\begin{eqnarray}
\langle\Lambda^{(\Sigma)}_\ell\rangle\ll\langle\Pi_\ell\rangle&=&O(1)\quad\text{for}\quad\lambda\ll\ell\ll\ell_{\mathrm{large}},\\
\langle\Pi_\ell\rangle\ll\langle\Lambda^{(\Sigma)}_\ell\rangle&=&O(1)\quad\text{for}\quad\ell_{\mathrm{small}}\ll\ell\ll\lambda.
\end{eqnarray}
The crossover scale $\lambda$ may be determined, for instance, as follows:
From the expressions for deformation work (\ref{eq:Deformation work}) and quantum baropycnal work (\ref{eq:Quantum baropycnal work}), and the definition of the quantum circulation $\kappa=h/m$, one obtains
\begin{equation}
\Lambda^{(\Sigma)}_\ell\sim\kappa^2\ell^{-3}\rho_0v_0,
\end{equation}
and 
\begin{equation}
\Pi_\ell\sim\rho_0\ell^{-1}v^3_0.
\end{equation}
Assuming that $\Pi_{\lambda}\sim\Lambda^{(\Sigma)}_{\lambda}$ for some intermediate scale $\lambda$, we obtain $\lambda\sim\kappa/v_0$.
Note that $\lambda$ is of the order of the mean intervortex distance $\ell_i$, i.e., $\lambda\sim\ell_i$.

Thus, a double-cascade occurs in quantum turbulence: one in $\ell_i\ll\ell\ll\ell_{\mathrm{large}}$ and the other in $\ell_{\mathrm{small}}\ll\ell\ll\ell_i$.
The former is the Richardson cascade, induced by deformation work, as in the case of classical turbulence.
The latter is the quantum stress cascade, induced by quantum baropycnal work, the existence of which is specific to quantum turbulence.

\subsubsection{Velocity power spectrum}
We consider the $p$th-order (absolute) structure function for the velocity field
\begin{equation}
S^v_p(\ell):=\langle|\delta{\bf v}(\ell)|^p\rangle,
\label{S_p}
\end{equation} 
with assumed scaling exponent $\zeta_p$:
\begin{equation}
S^v_p(\ell)\sim C_pv^p_0\left(\dfrac{\ell}{L}\right)^{\zeta_p}\quad\text{as}\quad\ell/L\rightarrow0,
\label{S_p scaling}
\end{equation}
where $C_p$ is a dimensionless constant.
Note that the second-order structure function $S^v_2(\ell)\propto\ell^{\zeta_2}$ is also related to the spectrum of the velocity field $E^v(k)\propto k^{-\zeta_2-1}$, if isotropy is assumed.

In $\ell_i\ll\ell\ll \ell_{\mathrm{large}}$, where the Richardson cascade is dominant, the velocity power spectrum is expected to follow the Kolmogorov spectrum, i.e., $E^v(k)\propto k^{-5/3}$.
In fact, as $h=1/3$ in $\ell_i\ll\ell\ll \ell_{\mathrm{large}}$, we obtain $\zeta_2\approx 2h=2/3$.
Therefore, the velocity power spectrum exhibits the following asymptotic behavior:
\begin{equation}
E^v(k)\sim C_{\mathrm{large}} k^{-5/3}\quad\text{for}\quad \ell^{-1}_{\mathrm{large}}\ll k\ll\ell^{-1}_i,
\label{eq:energy spectrum bound}
\end{equation}
where $C_{\mathrm{large}}$ is a positive constant.

In $\ell_{\mathrm{small}}\ll\ell\ll\ell_i$, where the quantum stress cascade becomes dominant, we obtain $\zeta_2=2$ because $h=1$ in this scale range.
This result implies that the velocity power spectrum exhibits the following asymptotic behavior:
\begin{equation}
E^v(k)\sim C_{\mathrm{small}}k^{-3}\quad\text{for}\quad \ell^{-1}_i\ll k\ll\ell^{-1}_{\mathrm{small}},
\label{eq:energy spectrum bound_2}
\end{equation}
where $C_{\mathrm{small}}$ is a positive constant.

\section{Concluding remarks\label{Concluding remarks}}
\subsection{Summary of results}
In this paper, we investigated the contribution of quantum stress to energy transfer across scales in quantum turbulence by applying the Onsager ``ideal turbulence'' theory.
First, we showed that two types of scale-to-scale energy flux contribute to the energy cascade: the energy flux that is the same as in classical turbulence, i.e., deformation work, and that due to quantum stress, i.e., quantum baropycnal work.
We then derived ``Kolmogorov's $4/5$-law'' for quantum turbulence with the proper definition of the ``inertial range'' $\ell_{\mathrm{small}}\ll\ell\ll\ell_{\mathrm{large}}$, where $\ell_{\mathrm{large}}$ is determined through pressure-dilatation and $\ell_{\mathrm{small}}$ through quantum-stress--strain.
Using the ``$4/5$-law'' and the scale dependence of the two energy fluxes, we established a double-cascade scenario comprising the Richardson cascade and the quantum stress cascade; the Richardson cascade, induced by deformation work, becomes dominant in $\ell_i\ll\ell\ll\ell_{\mathrm{large}}$, whereas the quantum stress cascade, induced by quantum baropycnal work, develops in $\ell_{\mathrm{small}}\ll\ell\ll\ell_i$.
The advantage of the analysis presented in this paper is that it can comprehensively and rigorously discuss the double-cascade process consisting of the Richardson cascade and the Kelvin-wave cascade, the existence of which has been predicted.

\subsection{Implication and discussion of results}
\subsubsection{The role of Kelvin waves}
It has been proposed that at scales smaller than the mean intervortex distance, the Kelvin-wave cascade occurs. 
Because both the quantum stress cascade and Kelvin-wave cascade occur at scales sufficiently smaller than the mean intervortex distance and Kelvin waves are accompanied by rapid density changes, we expect that the quantum stress cascade is related to the Kelvin-wave cascade.
We emphasize that the present analysis does not use any property of the Kelvin wave.
In fact, as can be seen from the calculation in Section \ref{Quantum baropycnal work}, the quantum stress cascade can occur if at least the quantum stress is somewhat singular, i.e., the spatial derivative of the density field is singular.
This implies that the existence of Kelvin waves may not be a prerequisite for the existence of the quantum stress cascade.
However, there is still a possibility that the upper bound of the quantum baropycnal work (\ref{eq:Estimation of the last term_naive}) can be optimized further considering Kelvin-wave oscillations.
Such an effect due to Kelvin-wave oscillations may be an example of \textit{depletion of nonlinearity} that will be explained in Section \ref{Note on the spectral index of quantum stress cascade}.
Thus, there is scope for discussion on whether Kelvin waves play a key role in energy transfer at small scales, and on the elementary processes other than Kelvin waves that contribute to the quantum stress cascade.

\subsubsection{Classical-quantum crossover}
Elucidating the structure of the crossover range between the Richardson and Kelvin-wave cascades is an important research topic.
L'vov \textit{et al.}\ conjecture that a bottleneck effect exists between the two cascades, which affects the shape of the energy spectrum around the mean intervortex distance $k\approx\ell^{-1}_i$ \cite{Lvov_etal_2007,L'vov_etal_2008}.
The bottleneck effect results from the faster energy transfer by the three-dimensional Richardson cascade than by the one-dimensional Kalvin-wave cascade.
While L'vov \textit{et al.}\ ignored the effects of the reconnections both on the cascade and on the bottleneck in the crossover region, Kozik \textit{et al.}\ hypothesized that the reconnection is the key process in the crossover region and concluded that the region can be classified into three relatively narrow scale regions \cite{Kozik_etal_2008}.
It is interesting to consider whether these arguments would be modified if the Kelvin-wave cascade were to be replaced by the quantum stress cascade, and whether our theoretical approach can reveal the structure of the crossover region.

\subsubsection{Scale locality of energy cascade}
The fundamental property underlying an energy cascade is scale locality.
This means that only modes near a given scale make major contributions to the transfer of energy at that scale.
Scale locality is fundamental in the sense that it justifies the concept of scaling and universality in the inertial range.
In the theoretical framework using the smooth coarse-graining approach employed in this paper, the definition and proof of scale locality are given for the classical incompressible \cite{Eyink_2005,Eyink_Aluie_2009-1,Aluie_Eyink_2009-2} and classical compressible turbulence cases \cite{Aluie_scale_locality,Aluie_PRL}.
Following the calculations in these works, we can show that the Richardson cascade in quantum turbulence is also scale local.
However, the quantum stress cascade is only \textit{ultraviolet local} and does not satisfy the sufficient condition to be \textit{infrared local} (Appendix \ref{Scale locality of quantum baropycnal work}).
Thus, the contributions of large-scale velocity increments could be non-negligible and could contribute to quantum baropycnal work.
Note that this situation is similar to the enstrophy cascade in two-dimensional incompressible turbulence \cite{Kraichnan_1967,Kraichnan_1971,Eyink_2005}.

\subsubsection{Note on the spectral index of quantum stress cascade\label{Note on the spectral index of quantum stress cascade}}
Although the spectral index corresponding to the quantum stress cascade is naively expected as $-3$ from the upper bound of quantum baropycnal work (\ref{eq:Estimation of the last term_naive}), there is a possibility that the spectrum can become shallower; in other words, the upper bound (\ref{eq:Estimation of the last term_naive}) may be overestimated.
There are, for example, two possibilities leading to such consequences: \textit{depletion of nonlinearity} and regularity of the density gradient field.
The depletion of nonlinearity is a phenomenon that the nonlinearity is ``reduced'' because of cancellations due to wave oscillations \cite{Eyink_2018} or \textit{dynamical alignment} \cite{Boldyrev_2005,Mason_etal_2006}.
In our case, Kelvin-wave oscillations could lead to cancellations in quantum baropycnal work, so that the upper bound (\ref{eq:Estimation of the last term_naive}) is no longer optimal.
To study such cancellation effects, ``Kolmogorov's $4/5$-law'' for quantum turbulence (\ref{4/5-law}) provides the correct starting point.

Another possibility to make the spectrum shallower is the regularity of the density gradient field.
In the estimation of the upper bound of the quantum baropycnal work, we implicitly assumed that the spatial derivative of the density field satisfies
\begin{equation}
\delta(\nabla\rho)({\bf r};{\bf x})=O(|{\bf r}|^\beta),\quad \beta=0.
\end{equation}
This assumption appears reasonable considering that in the simple case in which only one straight vortex line exists, the density profile behaves like $\sqrt{\rho}(r)\propto r$ for $r\ll\xi$ and $\sqrt{\rho}(r)\simeq\mathrm{const}$ for $r\gg\xi$, where $r$ is the radial coordinate from the vortex center.
However, it may also be possible that the spatial derivative of the density field satisfies the H\"older condition
\begin{equation}
\delta(\nabla\rho)({\bf r};{\bf x})=O(|{\bf r}|^\beta),\quad \beta\in(0,1).
\end{equation}
In this case, the upper bound of the quantum baropycnal work (\ref{eq:Estimation of the last term_naive}) can be further sharpened as follows:
\begin{equation}
\Lambda^{(\Sigma)}_\ell=O(\ell^{h+\beta-1}).
\end{equation}
Therefore, the upper bound becomes scale-independent in the case of $h=1-\beta$.
Hence, the asymptotic behavior of the velocity power spectrum can be estimated as
\begin{equation}
E^v(k)\propto k^{-3+2\beta}\quad\text{for}\quad\ell^{-1}_i\ll k\ll\ell^{-1}_{\mathrm{small}}.
\end{equation}
Thus the spectral index depends on exponent $\beta$ of the H\"older condition of the spatial derivative of the density field.
Therefore, a potentially interesting research direction could involve investigating not only the energy spectrum but also the singularity of the density field through numerical calculations.

We note that there are previous studies that claim the spectral index $-3$ at scales smaller than the mean intervortex distance.
For example, Narita theoretically estimated the value by constructing the phenomenological model and using the method in magneto-fluid turbulence theory \cite{Narita_2017}.
Yepez \textit{et al.}\ numerically simulated quantum turbulence described by the GP model using a unitary quantum lattice gas algorithm and obtained the value $-3$ \cite{Yepez}.
However, it should be noted that the result of this numerical simulation is disputed because (i) the number of vortices in the initial condition is too small to consider it as fully developed turbulence and (ii) they confuse the mean intervortex distance with the vortex core radius \cite{Comment_Krstulovic_2010,Comment_L'vov_2010}.

More recent direct numerical calculation results using the GP model show values closer to $-5/3$ in decaying turbulence with no artificial dissipation \cite{Leoni}, and $-7/5$ in steady turbulence with an artificial dissipation \cite{Kobayashi-Ueda}.
However, the scale range of the Kelvin-wave cascade in these numerical calculations is not wide enough to determine the spectral index.
Therefore, high-resolution numerical calculations capable of accurately resolving the scale range $\ell_{\mathrm{small}}\ll\ell\ll\ell_i$ must be performed.

\subsubsection{Relation between quantum and classical turbulence}
On the basis of the fact that pure quantum turbulence consists of discrete vortices, quantum turbulence is sometimes described as a prototype or ``skeleton'' of turbulence that provides the simplest way of treating the turbulence problem \cite{Kobayashi_Tsubota,*Kobayashi_Tsubota_PRL,Hanninen_2014}.
Furthermore, as quantum turbulence has zero viscosity, it is naively expected to be a concrete example of Onsager's ``ideal turbulence.''
However, the present analysis reveals that it is not appropriate to describe quantum turbulence as a prototype or ``ideal turbulence.''
This is because a double-cascade process occurs and the inertial range cannot extend to infinitesimal length scales owing to the effect of the quantum-stress--strain that converts kinetic energy into interaction and quantum energies.
Therefore, it seems reasonable to consider that quantum turbulence is more exotic than classical turbulence.

At finite temperature, there is a situation where quantum turbulence is described in terms of classical turbulence \cite{Lvov_etal_2004,Volovik_2004,Baggaley_2012,Nemirovskii_2020}.
In that case, a coarse-grained description of the turbulent motion of a superfluid, such as the HVBK method, is possible.
For future research, it may be interesting to study quantum turbulence at finite temperature with the approach of this study.

\begin{acknowledgements}
The author thanks G. Eyink and S.-i. Sasa for fruitful discussions and reading of the manuscript.
In particular, G. E. suggested emphasizing the result of the ``$4/5$-law'' and pointed out the possibility of the depletion of nonlinearity.
The present study was supported by JSPS KAKENHI Grant No. 20J20079, a Grant-in-Aid for JSPS Fellows.
\end{acknowledgements}

\appendix
\section{Commutator estimates\label{Commutator estimates}}
In this section, we present results concerning the cumulant estimation obtained by Drivas and Eyink \cite{Drivas_Eyink}, which are modified to suit for our analysis.
Hereafter, we take the $d$-dimensional bounded domain $\Omega$ and consider coarse-graining of functions $f_i\in L^\infty(\Omega), i=1,2,3,\cdots$.
As $L^\infty(\Omega)\subset L^p(\Omega)$ for $p\ge1$, $f_i\in L^p(\Omega)$.

\textit{Coarse-graining cumulants} $\{\bar{\tau}_\ell(f_1,f_2,\cdots,f_n)\}_n$ are defined iteratively in $n$ by $\bar{\tau}_\ell(f):=\bar{f}_\ell$ and
\begin{equation}
\overline{(f_1f_2\cdots f_n)}_\ell=\sum_\Pi\prod^{|\Pi|}_{p=1}\bar{\tau}_\ell\left(f_{i^{(p)}_1},\cdots,f_{i^{(p)}_{n_p}}\right),
\end{equation} 
where the sum is over all the partitions $\Pi$ of the set $\{1,2,\cdots,n\}$ into $|\Pi|$ disjoint subsets $\{i^{(p)}_1,\cdots,i^{(p)}_{n_p}\}$, $p=1,\cdots,|\Pi|$.
For example, when $n=2$,
\begin{equation}
\overline{(fg)}_\ell=\bar{f}_\ell\bar{g}_\ell+\bar{\tau}_\ell(f,g).
\end{equation}

\begin{lem}
For $n>1$, the coarse-graining cumulants are related to the cumulants of difference fields $\delta f({\bf r};{\bf x}):=f({\bf x}+{\bf r})-f({\bf x})$ as follows:
\begin{equation}
\bar{\tau}_\ell(f_1,f_2,\cdots,f_n)=\langle\delta f_1,\cdots,\delta f_n\rangle^c_\ell,
\end{equation}
where $\langle\cdot\rangle_\ell$ denotes the average over displacement vector ${\bf r}$ with density $G_\ell({\bf r})$ and superscript $c$ indicates the cumulant with respect to this average.
\label{lem:2}
\end{lem}

From the above lemma, the propositions below immediately follow:
\begin{prop}[Cumulant estimates]
For $n>1$,
\begin{eqnarray}
&&|\bar{\tau}_\ell(f_1,f_2,\cdots,f_n)|=O\left(\prod^n_{i=1}|\delta f_i(\ell)|\right),
\end{eqnarray}
where $|\delta f(\ell)|:=\sup_{|{\bf r}|<\ell}|\delta f({\bf r};\cdot)|$.
\label{prop:1}
\end{prop}

This proposition can be extended to the case of $L^p$-norm.
For the $L^p$-norm, for $p\in[1,\infty]$ and $n>1$,
\begin{eqnarray}
&&\|\bar{\tau}_\ell(f_1,f_2,\cdots,f_n)\|_p=O\left(\prod^n_{i=1}\|\delta f_i(\ell)\|_{p_i}\right)\\
&&\mathrm{with}\quad\dfrac{1}{p}=\sum^n_{i=1}\dfrac{1}{p_i},\notag
\label{eq:prop cumulant estimates 1}
\end{eqnarray}
where $\|\delta f(\ell)\|_p:=\sup_{|{\bf r}|<\ell}\|\delta f({\bf r};\cdot)\|_p$.

\begin{prop}[Cumulant-derivative estimates]
For $n>1$ and $\partial_k=\partial/\partial x_k$, $k=1,2,\cdots,d$,
\begin{eqnarray}
&&|\partial_{k_1}\cdots\partial_{k_m}\bar{\tau}_\ell(f_1,f_2,\cdots,f_n)|\notag\\
&&=O\left(\ell^{-m}\prod^n_{i=1}|\delta f_i(\ell)|\right).
\end{eqnarray}
\label{prop:2}
\end{prop}
This proposition can also be extended to the case of $L^p$-norm. 
For $n>1$ and $\partial_k=\partial/\partial x_k$, $k=1,2,\cdots,d$,
\begin{eqnarray}
&&\|\partial_{k_1}\cdots\partial_{k_m}\bar{\tau}_\ell(f_1,f_2,\cdots,f_n)\|_p\notag\\
&&=O\left(\ell^{-m}\prod^n_{i=1}\|\delta f_i(\ell)\|_{p_i}\right)\quad \mathrm{with} \quad\dfrac{1}{p}=\sum^n_{i=1}\dfrac{1}{p_i}.
\end{eqnarray}

\section{More sophisticated analysis using Besov spaces\label{More sophisticated analysis using Besov spaces}}
In this section, we formulate the statement described in this paper more precisely using the $L^p$-norm.
In particular, we refine Assumption 1 in Section \ref{Assumption 1}, the investigation of the scale dependence of the energy fluxes described in Section \ref{Scale dependence of the energy fluxes}, and the estimation of the spectral index of the velocity power spectrum in Section \ref{Proof of the existence of a double-cascade process}.

\subsection{Assumption 1: Regularity of velocity and density fields\label{Appendix: Assumption 1: Regularity of velocity and density fields}}
Instead of Assumption 1 imposed in Section \ref{Assumption 1}, we assume that 
\begin{equation}
\|\delta{\bf v}({\bf r};\cdot)\|_p\sim A_pv_{\textrm{rms}}\left(\dfrac{|{\bf r}|}{L}\right)^{\sigma_p} \quad\text{as}\quad |{\bf r}|/L\rightarrow0,
\label{assumption_v}
\end{equation}
with a dimensionless constant $A_p$ for $p\in[1,\infty]$ and $\sigma_p\in(0,1]$.
Here, $\delta{\bf a}({\bf r};{\bf x}):={\bf a}({\bf x}+{\bf r})-{\bf a}({\bf x})$ for any field ${\bf a}({\bf x})$, and $a_{\textrm{rms}}:=\langle |{\bf a}|^2\rangle^{1/2}$ denotes the root-mean-square of a field ${\bf a}({\bf x})$, where the symbol $\langle\cdot\rangle$ denotes the volume average.
Additionally, the symbol $\sim$ denotes ``asymptotically equivalent,'' i.e., $f(\xi)\sim g(\xi)$ for $\xi\rightarrow0$ if and only if $\lim_{\xi\rightarrow0}f(\xi)/g(\xi)=1$; $\|\cdot\|_p$ denotes the $L^p$-norm, i.e.,
 \begin{eqnarray}
\|{\bf a}\|_p&:=&\left(\dfrac{1}{\mathcal{L}^d}\int_{\Omega}|{\bf a}({\bf x})|^pd^d{\bf x}\right)^{1/p}\notag\\
&=&\langle|{\bf a}|^p\rangle^{1/p}
\end{eqnarray} 
for $p\in[0,\infty)$, and
\begin{equation}
\|{\bf a}\|_\infty:=\inf I_{{\bf a}}=\mathrm{ess}\sup|{\bf a}|
\end{equation}
for $p=\infty$, where $I_{{\bf a}}$ denotes the semi-infinite interval
\begin{equation}
I_{{\bf a}}:=\{c\in[0,\infty)|\text{the set $\{|{\bf a}|>c\}$ is of measure zero}\}.
\end{equation}
For a matrix $A=(a_{ij})$, we define its $L^p$-norm $\|A\|_p=\langle|A|^p\rangle^{1/p}$ via the Frobenius norm $|A|:=\sqrt{\sum_i\sum_j|a_{ij}|^2}$.
We note that the relation (\ref{assumption_v}) can be interpreted as the H\"older condition with index $\sigma_p$, not pointwise but in the sense of spatial $p$th-order moments.
This relation corresponds to the \textit{Besov regularity} \cite{Eyink_1995,Perrier_1996}.
According to recent experiments \cite{Rusaouen_etal_2017}, the condition (\ref{assumption_v}) is expected to hold up to $p=6$ at least in the scale range $\ell_i\ll\ell\ll\ell_{\mathrm{large}}$.

In addition, we impose the following conditions for the density field:
\begin{eqnarray}
\|\delta\rho({\bf r};\cdot)\|_p&=&O\left(|{\bf r}|/L\right)\quad\text{as}\quad |{\bf r}|/L\rightarrow0,\label{eq:delta_rho}\\
\|\nabla\rho\|_\infty&<&\infty,\label{eq:nabla_sqrt_rho}\\
\|\rho\|_\infty&<&\infty,\label{eq:rho}\\
\|1/\bar{\rho}_\ell\|_\infty&\le&M<\infty \quad\text{for}\quad\ell\ge\xi, \label{eq:inverse_rho}
\end{eqnarray}
where $M$ is a positive constant.
The requirements (\ref{eq:delta_rho}) and (\ref{eq:nabla_sqrt_rho}) are reasonable because the energy density of the system (\ref{total energy}) contains the density gradient term $\propto|\nabla\rho|^2$.
In the last condition (\ref{eq:inverse_rho}), $M$ can be defined, for instance, as $M:=\sup_{\ell\ge\xi}\|1/\bar{\rho}_\ell\|_\infty$.
Note that this requirement is not strong enough to prohibit the existence of vacuum regions, $\{{\bf x}\in\Omega|\rho({\bf x})=0\}$, where the quantized vortices are located.

\subsection{Scale dependence of the energy fluxes}

\subsubsection{Deformation work}
We examine the scale $\ell$ dependence of the deformation work $\Pi_\ell$ by calculating its norm.
Using the Cauchy-Schwarz and H\"older inequalities, we obtain
\begin{eqnarray}
\|\Pi_\ell\|_{p/3}&=&\|\bar{\rho}_\ell\nabla\tilde{{\bf v}}_\ell:\tilde{\tau}_\ell({\bf v},{\bf v})\|_{p/3}\notag\\
&\le&\|\rho\|_\infty\|\nabla\tilde{{\bf v}}_\ell\|_p\|\tilde{\tau}_\ell({\bf v},{\bf v})\|_{p/2}.
\label{Pi_factors}
\end{eqnarray}

For the second factor on the right-hand side of (\ref{Pi_factors}), $\|\nabla\tilde{{\bf v}}_\ell\|_p$, using the relation
\begin{equation}
\tilde{{\bf v}}_\ell=\bar{{\bf v}}_\ell+\dfrac{\bar{\tau}_\ell(\rho,{\bf v})}{\bar{\rho}_\ell}
\label{eq:tilde_u decompose}
\end{equation}
and the Minkowski inequality, one obtains
\begin{eqnarray}
\|\nabla\tilde{{\bf v}}_\ell\|_p&=&\left\|\nabla\left(\bar{{\bf v}}_\ell+\dfrac{\bar{\tau}_\ell(\rho,{\bf v})}{\bar{\rho}_\ell}\right)\right\|_p \notag\\
&\le&\left\|\nabla\bar{{\bf v}}_\ell\right\|_p+\left\|\dfrac{1}{\bar{\rho}_\ell}\nabla\bar{\tau}_\ell(\rho,{\bf v})\right\|_p\notag\\
&&+\left\|\dfrac{\bar{\tau}_\ell(\rho,{\bf v})}{\bar{\rho}^2_\ell}\nabla\bar{\rho}_\ell\right\|_p.
\label{eq:nabla tilde u}
\end{eqnarray}
For the first term of (\ref{eq:nabla tilde u}), as $\int d^d{\bf r}\nabla G({\bf r})={\bf 0}$, note that for any locally integrable function ${\bf a}({\bf x})$,
\begin{equation}
\nabla\bar{{\bf a}}_\ell({\bf x})=-\dfrac{1}{\ell}\int_{\Omega}d^{d}{\bf r}(\nabla G)_\ell({\bf r})\delta{\bf a}({\bf r};{\bf x}).
\end{equation}
Subsequently, the triangle inequality gives
\begin{eqnarray}
\|\nabla\bar{{\bf a}}_\ell\|_p&=&\left\|\dfrac{1}{\ell}\int_{\Omega}d^{d}{\bf r}(\nabla G)_\ell({\bf r})\delta{\bf a}({\bf r};\cdot)\right\|_p \notag\\
&\le&\dfrac{1}{\ell}\int_{\Omega}d^{d}{\bf r}|(\nabla G)_\ell({\bf r})|\|\delta{\bf a}({\bf r};\cdot)\|_p \notag\\
&\le&\dfrac{\mathrm{(const)}}{\ell}\sup_{|{\bf r}|<\ell}\|\delta{\bf a}({\bf r};\cdot)\|_p,
\label{eq:estimate_nabla_a}
\end{eqnarray}
and hence
\begin{equation}
\|\nabla\bar{{\bf v}}_\ell\|_p=O\left(\dfrac{\|\delta{\bf v}(\ell)\|_p}{\ell}\right),
\label{eq:estimate_nabla_v_tilde_1}
\end{equation}
where $\|\delta{\bf a}(\ell)\|_p:=\sup_{|{\bf r}|<\ell}\|\delta{\bf a}({\bf r};\cdot)\|_p$.
For the second and last terms of (\ref{eq:nabla tilde u}),
using the assumption (\ref{eq:inverse_rho}) and Propositions \ref{prop:1} and \ref{prop:2} in Appendix \ref{Commutator estimates}, one obtains
\begin{eqnarray}
\left\|\dfrac{1}{\bar{\rho}_\ell}\nabla\bar{\tau}_\ell(\rho,{\bf v})\right\|_p&\le&\dfrac{\mathrm{(const)}}{\ell}\|1/\bar{\rho}_\ell\|_\infty\|\delta\rho(\ell)\|_\infty\|\delta{\bf v}(\ell)\|_p \notag\\
&\le&\dfrac{\mathrm{(const)}}{\ell}M\|\rho\|_\infty\|\delta{\bf v}(\ell)\|_p,
\label{eq:estimate_nabla_v_tilde_2}
\end{eqnarray}
\begin{eqnarray}
\left\|\dfrac{\bar{\tau}_\ell(\rho,{\bf v})}{\bar{\rho}^2_\ell}\nabla\bar{\rho}_\ell\right\|_p&\le&M^2\|\nabla\bar{\rho}_\ell\|_\infty\|\bar{\tau}_\ell(\rho,{\bf v})\|_p \notag\\
&\le&\dfrac{\mathrm{(const)}}{\ell}M^2\|\delta\rho(\ell)\|^2_\infty\|\delta{\bf v}(\ell)\|_p\notag\\
&\le&\mathrm{(const)}M^2\|\rho\|^2_\infty\dfrac{\|\delta{\bf v}(\ell)\|_p}{\ell}.
\label{eq:estimate_nabla_v_tilde_3}
\end{eqnarray}
Thus, combining the results (\ref{eq:estimate_nabla_v_tilde_1}), (\ref{eq:estimate_nabla_v_tilde_2}), and (\ref{eq:estimate_nabla_v_tilde_3}), we obtain
\begin{eqnarray}
\|\nabla\tilde{{\bf v}}_\ell\|_p&=&\dfrac{\|\delta{\bf v}(\ell)\|_p}{\ell}\notag\\
&&\times\Bigl[O(1)+O(M\|\rho\|_\infty)+O(M^2\|\rho\|^2_\infty)\Bigr] \notag\\
&=&O\left(\dfrac{\|\delta {\bf v}(\ell)\|_p}{\ell}\right).
\label{eq:estimate nabla_tilde_u}
\end{eqnarray}

For the last factor on the right-hand side of (\ref{Pi_factors}), $\|\tilde{\tau}_\ell({\bf v},{\bf v})\|_{p/2}$, using the relation
\begin{equation}
\tilde{\tau}_\ell({\bf v},{\bf v})=\bar{\tau}_\ell({\bf v},{\bf v})+\dfrac{1}{\bar{\rho}_\ell}\bar{\tau}_\ell(\rho,{\bf v},{\bf v})-\dfrac{1}{\bar{\rho}^2_\ell}\bar{\tau}_\ell(\rho,{\bf v})\bar{\tau}_\ell(\rho,{\bf v})
\label{eq:tilde_u_u decompose}
\end{equation}
and the Minkowski inequality, we obtain
\begin{eqnarray}
\|\tilde{\tau}_\ell({\bf v},{\bf v})\|_{p/2}&\le&\|\bar{\tau}_\ell({\bf v},{\bf v})\|_{p/2}+\left\|\dfrac{1}{\bar{\rho}_\ell}\bar{\tau}_\ell(\rho,{\bf v},{\bf v})\right\|_{p/2}\notag\\
&&+\left\|\dfrac{1}{\bar{\rho}^2_\ell}\bar{\tau}_\ell(\rho,{\bf v})\bar{\tau}_\ell(\rho,{\bf v})\right\|_{p/2}.
\end{eqnarray}
Subsequently, using the assumption (\ref{eq:inverse_rho}) and Proposition \ref{prop:1} in Appendix \ref{Commutator estimates}, one obtains
\begin{eqnarray}
\|\tilde{\tau}_\ell({\bf v},{\bf v})\|_{p/2}&=&\|\delta{\bf v}(\ell)\|^2_p\notag\\
&&\times\Bigl[O(1)+O(M\|\rho\|_\infty)+O(M^2\|\rho\|^2_\infty)\Bigr] \notag\\
&=&O\left(\|\delta{\bf v}(\ell)\|^2_p\right), \quad p\ge2.
\label{eq:estimate tilde_tau_u_u}
\end{eqnarray}

Thus, from (\ref{Pi_factors}), (\ref{eq:estimate nabla_tilde_u}), (\ref{eq:estimate tilde_tau_u_u}), and the assumptions (\ref{assumption_v}) and (\ref{eq:rho}), we finally obtain
\begin{eqnarray}
\|\Pi_\ell\|_{p/3}&=&\|\bar{\rho}_\ell\nabla\tilde{{\bf v}}_\ell:\tilde{\tau}_\ell({\bf v},{\bf v})\|_{p/3}\notag\\
&=&O\left(\dfrac{\|\delta{\bf v}(\ell)\|^3_p}{\ell}\right) \notag\\
&=&O\left(\left(\dfrac{\ell}{L}\right)^{3\sigma_p-1}\right),\quad p\ge3,
\label{eq:Estimation of the second term}
\end{eqnarray}
as a rigorous upper bound.
Note that the upper bound of (\ref{eq:Estimation of the second term}) becomes independent of $\ell$ in the case of $\sigma_p=1/3$.

\subsubsection{Baropycnal work}
Next, we investigate the scale $\ell$ dependence of baropycnal work by calculating the norm.
Using the assumption (\ref{eq:inverse_rho}) and Cauchy-Schwarz and H\"older inequalities, we obtain 
\begin{eqnarray}
\|\Lambda^{(p)}_\ell\|_{p/3}&=&\|(1/\bar{\rho}_\ell)\nabla\bar{p}_\ell\cdot\bar{\tau}_\ell(\rho,{\bf v})\|_{p/3}\notag\\
&\le&M\|\nabla\bar{p}_\ell\|_p\|\bar{\tau}_\ell(\rho,{\bf v})\|_{p/2}.
\end{eqnarray}
For $\|\nabla\bar{p}_\ell\|_p$, from the inequality (\ref{eq:estimate_nabla_a}), 
\begin{eqnarray}
\|\nabla\bar{p}_\ell\|_p
&\le&\dfrac{\mathrm{(const)}}{\ell}\|\delta p(\ell)\|_p\notag\\
&=&O\left(\dfrac{\|\delta\rho(\ell)\|_p}{\ell}\right),
\end{eqnarray}
where we have used the fact that
\begin{equation}
\|\delta p(\ell)\|_p=O(\|\delta\rho(\ell)\|_p),
\end{equation}
which follows from the definition of $p$ and the mean value theorem.
For $\|\bar{\tau}_\ell(\rho,{\bf v})\|_{p/2}$, using Proposition \ref{prop:1} in Appendix \ref{Commutator estimates}, we obtain
\begin{equation}
\|\bar{\tau}_\ell(\rho,{\bf v})\|_{p/2}=O(\|\delta\rho(\ell)\|_p\|\delta{\bf v}(\ell)\|_p).
\label{eq:estimate tau_rho_u}
\end{equation}
From the assumptions (\ref{assumption_v}) and (\ref{eq:delta_rho}), we thus obtain
\begin{eqnarray}
\|\Lambda^{(p)}_\ell\|_{p/3}&=&\|(1/\bar{\rho}_\ell)\nabla\bar{p}_\ell\cdot\bar{\tau}_\ell(\rho,{\bf v})\|_{p/3} \notag\\
&=&O\left(\dfrac{1}{\ell}\|\delta\rho(\ell)\|_p\|\delta\rho(\ell)\|_p\|\delta{\bf v}(\ell)\|_p\right) \notag\\
&=&O\left(\left(\dfrac{\ell}{L}\right)^{\sigma_p+1}\right), \quad p\ge3.
\label{eq:Estimation of baropycnal work}
\end{eqnarray}

\subsubsection{Quantum baropycnal work}
We investigate the scale $\ell$ dependence of quantum baropycnal work by calculating its norm.
From the assumption (\ref{eq:inverse_rho}) and the Cauchy-Schwarz and H\"older inequalities, we obtain
\begin{eqnarray}
\|\Lambda^{(\Sigma)}_\ell\|_{p/3}&=&\left\|(1/\bar{\rho}_\ell)\nabla\cdot\bar{{\bf \Sigma}}_\ell\cdot\bar{\tau}_\ell(\rho,{\bf v})\right\|_{p/3}\notag\\
&\le&M\|\nabla\cdot\bar{{\bf \Sigma}}_\ell\|_p\|\bar{\tau}_\ell(\rho,{\bf v})\|_{p/2}.
\label{Lambda^Sigma_factors}
\end{eqnarray}
For $\|\nabla\cdot\bar{{\bf \Sigma}}_\ell\|_p$, using the Minkowski inequality and the inequality (\ref{eq:estimate_nabla_a}), one obtains
\begin{widetext}
\begin{eqnarray}
\|\nabla\cdot\bar{{\bf \Sigma}}_\ell\|_p
&\le&\left\|\dfrac{\hbar^2}{4m^2}\nabla\Delta\bar{\rho}_\ell\right\|_p+\left\|\dfrac{\hbar^2}{m^2}\nabla\cdot\overline{(\nabla\sqrt{\rho}\nabla\sqrt{\rho})}_\ell\right\|_p \notag\\
&\le&\dfrac{\mathrm{(const)}}{\ell^3}\|\delta\rho(\ell)\|_p +\dfrac{\mathrm{(const)}}{\ell}\|\delta\rho(\ell)\|_p\|\nabla\rho\|^2_\infty.
\label{eq:estimate nabla_sigma}
\end{eqnarray}
\end{widetext}
From (\ref{eq:estimate tau_rho_u}), (\ref{Lambda^Sigma_factors}), (\ref{eq:estimate nabla_sigma}), and the assumptions (\ref{assumption_v}) and (\ref{eq:delta_rho}) (\ref{eq:nabla_sqrt_rho}) (\ref{eq:inverse_rho}), we thus obtain
\begin{eqnarray}
\|\Lambda^{(\Sigma)}_\ell\|_{p/3}&=&\left\|(1/\bar{\rho}_\ell)\nabla\cdot\bar{{\bf \Sigma}}_\ell\cdot\bar{\tau}_\ell(\rho,{\bf v})\right\|_{p/3} \notag\\
&=&O\left(\dfrac{1}{\ell^3}\|\delta\rho(\ell)\|^2_p\|\delta{\bf v}(\ell)\|_p\right)\notag\\
&&+O\left(\dfrac{1}{\ell}\|\delta\rho(\ell)\|^2_p\|\delta {\bf v}(\ell)\|_p\right) \notag\\
&=&O\left(\left(\dfrac{\ell}{L}\right)^{\sigma_p-1}\right)+O\left(\left(\dfrac{\ell}{L}\right)^{\sigma_p}\right)\notag\\
&=&O\left(\left(\dfrac{\ell}{L}\right)^{\sigma_p-1}\right),\quad p\ge3.
\label{eq:Estimation of the last term}
\end{eqnarray}
Note that the upper bound (\ref{eq:Estimation of the last term}) becomes independent of $\ell$ in the case of $\sigma_p=1$.

\subsection{Proof of the existence of a double-cascade process\label{Existence of a double-cascade process}}
From (\ref{eq:Estimation of the second term}) and (\ref{eq:Estimation of the last term}), it immediately follows that the upper bounds of the mean deformation work $\langle\Pi_\ell\rangle$ and mean quantum baropycnal work $\langle\Lambda^{(\Sigma)}_\ell\rangle$ have different $\ell$ dependences.
In the case of $\sigma_3=1/3$,
\begin{equation}
\langle\Pi_\ell\rangle\le\langle|\Pi_\ell|\rangle=\|\Pi_\ell\|_1=O(1),
\end{equation}
\begin{equation}
\langle\Lambda^{(\Sigma)}_\ell\rangle\le\langle|\Lambda^{(\Sigma)}_\ell|\rangle=\|\Lambda^{(\Sigma)}_\ell\|_1=O\left(\left(\dfrac{\ell}{L}\right)^{-2/3}\right),
\end{equation}
whereas in the case of $\sigma_3=1$,
\begin{equation}
\langle\Pi_\ell\rangle\le\langle|\Pi_\ell|\rangle=\|\Pi_\ell\|_1=O\left(\left(\dfrac{\ell}{L}\right)^2\right),
\end{equation}
\begin{equation}
\langle\Lambda^{(\Sigma)}_\ell\rangle\le\langle|\Lambda^{(\Sigma)}_\ell|\rangle=\|\Lambda^{(\Sigma)}_\ell\|_1=O(1).
\end{equation}

Therefore, we can show that the double-cascade process exists using the same argument as in Section \ref{Proof of the existence of a double-cascade process}.

\subsection{Estimation of the spectral index of the velocity power spectrum\label{Estimation of the spectral index of the velocity power spectrum}}
In compressible turbulence, we can consider the spectra of the velocity ${\bf v}$ and the density-weighted velocity, such as $\sqrt{\rho}{\bf v}$ \cite{Kida_1990}.
In classical compressible turbulence, high-resolution numerical simulations exhibited the Kolmogorov spectrum for both the velocity \cite{Aluie_2012} and density-weighted velocity power spectra \cite{Wang-Yang-Shi} in the case in which the pressure-dilatation co-spectrum assumption is satisfied.
In this subsection, we consider the spectra of both the velocity ${\bf v}$ and density-weighted velocity $\sqrt{\rho}{\bf v}$.

\subsubsection{Velocity power spectrum}
We consider the $p$th-order (absolute) structure function for the velocity field,
\begin{equation}
S^v_p(\ell):=\langle|\delta{\bf v}(\ell)|^p\rangle=\|\delta{\bf v}(\ell)\|^p_p
\label{S_p}
\end{equation} 
with assumed scaling exponent $\zeta_p$:
\begin{equation}
S^v_p(\ell)\sim C_pv^p_0\left(\dfrac{\ell}{L}\right)^{\zeta_p}\quad\text{as}\quad\ell/L\rightarrow0,
\label{S_p scaling}
\end{equation}
where $C_p$ is a dimensionless constant.
Using the H\"older inequality, it can be shown that $\zeta_p$ is a concave function of $p\in[0,\infty)$ \cite{Frisch,Eyink_lecture}.
From this property, it immediately follows that $\sigma_p=\zeta_p/p$ is a non-increasing function of $p$ \cite{Eyink_lecture}.
Note that the second-order structure function $S^v_2(\ell)\propto\ell^{\zeta_2}$ is related to the velocity spectrum $E^v(k)\propto k^{-\zeta_2-1}$, assuming isotropy.

Because $\sigma_3=1/3$ in $\ell_i\ll\ell\ll \ell_{\mathrm{large}}$ and $\sigma_p$ is a non-increasing function of $p$, it follows that $\sigma_2\ge1/3$ in this scale range.
Hence, we can write $\zeta_2=2\sigma_2\equiv2/3+\mu/9$, where $\mu$ is a positive constant.
This additional constant $\mu$ formally corresponds to the so-called \textit{intermittency exponent} \cite{Frisch}.
Therefore, the velocity power spectrum exhibits the following asymptotic behavior:
\begin{equation}
E^v(k)\sim C_{\mathrm{large}} k^{-5/3-\mu/9}\quad\text{for}\quad \ell^{-1}_{\mathrm{large}}\ll k\ll\ell^{-1}_i,
\label{eq:energy spectrum bound}
\end{equation}
where $C_{\mathrm{large}}$ is a positive constant.

In $\ell_{\mathrm{small}}\ll\ell\ll\ell_i$, where the quantum stress cascade becomes dominant, $\sigma_3=1$.
Because $\sigma_p$ is a non-increasing function of $p$, it follows that $\sigma_2=1$.
This result implies that the velocity power spectrum exhibits the following asymptotic behavior:
\begin{equation}
E^v(k)\sim C_{\mathrm{small}}k^{-3}\quad\text{for}\quad \ell^{-1}_i\ll k\ll\ell^{-1}_{\mathrm{small}},
\label{eq:energy spectrum bound_2}
\end{equation}
where $C_{\mathrm{small}}$ is a positive constant.

\subsubsection{Density-weighted velocity power spectrum}
Next, we consider the spectrum of the density-weighted velocity $\sqrt{\rho}{\bf v}$.
This quantity has been thoroughly investigated in numerical simulations of quantum turbulence described by the GP model \cite{Kobayashi_Tsubota,Kobayashi_Tsubota_PRL,Kobayashi-Ueda,Leoni}.
We consider the $p$th-order (absolute) structure function for density-weighted velocity
\begin{equation}
S^{\sqrt{\rho}v}_p(\ell):=\langle|\delta(\sqrt{\rho}{\bf v})(\ell)|^p\rangle=\|\delta(\sqrt{\rho}{\bf v})(\ell)\|^p_p,
\end{equation} 
with assumed scaling exponent $\tilde{\zeta}_p$:
\begin{equation}
S^{\sqrt{\rho}v}_p(\ell)\sim \tilde{C}_p\rho^{p/2}v^p_0\left(\dfrac{\ell}{L}\right)^{\tilde{\zeta}_p}\quad\text{as}\quad\ell/L\rightarrow0,
\end{equation}
where $\tilde{C}_p$ is a dimensionless constant.
Using the H\"older inequality, one can also show that $\tilde{\zeta}_p$ is a concave function of $p\in[0,\infty)$ \cite{Frisch,Eyink_lecture}.
Note that the second-order structure function $S^{\sqrt{\rho}v}_2(\ell)\propto\ell^{\tilde{\zeta}_2}$ is also related to the density-weighted velocity spectrum $E(k)\propto k^{-\tilde{\zeta}_2-1}$, assuming isotropy.

In this case we cannot determine the exact value of $\tilde{\zeta}_2$ because, from the mean value theorem, Minkowski inequality, and assumptions (\ref{assumption_v}) and (\ref{eq:delta_rho}),
\begin{eqnarray}
\|\delta(\sqrt{\rho}{\bf v})({\bf r};\cdot)\|_p&\le&B_1\|\delta\rho({\bf r};\cdot)\|_p+B_2\|\delta{\bf v}({\bf r};\cdot)\|_p\notag\\
&\sim&B_3v_0\left(\dfrac{|{\bf r}|}{L}\right)^{\sigma_p}\quad\text{as}\quad|{\bf r}|/L\rightarrow0,\notag
\end{eqnarray}
where $B_1$, $B_2$, and $B_3$ are constants.
Hence, 
\begin{equation}
S^{\sqrt{\rho}v}_p(\ell)=\|\delta(\sqrt{\rho}{\bf v})(\ell)\|^p_p=O\left(\left(\dfrac{\ell}{L}\right)^{p\sigma_p}\right),
\end{equation}
and we cannot conclude that $\tilde{\zeta}_p=p\sigma_p$ in general.
If we assume that $\zeta_2\approx\tilde{\zeta}_2$ as in classical compressible turbulence, the asymptotic behavior of the kinetic energy spectrum $E(k)$ can be obtained as
\begin{equation}
E(k)\sim 
\begin{cases}
\tilde{C}_{\mathrm{large}} k^{-5/3-\mu/9}\quad\text{for}\quad \ell^{-1}_{\mathrm{large}}\ll k\ll\ell^{-1}_i,\\
\tilde{C}_{\mathrm{small}}k^{-3}\quad\text{for}\quad \ell^{-1}_i\ll k\ll\ell^{-1}_{\mathrm{small}},
\end{cases}
\end{equation}
where $\tilde{C}_{\mathrm{large}}$ and $\tilde{C}_{\mathrm{small}}$ are positive constants.

\section{Scale locality of quantum baropycnal work\label{Scale locality of quantum baropycnal work}}
In proving the scale locality of the quantum baropycnal work $\Lambda^{(\Sigma)}_\ell=-(1/\bar{\rho}_\ell)\nabla\cdot\bar{{\bf \Sigma}}_\ell\cdot\bar{\tau}_\ell(\rho,{\bf v})$, it is not necessary to consider $\nabla\sqrt{\rho}\nabla\sqrt{\rho}$ in the quantum stress because its contribution to the energy flux vanishes as $\ell/L\rightarrow0$ (see Section \ref{Quantum baropycnal work}).
Therefore, it is sufficient to prove the scale locality of 
\begin{equation}
Z_\ell(\rho,\rho,\rho,{\bf v}):=-(\hbar^2/4m^2\bar{\rho}_\ell)\nabla\Delta\bar{\rho}_\ell\cdot\bar{\tau}_\ell(\rho,{\bf v}),
\end{equation}
where the first density argument corresponds to the factor $1/\bar{\rho}_\ell$, and the second to the factor $\nabla\Delta\bar{\rho}_\ell$.
Following Eyink \cite{Eyink_2005}, we describe the energy flux $Z_\ell(\rho,\rho,\rho,{\bf v})$ as \textit{ultraviolet local} if $\langle Z_\ell(\rho'_\delta,\rho'_\delta,\rho'_\delta,{\bf v}'_\delta)\rangle$ decays as fast as $(\delta/\ell)^\alpha$, for $\alpha>0$, whenever $\delta\ll\ell$.
Similarly, we describe the energy flux $Z_\ell(\rho,\rho,\rho,{\bf v})$ as \textit{infrared local} if $\langle Z_\ell(\rho,\rho,\rho,\bar{{\bf v}}_\Delta)\rangle$ decays as fast as $(\Delta/\ell)^{-\alpha}$, for $\alpha>0$, whenever $\Delta\gg\ell$.
As observed by Aluie \cite{Aluie_scale_locality,Aluie_PRL}, in defining the infrared locality of the energy flux, the condition of negligible contribution of the large-scale density field to the flux is not necessary.
This is reasonable considering that, in the case of incompressible turbulence, the energy flux directly depends on the large-scale (${\bf k}={\bf 0}$) density field.

\subsection{Ultraviolet locality}
It is obvious that $1/\bar{\rho}_\ell$ has a vanishing contribution from small scales $\delta\ll\ell$ because its Fourier amplitudes decay faster than any power $n$ of wavenumber $k^{-n}$ as $k\rightarrow\infty$ as a direct consequence of the Riemann-Lebesgue lemma.
For the remaining three arguments, using the assumptions (\ref{assumption_v}), (\ref{eq:delta_rho}), and (\ref{eq:inverse_rho}) and the H\"older inequality, we obtain
\begin{eqnarray}
&&\|Z_\ell(\rho,\rho'_\delta,\rho'_\delta,{\bf v}'_\delta)\|_{p/3}\notag\\
&=&\left\|(\hbar^2/4m^2\bar{\rho}_\ell)\nabla\Delta\overline{(\rho'_\delta)}_\ell\cdot\bar{\tau}_\ell(\rho'_\delta,{\bf v}'_\delta)\right\|_{p/3}\notag\\
&\le&(\mathrm{const})\left(\dfrac{1}{\ell^3}\int_\Omega d^d{\bf r}|(\nabla\Delta G)_\ell({\bf r})|\|\rho'_\delta\|_p\right)\|\bar{\tau}_\ell(\rho'_\delta,{\bf v}'_\delta)\|_{p/2}\notag\\
&\le&(\mathrm{const})\dfrac{1}{\ell^3}\|\rho'_\delta\|_p\|\rho'_\delta\|_p\|{\bf v}'_\delta\|_p\notag\\
&=&O\left(\left(\dfrac{\ell}{L}\right)^{\sigma_p-1}\left(\dfrac{\delta}{\ell}\right)^{\sigma_p+2}\right), \quad p\ge3.
\end{eqnarray}
Here, we have used the fact that
\begin{eqnarray}
\|{\bf a}'_\delta\|_p&\le&\int_\Omega d^d{\bf r}G_\delta({\bf r})\|\delta{\bf a}({\bf r};\cdot)\|_p\notag\\
&\le&\int_\Omega d^d{\bf r}G_\delta({\bf r})\|\delta{\bf a}(\delta)\|_p.
\end{eqnarray}
Therefore, $Z_\ell$ is ultraviolet local.

\subsection{Infrared locality}
Using the assumptions (\ref{assumption_v}), (\ref{eq:delta_rho}), and (\ref{eq:inverse_rho}) and the H\"older inequality, we obtain
\begin{eqnarray}
&&\|Z_\ell(\rho,\rho,\rho,\bar{{\bf v}}_\Delta)\|_{p/3}\notag\\
&=&\left\|(\hbar^2/4m^2\bar{\rho}_\ell)\nabla\Delta\bar{\rho}_\ell\cdot\bar{\tau}_\ell(\rho,\bar{{\bf v}}_\Delta))\right\|_{p/3}\notag\\
&\le&(\mathrm{const})\dfrac{1}{\ell^3}\|\delta\rho(\ell)\|_p\|\delta\rho(\ell)\|_p\|\delta\bar{{\bf v}}_\Delta(\ell)\|_p\notag\\
&=&O\left(\left(\dfrac{\ell}{L}\right)^{\sigma_p-1}\left(\dfrac{\Delta}{\ell}\right)^{\sigma_p-1}\right),\quad p\ge3.
\end{eqnarray}
Here, we have used the following evaluation \cite{Eyink_2005}: 
\begin{widetext}
\begin{eqnarray}
\|\delta\bar{{\bf v}}_\Delta(\ell)\|_p&\le&\sup_{|{\bf \rho}|<\ell}\left\|\int_\Omega d^d{\bf r}G_\Delta({\bf r})\left({\bf v}({\bf x}+{\bf \rho}+{\bf r})-{\bf v}({\bf x}+{\bf r})\right)\right\|_p\notag\\
&=&\sup_{|{\bf \rho}|<\ell}\left\|\int_\Omega d^d{\bf r}\left(G_\Delta({\bf r}-{\bf \rho})-G_\Delta({\bf r})\right){\bf v}({\bf x}+{\bf r})\right\|_p\notag\\
&=&\sup_{|{\bf \rho}|<\ell}\left\|\dfrac{1}{\Delta}\int^1_0d\theta\int_\Omega d^d{\bf r}{\bf \rho}\cdot(\nabla G)_\Delta({\bf r}-\theta{\bf \rho})\delta{\bf v}({\bf r};{\bf x})\right\|_p\notag\\
&\le&\dfrac{\ell}{\Delta}\int^1_0d\theta\int_\Omega d^d{\bf r}|(\nabla G)_\Delta({\bf r})|\|\delta{\bf v}(\Delta)\|_p\notag\\
&=&O\left(\left(\dfrac{\ell}{L}\right)\left(\dfrac{\Delta}{L}\right)^{\sigma_p-1}\right).
\end{eqnarray}
\end{widetext}
Therefore, in the scale range $\ell_{\mathrm{small}}\ll\ell\ll\Delta\ll\ell_i$,
\begin{eqnarray}
\langle Z_\ell(\rho,\rho,\rho,\bar{{\bf v}}_\Delta)\rangle&\le&\|Z_\ell(\rho,\rho,\rho,\bar{{\bf v}}_\Delta)\|_1\notag\\
&=&O(1)\quad\text{for}\quad \ell\ll\Delta,
\end{eqnarray}
because $\sigma_3=1$ in this range.
Thus, $Z_\ell$ does not satisfy the sufficient condition to be infrared local.

\bibliography{Theoretical_analysis_of_quantum_turbulence_using_the_Onsager_ideal_turbulence_theory}

\begin{thebibliography}{94}%
\makeatletter
\providecommand \@ifxundefined [1]{%
 \@ifx{#1\undefined}
}%
\providecommand \@ifnum [1]{%
 \ifnum #1\expandafter \@firstoftwo
 \else \expandafter \@secondoftwo
 \fi
}%
\providecommand \@ifx [1]{%
 \ifx #1\expandafter \@firstoftwo
 \else \expandafter \@secondoftwo
 \fi
}%
\providecommand \natexlab [1]{#1}%
\providecommand \enquote  [1]{``#1''}%
\providecommand \bibnamefont  [1]{#1}%
\providecommand \bibfnamefont [1]{#1}%
\providecommand \citenamefont [1]{#1}%
\providecommand \href@noop [0]{\@secondoftwo}%
\providecommand \href [0]{\begingroup \@sanitize@url \@href}%
\providecommand \@href[1]{\@@startlink{#1}\@@href}%
\providecommand \@@href[1]{\endgroup#1\@@endlink}%
\providecommand \@sanitize@url [0]{\catcode `\\12\catcode `\$12\catcode
  `\&12\catcode `\#12\catcode `\^12\catcode `\_12\catcode `\%12\relax}%
\providecommand \@@startlink[1]{}%
\providecommand \@@endlink[0]{}%
\providecommand \url  [0]{\begingroup\@sanitize@url \@url }%
\providecommand \@url [1]{\endgroup\@href {#1}{\urlprefix }}%
\providecommand \urlprefix  [0]{URL }%
\providecommand \Eprint [0]{\href }%
\providecommand \doibase [0]{https://doi.org/}%
\providecommand \selectlanguage [0]{\@gobble}%
\providecommand \bibinfo  [0]{\@secondoftwo}%
\providecommand \bibfield  [0]{\@secondoftwo}%
\providecommand \translation [1]{[#1]}%
\providecommand \BibitemOpen [0]{}%
\providecommand \bibitemStop [0]{}%
\providecommand \bibitemNoStop [0]{.\EOS\space}%
\providecommand \EOS [0]{\spacefactor3000\relax}%
\providecommand \BibitemShut  [1]{\csname bibitem#1\endcsname}%
\let\auto@bib@innerbib\@empty
\bibitem [{\citenamefont {Onsager}(1949)}]{Onsager_1949}%
  \BibitemOpen
  \bibfield  {author} {\bibinfo {author} {\bibfnamefont {L.}~\bibnamefont
  {Onsager}},\ }\bibfield  {title} {\bibinfo {title} {Statistical
  {H}ydrodynamics},\ }\href@noop {} {\bibfield  {journal} {\bibinfo  {journal}
  {Nuovo Cimento Suppl.}\ }\textbf {\bibinfo {volume} {6}},\ \bibinfo {pages}
  {279} (\bibinfo {year} {1949})}\BibitemShut {NoStop}%
\bibitem [{\citenamefont {Eyink}(2018{\natexlab{a}})}]{Eyink_Review}%
  \BibitemOpen
  \bibfield  {author} {\bibinfo {author} {\bibfnamefont {G.~L.}\ \bibnamefont
  {Eyink}},\ }\bibfield  {title} {\bibinfo {title} {Review of the {O}nsager
  ``{I}deal {T}urbulence'' {T}heory},\ }\href@noop {} {\bibfield  {journal}
  {\bibinfo  {journal} {arXiv preprint arXiv:1803.02223}\ } (\bibinfo {year}
  {2018}{\natexlab{a}})}\BibitemShut {NoStop}%
\bibitem [{\citenamefont {Eyink}\ and\ \citenamefont
  {Sreenivasan}(2006)}]{Eyink_Sreenivasan}%
  \BibitemOpen
  \bibfield  {author} {\bibinfo {author} {\bibfnamefont {G.~L.}\ \bibnamefont
  {Eyink}}\ and\ \bibinfo {author} {\bibfnamefont {K.~R.}\ \bibnamefont
  {Sreenivasan}},\ }\bibfield  {title} {\bibinfo {title} {Onsager and the
  theory of hydrodynamic turbulence},\ }\href@noop {} {\bibfield  {journal}
  {\bibinfo  {journal} {Rev. Mod. Phys.}\ }\textbf {\bibinfo {volume} {78}},\
  \bibinfo {pages} {87} (\bibinfo {year} {2006})}\BibitemShut {NoStop}%
\bibitem [{\citenamefont {Richardson}(1922)}]{Richardson_1922}%
  \BibitemOpen
  \bibfield  {author} {\bibinfo {author} {\bibfnamefont {L.~F.}\ \bibnamefont
  {Richardson}},\ }\href@noop {} {\emph {\bibinfo {title} {Weather prediction
  by numerical process}}}\ (\bibinfo  {publisher} {Cambridge university
  press},\ \bibinfo {year} {1922})\BibitemShut {NoStop}%
\bibitem [{\citenamefont {Kolmogorov}(1941{\natexlab{a}})}]{K41_a}%
  \BibitemOpen
  \bibfield  {author} {\bibinfo {author} {\bibfnamefont {A.~N.}\ \bibnamefont
  {Kolmogorov}},\ }\bibfield  {title} {\bibinfo {title} {The local structure of
  turbulence in incompressible viscous fluid for very large {R}eynolds
  numbers},\ }\href@noop {} {\bibfield  {journal} {\bibinfo  {journal} {Dokl.
  Akad. Nauk SSSR}\ }\textbf {\bibinfo {volume} {30}},\ \bibinfo {pages} {9}
  (\bibinfo {year} {1941}{\natexlab{a}})}\BibitemShut {NoStop}%
\bibitem [{\citenamefont {Kolmogorov}(1941{\natexlab{b}})}]{K41_b}%
  \BibitemOpen
  \bibfield  {author} {\bibinfo {author} {\bibfnamefont {A.~N.}\ \bibnamefont
  {Kolmogorov}},\ }\bibfield  {title} {\bibinfo {title} {On degeneration of
  isotropic turbulence in an incompressible viscous liquid},\ }\href@noop {}
  {\bibfield  {journal} {\bibinfo  {journal} {Dokl. Akad. Nauk SSSR}\ }\textbf
  {\bibinfo {volume} {31}},\ \bibinfo {pages} {538} (\bibinfo {year}
  {1941}{\natexlab{b}})}\BibitemShut {NoStop}%
\bibitem [{\citenamefont {Kolmogorov}(1941{\natexlab{c}})}]{K41_c}%
  \BibitemOpen
  \bibfield  {author} {\bibinfo {author} {\bibfnamefont {A.~N.}\ \bibnamefont
  {Kolmogorov}},\ }\bibfield  {title} {\bibinfo {title} {Dissipation of energy
  in locally isotropic turbulence},\ }\href@noop {} {\bibfield  {journal}
  {\bibinfo  {journal} {Dokl. Akad. Nauk SSSR}\ }\textbf {\bibinfo {volume}
  {32}},\ \bibinfo {pages} {16} (\bibinfo {year}
  {1941}{\natexlab{c}})}\BibitemShut {NoStop}%
\bibitem [{\citenamefont {Taylor}(1917)}]{Taylor_1917}%
  \BibitemOpen
  \bibfield  {author} {\bibinfo {author} {\bibfnamefont {G.~I.}\ \bibnamefont
  {Taylor}},\ }\bibfield  {title} {\bibinfo {title} {Observations and
  {S}peculations on the {N}ature of {T}urbulent {M}otion},\ }in\ \href@noop {}
  {\emph {\bibinfo {booktitle} {Scientific Papers}}},\ Vol.~\bibinfo {volume}
  {2},\ \bibinfo {editor} {edited by\ \bibinfo {editor} {\bibfnamefont {G.~K.}\
  \bibnamefont {Batchelor}}}\ (\bibinfo  {publisher} {Cambridge University
  Press},\ \bibinfo {year} {1917})\ p.~\bibinfo {pages} {69}\BibitemShut
  {NoStop}%
\bibitem [{\citenamefont {Eyink}(1995{\natexlab{a}})}]{Eyink_1995_2D}%
  \BibitemOpen
  \bibfield  {author} {\bibinfo {author} {\bibfnamefont {G.~L.}\ \bibnamefont
  {Eyink}},\ }\bibfield  {title} {\bibinfo {title} {Exact {R}esults on
  {S}caling {E}xponents in the 2{D} {E}nstrophy {C}ascade},\ }\href@noop {}
  {\bibfield  {journal} {\bibinfo  {journal} {Phys. Rev. Lett.}\ }\textbf
  {\bibinfo {volume} {74}},\ \bibinfo {pages} {3800} (\bibinfo {year}
  {1995}{\natexlab{a}})}\BibitemShut {NoStop}%
\bibitem [{\citenamefont {Eyink}(2001)}]{Eyink_2001_2D}%
  \BibitemOpen
  \bibfield  {author} {\bibinfo {author} {\bibfnamefont {G.~L.}\ \bibnamefont
  {Eyink}},\ }\bibfield  {title} {\bibinfo {title} {Dissipation in turbulent
  solutions of 2{D} {E}uler equations},\ }\href@noop {} {\bibfield  {journal}
  {\bibinfo  {journal} {Nonlinearity}\ }\textbf {\bibinfo {volume} {14}},\
  \bibinfo {pages} {787} (\bibinfo {year} {2001})}\BibitemShut {NoStop}%
\bibitem [{\citenamefont {Chae}(2003)}]{Chae_2003}%
  \BibitemOpen
  \bibfield  {author} {\bibinfo {author} {\bibfnamefont {D.}~\bibnamefont
  {Chae}},\ }\bibfield  {title} {\bibinfo {title} {Remarks on the {H}elicity of
  the 3-{D} {I}ncompressible {E}uler {E}quations},\ }\href@noop {} {\bibfield
  {journal} {\bibinfo  {journal} {Commun. Math. Phys.}\ }\textbf {\bibinfo
  {volume} {240}},\ \bibinfo {pages} {501} (\bibinfo {year}
  {2003})}\BibitemShut {NoStop}%
\bibitem [{\citenamefont {Chen}\ \emph {et~al.}(2003)\citenamefont {Chen},
  \citenamefont {Chen},\ and\ \citenamefont {Eyink}}]{Chen_etal_2003}%
  \BibitemOpen
  \bibfield  {author} {\bibinfo {author} {\bibfnamefont {Q.}~\bibnamefont
  {Chen}}, \bibinfo {author} {\bibfnamefont {S.}~\bibnamefont {Chen}},\ and\
  \bibinfo {author} {\bibfnamefont {G.~L.}\ \bibnamefont {Eyink}},\ }\bibfield
  {title} {\bibinfo {title} {The {J}oint {C}ascade of {E}nergy and {H}elicity
  in {T}hree-{D}imensional {T}urbulence},\ }\href@noop {} {\bibfield  {journal}
  {\bibinfo  {journal} {Phys. Fluids}\ }\textbf {\bibinfo {volume} {15}},\
  \bibinfo {pages} {361} (\bibinfo {year} {2003})}\BibitemShut {NoStop}%
\bibitem [{\citenamefont {Caflisch}\ \emph {et~al.}(1997)\citenamefont
  {Caflisch}, \citenamefont {Klapper},\ and\ \citenamefont
  {Steele}}]{Caflisch_etal_1997}%
  \BibitemOpen
  \bibfield  {author} {\bibinfo {author} {\bibfnamefont {R.~E.}\ \bibnamefont
  {Caflisch}}, \bibinfo {author} {\bibfnamefont {I.}~\bibnamefont {Klapper}},\
  and\ \bibinfo {author} {\bibfnamefont {G.}~\bibnamefont {Steele}},\
  }\bibfield  {title} {\bibinfo {title} {Remarks on {S}ingularities,
  {D}imension and {E}nergy {D}issipation for {I}deal {H}ydrodynamics and
  {MHD}},\ }\href@noop {} {\bibfield  {journal} {\bibinfo  {journal} {Commun.
  Math. Phys.}\ }\textbf {\bibinfo {volume} {184}},\ \bibinfo {pages} {443}
  (\bibinfo {year} {1997})}\BibitemShut {NoStop}%
\bibitem [{\citenamefont {Aluie}\ and\ \citenamefont
  {Eyink}(2010)}]{Aluie_Eyink_2010_MHD}%
  \BibitemOpen
  \bibfield  {author} {\bibinfo {author} {\bibfnamefont {H.}~\bibnamefont
  {Aluie}}\ and\ \bibinfo {author} {\bibfnamefont {G.~L.}\ \bibnamefont
  {Eyink}},\ }\bibfield  {title} {\bibinfo {title} {Scale {L}ocality of
  {M}agnetohydrodynamic {T}urbulence},\ }\href@noop {} {\bibfield  {journal}
  {\bibinfo  {journal} {Phys. Rev. Lett.}\ }\textbf {\bibinfo {volume} {104}},\
  \bibinfo {pages} {081101} (\bibinfo {year} {2010})}\BibitemShut {NoStop}%
\bibitem [{\citenamefont {Aluie}(2017)}]{Aluie_2017}%
  \BibitemOpen
  \bibfield  {author} {\bibinfo {author} {\bibfnamefont {H.}~\bibnamefont
  {Aluie}},\ }\bibfield  {title} {\bibinfo {title} {Coarse-grained
  incompressible magnetohydrodynamics: analyzing the turbulent cascades},\
  }\href@noop {} {\bibfield  {journal} {\bibinfo  {journal} {New J. Phys.}\
  }\textbf {\bibinfo {volume} {19}},\ \bibinfo {pages} {025008} (\bibinfo
  {year} {2017})}\BibitemShut {NoStop}%
\bibitem [{\citenamefont {Feireisl}\ \emph {et~al.}(2017)\citenamefont
  {Feireisl}, \citenamefont {Gwiazda}, \citenamefont
  {{\'S}wierczewska-Gwiazda},\ and\ \citenamefont {Wiedemann}}]{Feireisl_2017}%
  \BibitemOpen
  \bibfield  {author} {\bibinfo {author} {\bibfnamefont {E.}~\bibnamefont
  {Feireisl}}, \bibinfo {author} {\bibfnamefont {P.}~\bibnamefont {Gwiazda}},
  \bibinfo {author} {\bibfnamefont {A.}~\bibnamefont
  {{\'S}wierczewska-Gwiazda}},\ and\ \bibinfo {author} {\bibfnamefont
  {E.}~\bibnamefont {Wiedemann}},\ }\bibfield  {title} {\bibinfo {title}
  {Regularity and {E}nergy {C}onservation for the {C}ompressible {E}uler
  {E}quations},\ }\href@noop {} {\bibfield  {journal} {\bibinfo  {journal}
  {Arch. Ration. Mech. An.}\ }\textbf {\bibinfo {volume} {223}},\ \bibinfo
  {pages} {1375} (\bibinfo {year} {2017})}\BibitemShut {NoStop}%
\bibitem [{\citenamefont {Eyink}\ and\ \citenamefont
  {Drivas}(2018{\natexlab{a}})}]{Eyink_Drivas}%
  \BibitemOpen
  \bibfield  {author} {\bibinfo {author} {\bibfnamefont {G.~L.}\ \bibnamefont
  {Eyink}}\ and\ \bibinfo {author} {\bibfnamefont {T.~D.}\ \bibnamefont
  {Drivas}},\ }\bibfield  {title} {\bibinfo {title} {Cascades and {D}issipative
  {A}nomalies in {C}ompressible {F}luid {T}urbulence},\ }\href@noop {}
  {\bibfield  {journal} {\bibinfo  {journal} {Phys. Rev. X}\ }\textbf {\bibinfo
  {volume} {8}},\ \bibinfo {pages} {011022} (\bibinfo {year}
  {2018}{\natexlab{a}})}\BibitemShut {NoStop}%
\bibitem [{\citenamefont {Drivas}\ and\ \citenamefont
  {Eyink}(2018)}]{Drivas_Eyink}%
  \BibitemOpen
  \bibfield  {author} {\bibinfo {author} {\bibfnamefont {T.~D.}\ \bibnamefont
  {Drivas}}\ and\ \bibinfo {author} {\bibfnamefont {G.~L.}\ \bibnamefont
  {Eyink}},\ }\bibfield  {title} {\bibinfo {title} {An {O}nsager {S}ingularity
  {T}heorem for {T}urbulent {S}olutions of {C}ompressible {E}uler
  {E}quations},\ }\href@noop {} {\bibfield  {journal} {\bibinfo  {journal}
  {Commun. Math. Phys.}\ }\textbf {\bibinfo {volume} {359}},\ \bibinfo {pages}
  {733} (\bibinfo {year} {2018})}\BibitemShut {NoStop}%
\bibitem [{\citenamefont {Aluie}(2011)}]{Aluie_PRL}%
  \BibitemOpen
  \bibfield  {author} {\bibinfo {author} {\bibfnamefont {H.}~\bibnamefont
  {Aluie}},\ }\bibfield  {title} {\bibinfo {title} {Compressible {T}urbulence:
  {T}he {C}ascade and its {L}ocality},\ }\href@noop {} {\bibfield  {journal}
  {\bibinfo  {journal} {Phys. Rev. Lett.}\ }\textbf {\bibinfo {volume} {106}},\
  \bibinfo {pages} {174502} (\bibinfo {year} {2011})}\BibitemShut {NoStop}%
\bibitem [{\citenamefont {Aluie}(2010)}]{Aluie_scale_locality}%
  \BibitemOpen
  \bibfield  {author} {\bibinfo {author} {\bibfnamefont {H.}~\bibnamefont
  {Aluie}},\ }\bibfield  {title} {\bibinfo {title} {Scale locality and the
  inertial range in compressible turbulence},\ }\href@noop {} {\bibfield
  {journal} {\bibinfo  {journal} {arXiv preprint arXiv:1101.0150}\ } (\bibinfo
  {year} {2010})}\BibitemShut {NoStop}%
\bibitem [{\citenamefont {Aluie}(2013)}]{Aluie_scale_decomposition}%
  \BibitemOpen
  \bibfield  {author} {\bibinfo {author} {\bibfnamefont {H.}~\bibnamefont
  {Aluie}},\ }\bibfield  {title} {\bibinfo {title} {Scale decomposition in
  compressible turbulence},\ }\href@noop {} {\bibfield  {journal} {\bibinfo
  {journal} {Physica D}\ }\textbf {\bibinfo {volume} {247}},\ \bibinfo {pages}
  {54} (\bibinfo {year} {2013})}\BibitemShut {NoStop}%
\bibitem [{\citenamefont {Eyink}(2018{\natexlab{b}})}]{Eyink_2018}%
  \BibitemOpen
  \bibfield  {author} {\bibinfo {author} {\bibfnamefont {G.~L.}\ \bibnamefont
  {Eyink}},\ }\bibfield  {title} {\bibinfo {title} {Cascades and {D}issipative
  {A}nomalies in {N}early {C}ollisionless {P}lasma {T}urbulence},\ }\href@noop
  {} {\bibfield  {journal} {\bibinfo  {journal} {Phys. Rev. X}\ }\textbf
  {\bibinfo {volume} {8}},\ \bibinfo {pages} {041020} (\bibinfo {year}
  {2018}{\natexlab{b}})}\BibitemShut {NoStop}%
\bibitem [{\citenamefont {Eyink}\ and\ \citenamefont
  {Drivas}(2018{\natexlab{b}})}]{Eyink_Drivas_2018_relativistic}%
  \BibitemOpen
  \bibfield  {author} {\bibinfo {author} {\bibfnamefont {G.~L.}\ \bibnamefont
  {Eyink}}\ and\ \bibinfo {author} {\bibfnamefont {T.~D.}\ \bibnamefont
  {Drivas}},\ }\bibfield  {title} {\bibinfo {title} {Cascades and {D}issipative
  {A}nomalies in {R}elativistic {F}luid {T}urbulence},\ }\href@noop {}
  {\bibfield  {journal} {\bibinfo  {journal} {Phys. Rev. X}\ }\textbf {\bibinfo
  {volume} {8}},\ \bibinfo {pages} {011023} (\bibinfo {year}
  {2018}{\natexlab{b}})}\BibitemShut {NoStop}%
\bibitem [{\citenamefont {Feynman}(1955)}]{Feynman}%
  \BibitemOpen
  \bibfield  {author} {\bibinfo {author} {\bibfnamefont {R.~P.}\ \bibnamefont
  {Feynman}},\ }\bibfield  {title} {\bibinfo {title} {Application of {Q}uantum
  {M}echanics to {L}iquid {H}elium},\ }in\ \href@noop {} {\emph {\bibinfo
  {booktitle} {Progress in low temperature physics}}},\ Vol.~\bibinfo {volume}
  {1}\ (\bibinfo  {publisher} {Elsevier},\ \bibinfo {year} {1955})\ pp.\
  \bibinfo {pages} {17--53}\BibitemShut {NoStop}%
\bibitem [{\citenamefont {Tsubota}\ \emph {et~al.}(2013)\citenamefont
  {Tsubota}, \citenamefont {Kobayashi},\ and\ \citenamefont
  {Takeuchi}}]{Tsubota_etal_2013}%
  \BibitemOpen
  \bibfield  {author} {\bibinfo {author} {\bibfnamefont {M.}~\bibnamefont
  {Tsubota}}, \bibinfo {author} {\bibfnamefont {M.}~\bibnamefont {Kobayashi}},\
  and\ \bibinfo {author} {\bibfnamefont {H.}~\bibnamefont {Takeuchi}},\
  }\bibfield  {title} {\bibinfo {title} {Quantum hydrodynamics},\ }\href@noop
  {} {\bibfield  {journal} {\bibinfo  {journal} {Phys. Rep.}\ }\textbf
  {\bibinfo {volume} {522}},\ \bibinfo {pages} {191} (\bibinfo {year}
  {2013})}\BibitemShut {NoStop}%
\bibitem [{\citenamefont {Madeira}\ \emph {et~al.}(2020)\citenamefont
  {Madeira}, \citenamefont {Caracanhas}, \citenamefont {Dos~Santos},\ and\
  \citenamefont {Bagnato}}]{Madeira_etal_2020}%
  \BibitemOpen
  \bibfield  {author} {\bibinfo {author} {\bibfnamefont {L.}~\bibnamefont
  {Madeira}}, \bibinfo {author} {\bibfnamefont {M.}~\bibnamefont {Caracanhas}},
  \bibinfo {author} {\bibfnamefont {F.}~\bibnamefont {Dos~Santos}},\ and\
  \bibinfo {author} {\bibfnamefont {V.}~\bibnamefont {Bagnato}},\ }\bibfield
  {title} {\bibinfo {title} {Quantum {T}urbulence in {Q}uantum {G}ases},\
  }\href@noop {} {\bibfield  {journal} {\bibinfo  {journal} {Annu. Rev.
  Condens. Matter Phys.}\ }\textbf {\bibinfo {volume} {11}},\ \bibinfo {pages}
  {37} (\bibinfo {year} {2020})}\BibitemShut {NoStop}%
\bibitem [{\citenamefont {Tsubota}(2008)}]{Tsubota_2008}%
  \BibitemOpen
  \bibfield  {author} {\bibinfo {author} {\bibfnamefont {M.}~\bibnamefont
  {Tsubota}},\ }\bibfield  {title} {\bibinfo {title} {Quantum {T}urbulence},\
  }\href@noop {} {\bibfield  {journal} {\bibinfo  {journal} {J. Phys. Soc.
  Jpn}\ }\textbf {\bibinfo {volume} {77}},\ \bibinfo {pages} {111006} (\bibinfo
  {year} {2008})}\BibitemShut {NoStop}%
\bibitem [{\citenamefont {Barenghi}\ \emph {et~al.}(2014)\citenamefont
  {Barenghi}, \citenamefont {Skrbek},\ and\ \citenamefont
  {Sreenivasan}}]{QT_review}%
  \BibitemOpen
  \bibfield  {author} {\bibinfo {author} {\bibfnamefont {C.~F.}\ \bibnamefont
  {Barenghi}}, \bibinfo {author} {\bibfnamefont {L.}~\bibnamefont {Skrbek}},\
  and\ \bibinfo {author} {\bibfnamefont {K.~R.}\ \bibnamefont {Sreenivasan}},\
  }\bibfield  {title} {\bibinfo {title} {Introduction to quantum turbulence},\
  }\href@noop {} {\bibfield  {journal} {\bibinfo  {journal} {Proc. Natl. Acad.
  Sci. U. S. A.}\ }\textbf {\bibinfo {volume} {111}},\ \bibinfo {pages} {4647}
  (\bibinfo {year} {2014})}\BibitemShut {NoStop}%
\bibitem [{\citenamefont {Tsatsos}\ \emph {et~al.}(2016)\citenamefont
  {Tsatsos}, \citenamefont {Tavares}, \citenamefont {Cidrim}, \citenamefont
  {Fritsch}, \citenamefont {Caracanhas}, \citenamefont {dos Santos},
  \citenamefont {Barenghi},\ and\ \citenamefont {Bagnato}}]{Tsatsos_2016}%
  \BibitemOpen
  \bibfield  {author} {\bibinfo {author} {\bibfnamefont {M.~C.}\ \bibnamefont
  {Tsatsos}}, \bibinfo {author} {\bibfnamefont {P.~E.}\ \bibnamefont
  {Tavares}}, \bibinfo {author} {\bibfnamefont {A.}~\bibnamefont {Cidrim}},
  \bibinfo {author} {\bibfnamefont {A.~R.}\ \bibnamefont {Fritsch}}, \bibinfo
  {author} {\bibfnamefont {M.~A.}\ \bibnamefont {Caracanhas}}, \bibinfo
  {author} {\bibfnamefont {F.~E.~A.}\ \bibnamefont {dos Santos}}, \bibinfo
  {author} {\bibfnamefont {C.~F.}\ \bibnamefont {Barenghi}},\ and\ \bibinfo
  {author} {\bibfnamefont {V.~S.}\ \bibnamefont {Bagnato}},\ }\bibfield
  {title} {\bibinfo {title} {Quantum turbulence in trapped atomic
  {B}ose--{E}instein condensates},\ }\href@noop {} {\bibfield  {journal}
  {\bibinfo  {journal} {Phys. Rep.}\ }\textbf {\bibinfo {volume} {622}},\
  \bibinfo {pages} {1} (\bibinfo {year} {2016})}\BibitemShut {NoStop}%
\bibitem [{\citenamefont {Nore}\ \emph
  {et~al.}(1997{\natexlab{a}})\citenamefont {Nore}, \citenamefont {Abid},\ and\
  \citenamefont {Brachet}}]{Nore_etal_1997_PhysFluid}%
  \BibitemOpen
  \bibfield  {author} {\bibinfo {author} {\bibfnamefont {C.}~\bibnamefont
  {Nore}}, \bibinfo {author} {\bibfnamefont {M.}~\bibnamefont {Abid}},\ and\
  \bibinfo {author} {\bibfnamefont {M.~E.}\ \bibnamefont {Brachet}},\
  }\bibfield  {title} {\bibinfo {title} {Decaying {K}olmogorov turbulence in a
  model of superflow},\ }\href@noop {} {\bibfield  {journal} {\bibinfo
  {journal} {Phys. Fluids}\ }\textbf {\bibinfo {volume} {9}},\ \bibinfo {pages}
  {2644} (\bibinfo {year} {1997}{\natexlab{a}})}\BibitemShut {NoStop}%
\bibitem [{\citenamefont {Nore}\ \emph
  {et~al.}(1997{\natexlab{b}})\citenamefont {Nore}, \citenamefont {Abid},\ and\
  \citenamefont {Brachet}}]{Nore_etal_1997_PRL}%
  \BibitemOpen
  \bibfield  {author} {\bibinfo {author} {\bibfnamefont {C.}~\bibnamefont
  {Nore}}, \bibinfo {author} {\bibfnamefont {M.}~\bibnamefont {Abid}},\ and\
  \bibinfo {author} {\bibfnamefont {M.~E.}\ \bibnamefont {Brachet}},\
  }\bibfield  {title} {\bibinfo {title} {Kolmogorov {T}urbulence in
  {L}ow-{T}emperature {S}uperflows},\ }\href@noop {} {\bibfield  {journal}
  {\bibinfo  {journal} {Phys. Rev. Lett.}\ }\textbf {\bibinfo {volume} {78}},\
  \bibinfo {pages} {3896} (\bibinfo {year} {1997}{\natexlab{b}})}\BibitemShut
  {NoStop}%
\bibitem [{\citenamefont {Kobayashi}\ and\ \citenamefont
  {Tsubota}(2005{\natexlab{a}})}]{Kobayashi_Tsubota}%
  \BibitemOpen
  \bibfield  {author} {\bibinfo {author} {\bibfnamefont {M.}~\bibnamefont
  {Kobayashi}}\ and\ \bibinfo {author} {\bibfnamefont {M.}~\bibnamefont
  {Tsubota}},\ }\bibfield  {title} {\bibinfo {title} {Kolmogorov {S}pectrum of
  {Q}uantum {T}urbulence},\ }\href@noop {} {\bibfield  {journal} {\bibinfo
  {journal} {J. Phys. Soc. Jpn}\ }\textbf {\bibinfo {volume} {74}},\ \bibinfo
  {pages} {3248} (\bibinfo {year} {2005}{\natexlab{a}})}\BibitemShut {NoStop}%
\bibitem [{\citenamefont {Kobayashi}\ and\ \citenamefont
  {Tsubota}(2005{\natexlab{b}})}]{Kobayashi_Tsubota_PRL}%
  \BibitemOpen
  \bibfield  {author} {\bibinfo {author} {\bibfnamefont {M.}~\bibnamefont
  {Kobayashi}}\ and\ \bibinfo {author} {\bibfnamefont {M.}~\bibnamefont
  {Tsubota}},\ }\bibfield  {title} {\bibinfo {title} {Kolmogorov {S}pectrum of
  {S}uperfluid {T}urbulence: {N}umerical {A}nalysis of the {G}ross-{P}itaevskii
  {E}quation with a {S}mall-{S}cale {D}issipation},\ }\href@noop {} {\bibfield
  {journal} {\bibinfo  {journal} {Phys. Rev. Lett.}\ }\textbf {\bibinfo
  {volume} {94}},\ \bibinfo {pages} {065302} (\bibinfo {year}
  {2005}{\natexlab{b}})}\BibitemShut {NoStop}%
\bibitem [{\citenamefont {Salort}\ \emph {et~al.}(2010)\citenamefont {Salort},
  \citenamefont {Baudet}, \citenamefont {Castaing}, \citenamefont {Chabaud},
  \citenamefont {Daviaud}, \citenamefont {Didelot}, \citenamefont {Diribarne},
  \citenamefont {Dubrulle}, \citenamefont {Gagne}, \citenamefont {Gauthier}
  \emph {et~al.}}]{QT_experiments}%
  \BibitemOpen
  \bibfield  {author} {\bibinfo {author} {\bibfnamefont {J.}~\bibnamefont
  {Salort}}, \bibinfo {author} {\bibfnamefont {C.}~\bibnamefont {Baudet}},
  \bibinfo {author} {\bibfnamefont {B.}~\bibnamefont {Castaing}}, \bibinfo
  {author} {\bibfnamefont {B.}~\bibnamefont {Chabaud}}, \bibinfo {author}
  {\bibfnamefont {F.}~\bibnamefont {Daviaud}}, \bibinfo {author} {\bibfnamefont
  {T.}~\bibnamefont {Didelot}}, \bibinfo {author} {\bibfnamefont
  {P.}~\bibnamefont {Diribarne}}, \bibinfo {author} {\bibfnamefont
  {B.}~\bibnamefont {Dubrulle}}, \bibinfo {author} {\bibfnamefont
  {Y.}~\bibnamefont {Gagne}}, \bibinfo {author} {\bibfnamefont
  {F.}~\bibnamefont {Gauthier}}, \emph {et~al.},\ }\bibfield  {title} {\bibinfo
  {title} {Turbulent velocity spectra in superfluid flows},\ }\href@noop {}
  {\bibfield  {journal} {\bibinfo  {journal} {Phys. Fluids}\ }\textbf {\bibinfo
  {volume} {22}},\ \bibinfo {pages} {125102} (\bibinfo {year}
  {2010})}\BibitemShut {NoStop}%
\bibitem [{\citenamefont {H{\"a}nninen}\ and\ \citenamefont
  {Baggaley}(2014)}]{Hanninen_2014}%
  \BibitemOpen
  \bibfield  {author} {\bibinfo {author} {\bibfnamefont {R.}~\bibnamefont
  {H{\"a}nninen}}\ and\ \bibinfo {author} {\bibfnamefont {A.~W.}\ \bibnamefont
  {Baggaley}},\ }\bibfield  {title} {\bibinfo {title} {Vortex filament method
  as a tool for computational visualization of quantum turbulence},\
  }\href@noop {} {\bibfield  {journal} {\bibinfo  {journal} {Proc. Natl. Acad.
  Sci. U. S. A.}\ }\textbf {\bibinfo {volume} {111}},\ \bibinfo {pages} {4667}
  (\bibinfo {year} {2014})}\BibitemShut {NoStop}%
\bibitem [{\citenamefont {Thomson}(1880)}]{Thomson_1880}%
  \BibitemOpen
  \bibfield  {author} {\bibinfo {author} {\bibfnamefont {W.}~\bibnamefont
  {Thomson}},\ }\bibfield  {title} {\bibinfo {title} {Vibrations of a columnar
  vortex},\ }\href@noop {} {\bibfield  {journal} {\bibinfo  {journal} {Phil.
  Mag.}\ }\textbf {\bibinfo {volume} {10}},\ \bibinfo {pages} {155} (\bibinfo
  {year} {1880})}\BibitemShut {NoStop}%
\bibitem [{\citenamefont {Kozik}\ and\ \citenamefont {Svistunov}(2004)}]{KS}%
  \BibitemOpen
  \bibfield  {author} {\bibinfo {author} {\bibfnamefont {E.}~\bibnamefont
  {Kozik}}\ and\ \bibinfo {author} {\bibfnamefont {B.}~\bibnamefont
  {Svistunov}},\ }\bibfield  {title} {\bibinfo {title} {Kelvin-{W}ave {C}ascade
  and {D}ecay of {S}uperfluid {T}urbulence},\ }\href@noop {} {\bibfield
  {journal} {\bibinfo  {journal} {Phys. Rev. Lett.}\ }\textbf {\bibinfo
  {volume} {92}},\ \bibinfo {pages} {035301} (\bibinfo {year}
  {2004})}\BibitemShut {NoStop}%
\bibitem [{\citenamefont {L'vov}\ and\ \citenamefont
  {Nazarenko}(2010{\natexlab{a}})}]{LN}%
  \BibitemOpen
  \bibfield  {author} {\bibinfo {author} {\bibfnamefont {V.~S.}\ \bibnamefont
  {L'vov}}\ and\ \bibinfo {author} {\bibfnamefont {S.}~\bibnamefont
  {Nazarenko}},\ }\bibfield  {title} {\bibinfo {title} {Spectrum of
  {K}elvin-{W}ave {T}urbulence in {S}uperfluids},\ }\href@noop {} {\bibfield
  {journal} {\bibinfo  {journal} {JETP Lett.}\ }\textbf {\bibinfo {volume}
  {91}},\ \bibinfo {pages} {428} (\bibinfo {year}
  {2010}{\natexlab{a}})}\BibitemShut {NoStop}%
\bibitem [{\citenamefont {Vinen}(2000)}]{Vinen_2000}%
  \BibitemOpen
  \bibfield  {author} {\bibinfo {author} {\bibfnamefont {W.~F.}\ \bibnamefont
  {Vinen}},\ }\bibfield  {title} {\bibinfo {title} {Classical character of
  turbulence in a quantum liquid},\ }\href@noop {} {\bibfield  {journal}
  {\bibinfo  {journal} {Phys. Rev. B}\ }\textbf {\bibinfo {volume} {61}},\
  \bibinfo {pages} {1410} (\bibinfo {year} {2000})}\BibitemShut {NoStop}%
\bibitem [{\citenamefont {Vinen}\ and\ \citenamefont
  {Niemela}(2002)}]{Vinen_Niemela_2002}%
  \BibitemOpen
  \bibfield  {author} {\bibinfo {author} {\bibfnamefont {W.~F.}\ \bibnamefont
  {Vinen}}\ and\ \bibinfo {author} {\bibfnamefont {J.~J.}\ \bibnamefont
  {Niemela}},\ }\bibfield  {title} {\bibinfo {title} {Quantum {T}urbulence},\
  }\href@noop {} {\bibfield  {journal} {\bibinfo  {journal} {J. Low Temp.
  Phys.}\ }\textbf {\bibinfo {volume} {128}},\ \bibinfo {pages} {167} (\bibinfo
  {year} {2002})}\BibitemShut {NoStop}%
\bibitem [{\citenamefont {Nemirovskii}(2013)}]{Nemirovskii_2013}%
  \BibitemOpen
  \bibfield  {author} {\bibinfo {author} {\bibfnamefont {S.~K.}\ \bibnamefont
  {Nemirovskii}},\ }\bibfield  {title} {\bibinfo {title} {Energy {S}pectrum of
  the {3D} {V}elocity {F}ield, {I}nduced by {V}ortex {T}angle},\ }\href@noop {}
  {\bibfield  {journal} {\bibinfo  {journal} {J. Low Temp. Phys.}\ }\textbf
  {\bibinfo {volume} {171}},\ \bibinfo {pages} {504} (\bibinfo {year}
  {2013})}\BibitemShut {NoStop}%
\bibitem [{\citenamefont {Schwarz}(1985)}]{Schwarz_1985}%
  \BibitemOpen
  \bibfield  {author} {\bibinfo {author} {\bibfnamefont {K.~W.}\ \bibnamefont
  {Schwarz}},\ }\bibfield  {title} {\bibinfo {title} {Three-dimensional vortex
  dynamics in superfluid $^{4}${H}e: {L}ine-line and line-boundary
  interactions},\ }\href@noop {} {\bibfield  {journal} {\bibinfo  {journal}
  {Phys. Rev. B}\ }\textbf {\bibinfo {volume} {31}},\ \bibinfo {pages} {5782}
  (\bibinfo {year} {1985})}\BibitemShut {NoStop}%
\bibitem [{\citenamefont {Gross}(1961)}]{Gross_1961}%
  \BibitemOpen
  \bibfield  {author} {\bibinfo {author} {\bibfnamefont {E.~P.}\ \bibnamefont
  {Gross}},\ }\bibfield  {title} {\bibinfo {title} {Structure of a {Q}uantized
  {V}ortex in {B}oson {S}ystems},\ }\href@noop {} {\bibfield  {journal}
  {\bibinfo  {journal} {Nuovo Cimento Suppl.}\ }\textbf {\bibinfo {volume}
  {20}},\ \bibinfo {pages} {454} (\bibinfo {year} {1961})}\BibitemShut
  {NoStop}%
\bibitem [{\citenamefont {Pitaevskii}(1961)}]{Pitaevskii_1961}%
  \BibitemOpen
  \bibfield  {author} {\bibinfo {author} {\bibfnamefont {L.~P.}\ \bibnamefont
  {Pitaevskii}},\ }\bibfield  {title} {\bibinfo {title} {Vortex lines in an
  imperfect {B}ose gas},\ }\href@noop {} {\bibfield  {journal} {\bibinfo
  {journal} {Sov. Phys. JETP}\ }\textbf {\bibinfo {volume} {13}},\ \bibinfo
  {pages} {451} (\bibinfo {year} {1961})}\BibitemShut {NoStop}%
\bibitem [{\citenamefont {Kobayashi}\ and\ \citenamefont
  {Tsubota}(2007)}]{Kobayashi_Tsubota_2007}%
  \BibitemOpen
  \bibfield  {author} {\bibinfo {author} {\bibfnamefont {M.}~\bibnamefont
  {Kobayashi}}\ and\ \bibinfo {author} {\bibfnamefont {M.}~\bibnamefont
  {Tsubota}},\ }\bibfield  {title} {\bibinfo {title} {Quantum turbulence in a
  trapped {B}ose-{E}instein condensate},\ }\href@noop {} {\bibfield  {journal}
  {\bibinfo  {journal} {Phys. Rev. A}\ }\textbf {\bibinfo {volume} {76}},\
  \bibinfo {pages} {045603} (\bibinfo {year} {2007})}\BibitemShut {NoStop}%
\bibitem [{\citenamefont {Madelung}(1926)}]{Madelung_1926}%
  \BibitemOpen
  \bibfield  {author} {\bibinfo {author} {\bibfnamefont {E.}~\bibnamefont
  {Madelung}},\ }\bibfield  {title} {\bibinfo {title} {Eine anschauliche
  {D}eutung der {G}leichung von {S}chr{\"o}dinger},\ }\href@noop {} {\bibfield
  {journal} {\bibinfo  {journal} {NW}\ }\textbf {\bibinfo {volume} {14}},\
  \bibinfo {pages} {1004} (\bibinfo {year} {1926})}\BibitemShut {NoStop}%
\bibitem [{\citenamefont {Madelung}(1927)}]{Madelung_1927}%
  \BibitemOpen
  \bibfield  {author} {\bibinfo {author} {\bibfnamefont {E.}~\bibnamefont
  {Madelung}},\ }\bibfield  {title} {\bibinfo {title} {Quantentheorie in
  hydrodynamischer form},\ }\href@noop {} {\bibfield  {journal} {\bibinfo
  {journal} {Z. Phys}\ }\textbf {\bibinfo {volume} {40}},\ \bibinfo {pages}
  {322} (\bibinfo {year} {1927})}\BibitemShut {NoStop}%
\bibitem [{\citenamefont {Clark~di Leoni}\ \emph {et~al.}(2017)\citenamefont
  {Clark~di Leoni}, \citenamefont {Mininni},\ and\ \citenamefont
  {Brachet}}]{Leoni}%
  \BibitemOpen
  \bibfield  {author} {\bibinfo {author} {\bibfnamefont {P.}~\bibnamefont
  {Clark~di Leoni}}, \bibinfo {author} {\bibfnamefont {P.~D.}\ \bibnamefont
  {Mininni}},\ and\ \bibinfo {author} {\bibfnamefont {M.~E.}\ \bibnamefont
  {Brachet}},\ }\bibfield  {title} {\bibinfo {title} {Dual cascade and
  dissipation mechanisms in helical quantum turbulence},\ }\href@noop {}
  {\bibfield  {journal} {\bibinfo  {journal} {Phys. Rev. A}\ }\textbf {\bibinfo
  {volume} {95}},\ \bibinfo {pages} {053636} (\bibinfo {year}
  {2017})}\BibitemShut {NoStop}%
\bibitem [{\citenamefont {Fujimoto}\ and\ \citenamefont
  {Tsubota}(2015)}]{Fujimoto_2015}%
  \BibitemOpen
  \bibfield  {author} {\bibinfo {author} {\bibfnamefont {K.}~\bibnamefont
  {Fujimoto}}\ and\ \bibinfo {author} {\bibfnamefont {M.}~\bibnamefont
  {Tsubota}},\ }\bibfield  {title} {\bibinfo {title} {Bogoliubov-wave
  turbulence in {B}ose-{E}instein condensates},\ }\href@noop {} {\bibfield
  {journal} {\bibinfo  {journal} {Phys. Rev. A}\ }\textbf {\bibinfo {volume}
  {91}},\ \bibinfo {pages} {053620} (\bibinfo {year} {2015})}\BibitemShut
  {NoStop}%
\bibitem [{\citenamefont {Villois}\ \emph {et~al.}(2017)\citenamefont
  {Villois}, \citenamefont {Proment},\ and\ \citenamefont
  {Krstulovic}}]{Villois_2017}%
  \BibitemOpen
  \bibfield  {author} {\bibinfo {author} {\bibfnamefont {A.}~\bibnamefont
  {Villois}}, \bibinfo {author} {\bibfnamefont {D.}~\bibnamefont {Proment}},\
  and\ \bibinfo {author} {\bibfnamefont {G.}~\bibnamefont {Krstulovic}},\
  }\bibfield  {title} {\bibinfo {title} {Universal and nonuniversal aspects of
  vortex reconnections in superfluids},\ }\href@noop {} {\bibfield  {journal}
  {\bibinfo  {journal} {Phys. Rev. Fluids}\ }\textbf {\bibinfo {volume} {2}},\
  \bibinfo {pages} {044701} (\bibinfo {year} {2017})}\BibitemShut {NoStop}%
\bibitem [{\citenamefont {H{\"a}nninen}\ and\ \citenamefont
  {Schoepe}(2010)}]{Hanninen_Schoepe_2010}%
  \BibitemOpen
  \bibfield  {author} {\bibinfo {author} {\bibfnamefont {R.}~\bibnamefont
  {H{\"a}nninen}}\ and\ \bibinfo {author} {\bibfnamefont {W.}~\bibnamefont
  {Schoepe}},\ }\bibfield  {title} {\bibinfo {title} {Universal onset of
  quantum turbulence in oscillating flows and crossover to steady flows},\
  }\href@noop {} {\bibfield  {journal} {\bibinfo  {journal} {J. Low Temp.
  Phys.}\ }\textbf {\bibinfo {volume} {158}},\ \bibinfo {pages} {410} (\bibinfo
  {year} {2010})}\BibitemShut {NoStop}%
\bibitem [{\citenamefont {Wang}\ \emph {et~al.}(2013)\citenamefont {Wang},
  \citenamefont {Yang}, \citenamefont {Shi}, \citenamefont {Xiao},
  \citenamefont {He},\ and\ \citenamefont {Chen}}]{Wang-Yang-Shi}%
  \BibitemOpen
  \bibfield  {author} {\bibinfo {author} {\bibfnamefont {J.}~\bibnamefont
  {Wang}}, \bibinfo {author} {\bibfnamefont {Y.}~\bibnamefont {Yang}}, \bibinfo
  {author} {\bibfnamefont {Y.}~\bibnamefont {Shi}}, \bibinfo {author}
  {\bibfnamefont {Z.}~\bibnamefont {Xiao}}, \bibinfo {author} {\bibfnamefont
  {X.~T.}\ \bibnamefont {He}},\ and\ \bibinfo {author} {\bibfnamefont
  {S.}~\bibnamefont {Chen}},\ }\bibfield  {title} {\bibinfo {title} {Cascade of
  {K}inetic {E}nergy in {T}hree-{D}imensional {C}ompressible {T}urbulence},\
  }\href@noop {} {\bibfield  {journal} {\bibinfo  {journal} {Phys. Rev. Lett.}\
  }\textbf {\bibinfo {volume} {110}},\ \bibinfo {pages} {214505} (\bibinfo
  {year} {2013})}\BibitemShut {NoStop}%
\bibitem [{\citenamefont {Dunn}\ and\ \citenamefont
  {Serrin}(1986)}]{Dunn_Serrin}%
  \BibitemOpen
  \bibfield  {author} {\bibinfo {author} {\bibfnamefont {J.~E.}\ \bibnamefont
  {Dunn}}\ and\ \bibinfo {author} {\bibfnamefont {J.}~\bibnamefont {Serrin}},\
  }\bibfield  {title} {\bibinfo {title} {On the {T}hermomechanics of
  {I}nterstitial {W}orking},\ }in\ \href@noop {} {\emph {\bibinfo {booktitle}
  {The Breadth and Depth of Continuum Mechanics}}}\ (\bibinfo  {publisher}
  {Springer},\ \bibinfo {year} {1986})\ pp.\ \bibinfo {pages}
  {705--743}\BibitemShut {NoStop}%
\bibitem [{\citenamefont {Benzoni-Gavage}()}]{Gavage}%
  \BibitemOpen
  \bibfield  {author} {\bibinfo {author} {\bibfnamefont {S.}~\bibnamefont
  {Benzoni-Gavage}},\ }\href@noop {} {\bibinfo {title} {Propagating phase
  boundaries and capillary fluids}},\ \bibinfo {note}
  {\url{http://math.univ-lyon1.fr/~benzoni/Levico.pdf}}\BibitemShut {NoStop}%
\bibitem [{\citenamefont {Leonard}\ \emph {et~al.}(1974)\citenamefont {Leonard}
  \emph {et~al.}}]{Leonard_1974}%
  \BibitemOpen
  \bibfield  {author} {\bibinfo {author} {\bibfnamefont {A.}~\bibnamefont
  {Leonard}} \emph {et~al.},\ }\bibfield  {title} {\bibinfo {title} {Energy
  cascade in large-eddy simulations of turbulent fluid flows},\ }\href@noop {}
  {\bibfield  {journal} {\bibinfo  {journal} {Adv. Geophys. A}\ }\textbf
  {\bibinfo {volume} {18}},\ \bibinfo {pages} {237} (\bibinfo {year}
  {1974})}\BibitemShut {NoStop}%
\bibitem [{\citenamefont {Germano}(1992)}]{Germano_1992}%
  \BibitemOpen
  \bibfield  {author} {\bibinfo {author} {\bibfnamefont {M.}~\bibnamefont
  {Germano}},\ }\bibfield  {title} {\bibinfo {title} {Turbulence: the filtering
  approach},\ }\href@noop {} {\bibfield  {journal} {\bibinfo  {journal} {J.
  Fluid Mech.}\ }\textbf {\bibinfo {volume} {238}},\ \bibinfo {pages} {325}
  (\bibinfo {year} {1992})}\BibitemShut {NoStop}%
\bibitem [{\citenamefont {Favre}(1969)}]{Favre}%
  \BibitemOpen
  \bibfield  {author} {\bibinfo {author} {\bibfnamefont {A.}~\bibnamefont
  {Favre}},\ }\bibfield  {title} {\bibinfo {title} {Statistical equations of
  turbulent gases},\ }in\ \href@noop {} {\emph {\bibinfo {booktitle} {Problems
  of Hydrodynamics and Continuum Mechanics}}},\ \bibinfo {editor} {edited by\
  \bibinfo {editor} {\bibfnamefont {M.~A.}\ \bibnamefont {Lavrentiev}}}\
  (\bibinfo  {publisher} {SIAM, Philadelphia},\ \bibinfo {year} {1969})\ pp.\
  \bibinfo {pages} {37--44}\BibitemShut {NoStop}%
\bibitem [{\citenamefont {Parisi}\ and\ \citenamefont
  {Frisch}(1985)}]{Parisi-Frisch}%
  \BibitemOpen
  \bibfield  {author} {\bibinfo {author} {\bibfnamefont {G.}~\bibnamefont
  {Parisi}}\ and\ \bibinfo {author} {\bibfnamefont {U.}~\bibnamefont
  {Frisch}},\ }\bibfield  {title} {\bibinfo {title} {On the singularity
  structure of fully developed turbulence in turbulence and predictability in
  geophysical fluid dynamics and climate dynamics},\ }\href@noop {} {\bibfield
  {journal} {\bibinfo  {journal} {NTurbulence and Predictability of Geophysical
  Flows and Climate Dynamics}\ }\textbf {\bibinfo {volume} {88}} (\bibinfo
  {year} {1985})}\BibitemShut {NoStop}%
\bibitem [{\citenamefont {Frisch}(1995)}]{Frisch}%
  \BibitemOpen
  \bibfield  {author} {\bibinfo {author} {\bibfnamefont {U.}~\bibnamefont
  {Frisch}},\ }\href@noop {} {\emph {\bibinfo {title} {Turbulence}}}\ (\bibinfo
   {publisher} {Cambridge university press},\ \bibinfo {year}
  {1995})\BibitemShut {NoStop}%
\bibitem [{\citenamefont {Barenghi}(2008)}]{Barenghi_2008}%
  \BibitemOpen
  \bibfield  {author} {\bibinfo {author} {\bibfnamefont {C.~F.}\ \bibnamefont
  {Barenghi}},\ }\bibfield  {title} {\bibinfo {title} {Is the {R}eynolds number
  infinite in superfluid turbulence?},\ }\href@noop {} {\bibfield  {journal}
  {\bibinfo  {journal} {Physica D}\ }\textbf {\bibinfo {volume} {237}},\
  \bibinfo {pages} {2195} (\bibinfo {year} {2008})}\BibitemShut {NoStop}%
\bibitem [{\citenamefont {Bradley}\ \emph {et~al.}(2006)\citenamefont
  {Bradley}, \citenamefont {Clubb}, \citenamefont {Fisher}, \citenamefont
  {Gu{\'e}nault}, \citenamefont {Haley}, \citenamefont {Matthews},
  \citenamefont {Pickett}, \citenamefont {Tsepelin},\ and\ \citenamefont
  {Zaki}}]{Bradley_2006}%
  \BibitemOpen
  \bibfield  {author} {\bibinfo {author} {\bibfnamefont {D.~I.}\ \bibnamefont
  {Bradley}}, \bibinfo {author} {\bibfnamefont {D.~O.}\ \bibnamefont {Clubb}},
  \bibinfo {author} {\bibfnamefont {S.~N.}\ \bibnamefont {Fisher}}, \bibinfo
  {author} {\bibfnamefont {A.~M.}\ \bibnamefont {Gu{\'e}nault}}, \bibinfo
  {author} {\bibfnamefont {R.~P.}\ \bibnamefont {Haley}}, \bibinfo {author}
  {\bibfnamefont {C.~J.}\ \bibnamefont {Matthews}}, \bibinfo {author}
  {\bibfnamefont {G.~R.}\ \bibnamefont {Pickett}}, \bibinfo {author}
  {\bibfnamefont {V.}~\bibnamefont {Tsepelin}},\ and\ \bibinfo {author}
  {\bibfnamefont {K.}~\bibnamefont {Zaki}},\ }\bibfield  {title} {\bibinfo
  {title} {Decay of {P}ure {Q}uantum {T}urbulence in {S}uperfluid
  $^{3}${H}e-{B}},\ }\href@noop {} {\bibfield  {journal} {\bibinfo  {journal}
  {Phys. Rev. Lett.}\ }\textbf {\bibinfo {volume} {96}},\ \bibinfo {pages}
  {035301} (\bibinfo {year} {2006})}\BibitemShut {NoStop}%
\bibitem [{\citenamefont {Leadbeater}\ \emph {et~al.}(2001)\citenamefont
  {Leadbeater}, \citenamefont {Winiecki}, \citenamefont {Samuels},
  \citenamefont {Barenghi},\ and\ \citenamefont
  {Adams}}]{Leadbeater_etal_2001}%
  \BibitemOpen
  \bibfield  {author} {\bibinfo {author} {\bibfnamefont {M.}~\bibnamefont
  {Leadbeater}}, \bibinfo {author} {\bibfnamefont {T.}~\bibnamefont
  {Winiecki}}, \bibinfo {author} {\bibfnamefont {D.~C.}\ \bibnamefont
  {Samuels}}, \bibinfo {author} {\bibfnamefont {C.~F.}\ \bibnamefont
  {Barenghi}},\ and\ \bibinfo {author} {\bibfnamefont {C.~S.}\ \bibnamefont
  {Adams}},\ }\bibfield  {title} {\bibinfo {title} {Sound {E}mission due to
  {S}uperfluid {V}ortex {R}econnections},\ }\href@noop {} {\bibfield  {journal}
  {\bibinfo  {journal} {Phys. Rev. Lett.}\ }\textbf {\bibinfo {volume} {86}},\
  \bibinfo {pages} {1410} (\bibinfo {year} {2001})}\BibitemShut {NoStop}%
\bibitem [{\citenamefont {Zuccher}\ \emph {et~al.}(2012)\citenamefont
  {Zuccher}, \citenamefont {Caliari}, \citenamefont {Baggaley},\ and\
  \citenamefont {Barenghi}}]{Zuccher_etal_2012}%
  \BibitemOpen
  \bibfield  {author} {\bibinfo {author} {\bibfnamefont {S.}~\bibnamefont
  {Zuccher}}, \bibinfo {author} {\bibfnamefont {M.}~\bibnamefont {Caliari}},
  \bibinfo {author} {\bibfnamefont {A.~W.}\ \bibnamefont {Baggaley}},\ and\
  \bibinfo {author} {\bibfnamefont {C.~F.}\ \bibnamefont {Barenghi}},\
  }\bibfield  {title} {\bibinfo {title} {Quantum vortex reconnections},\
  }\href@noop {} {\bibfield  {journal} {\bibinfo  {journal} {Phys. Fluids}\
  }\textbf {\bibinfo {volume} {24}},\ \bibinfo {pages} {125108} (\bibinfo
  {year} {2012})}\BibitemShut {NoStop}%
\bibitem [{\citenamefont {Vinen}(2001)}]{Vinen_2001}%
  \BibitemOpen
  \bibfield  {author} {\bibinfo {author} {\bibfnamefont {W.~F.}\ \bibnamefont
  {Vinen}},\ }\bibfield  {title} {\bibinfo {title} {Decay of superfluid
  turbulence at a very low temperature: {T}he radiation of sound from a
  {K}elvin wave on a quantized vortex},\ }\href@noop {} {\bibfield  {journal}
  {\bibinfo  {journal} {Phys. Rev. B}\ }\textbf {\bibinfo {volume} {64}},\
  \bibinfo {pages} {134520} (\bibinfo {year} {2001})}\BibitemShut {NoStop}%
\bibitem [{\citenamefont {D{\k{e}}biec}\ \emph {et~al.}(2018)\citenamefont
  {D{\k{e}}biec}, \citenamefont {Gwiazda}, \citenamefont
  {{\'S}wierczewska-Gwiazda},\ and\ \citenamefont
  {Tzavaras}}]{Debiec_etal_2018}%
  \BibitemOpen
  \bibfield  {author} {\bibinfo {author} {\bibfnamefont {T.}~\bibnamefont
  {D{\k{e}}biec}}, \bibinfo {author} {\bibfnamefont {P.}~\bibnamefont
  {Gwiazda}}, \bibinfo {author} {\bibfnamefont {A.}~\bibnamefont
  {{\'S}wierczewska-Gwiazda}},\ and\ \bibinfo {author} {\bibfnamefont
  {A.}~\bibnamefont {Tzavaras}},\ }\bibfield  {title} {\bibinfo {title}
  {Conservation of energy for the {E}uler--{K}orteweg equations},\ }\href@noop
  {} {\bibfield  {journal} {\bibinfo  {journal} {Calc. Var. Partial Dif.}\
  }\textbf {\bibinfo {volume} {57}},\ \bibinfo {pages} {160} (\bibinfo {year}
  {2018})}\BibitemShut {NoStop}%
\bibitem [{\citenamefont {Kobayashi}\ and\ \citenamefont
  {Ueda}(2016)}]{Kobayashi-Ueda}%
  \BibitemOpen
  \bibfield  {author} {\bibinfo {author} {\bibfnamefont {M.}~\bibnamefont
  {Kobayashi}}\ and\ \bibinfo {author} {\bibfnamefont {M.}~\bibnamefont
  {Ueda}},\ }\bibfield  {title} {\bibinfo {title} {Topologically protected pure
  helicity cascade in non-{A}belian quantum turbulence},\ }\href@noop {}
  {\bibfield  {journal} {\bibinfo  {journal} {arXiv preprint arXiv:1606.07190}\
  } (\bibinfo {year} {2016})}\BibitemShut {NoStop}%
\bibitem [{\citenamefont {Bradley}\ and\ \citenamefont
  {Anderson}(2012)}]{Bradley_Anderson_2012}%
  \BibitemOpen
  \bibfield  {author} {\bibinfo {author} {\bibfnamefont {A.~S.}\ \bibnamefont
  {Bradley}}\ and\ \bibinfo {author} {\bibfnamefont {B.~P.}\ \bibnamefont
  {Anderson}},\ }\bibfield  {title} {\bibinfo {title} {Energy spectra of vortex
  distributions in two-dimensional quantum turbulence},\ }\href@noop {}
  {\bibfield  {journal} {\bibinfo  {journal} {Phys. Rev. X}\ }\textbf {\bibinfo
  {volume} {2}},\ \bibinfo {pages} {041001} (\bibinfo {year}
  {2012})}\BibitemShut {NoStop}%
\bibitem [{\citenamefont {Shukla}\ \emph {et~al.}(2019)\citenamefont {Shukla},
  \citenamefont {Mininni}, \citenamefont {Krstulovic}, \citenamefont
  {di~Leoni},\ and\ \citenamefont {Brachet}}]{Shukla_etal_2019}%
  \BibitemOpen
  \bibfield  {author} {\bibinfo {author} {\bibfnamefont {V.}~\bibnamefont
  {Shukla}}, \bibinfo {author} {\bibfnamefont {P.~D.}\ \bibnamefont {Mininni}},
  \bibinfo {author} {\bibfnamefont {G.}~\bibnamefont {Krstulovic}}, \bibinfo
  {author} {\bibfnamefont {P.~C.}\ \bibnamefont {di~Leoni}},\ and\ \bibinfo
  {author} {\bibfnamefont {M.~E.}\ \bibnamefont {Brachet}},\ }\bibfield
  {title} {\bibinfo {title} {Quantitative estimation of effective viscosity in
  quantum turbulence},\ }\href@noop {} {\bibfield  {journal} {\bibinfo
  {journal} {Phys. Rev. A}\ }\textbf {\bibinfo {volume} {99}},\ \bibinfo
  {pages} {043605} (\bibinfo {year} {2019})}\BibitemShut {NoStop}%
\bibitem [{\citenamefont {Aluie}\ \emph {et~al.}(2012)\citenamefont {Aluie},
  \citenamefont {Li},\ and\ \citenamefont {Li}}]{Aluie_2012}%
  \BibitemOpen
  \bibfield  {author} {\bibinfo {author} {\bibfnamefont {H.}~\bibnamefont
  {Aluie}}, \bibinfo {author} {\bibfnamefont {S.}~\bibnamefont {Li}},\ and\
  \bibinfo {author} {\bibfnamefont {H.}~\bibnamefont {Li}},\ }\bibfield
  {title} {\bibinfo {title} {Conservative cascade of kinetic energy in
  compressible turbulence},\ }\href@noop {} {\bibfield  {journal} {\bibinfo
  {journal} {Astrophys. J. Lett.}\ }\textbf {\bibinfo {volume} {751}},\
  \bibinfo {pages} {L29} (\bibinfo {year} {2012})}\BibitemShut {NoStop}%
\bibitem [{\citenamefont {Tennekes}\ and\ \citenamefont
  {Lumley}(1972)}]{A_first_course_in_turbulence}%
  \BibitemOpen
  \bibfield  {author} {\bibinfo {author} {\bibfnamefont {H.}~\bibnamefont
  {Tennekes}}\ and\ \bibinfo {author} {\bibfnamefont {J.~L.}\ \bibnamefont
  {Lumley}},\ }\href@noop {} {\emph {\bibinfo {title} {A first course in
  turbulence}}}\ (\bibinfo  {publisher} {MIT press},\ \bibinfo {year}
  {1972})\BibitemShut {NoStop}%
\bibitem [{\citenamefont {Lees}\ and\ \citenamefont
  {Aluie}(2019)}]{Lees_Aluie}%
  \BibitemOpen
  \bibfield  {author} {\bibinfo {author} {\bibfnamefont {A.}~\bibnamefont
  {Lees}}\ and\ \bibinfo {author} {\bibfnamefont {H.}~\bibnamefont {Aluie}},\
  }\bibfield  {title} {\bibinfo {title} {Baropycnal work: A mechanism for
  energy transfer across scales},\ }\href@noop {} {\bibfield  {journal}
  {\bibinfo  {journal} {Fluids}\ }\textbf {\bibinfo {volume} {4}},\ \bibinfo
  {pages} {92} (\bibinfo {year} {2019})}\BibitemShut {NoStop}%
\bibitem [{\citenamefont {Eyink}(2005)}]{Eyink_2005}%
  \BibitemOpen
  \bibfield  {author} {\bibinfo {author} {\bibfnamefont {G.~L.}\ \bibnamefont
  {Eyink}},\ }\bibfield  {title} {\bibinfo {title} {Locality of turbulent
  cascades},\ }\href@noop {} {\bibfield  {journal} {\bibinfo  {journal}
  {Physica D}\ }\textbf {\bibinfo {volume} {207}},\ \bibinfo {pages} {91}
  (\bibinfo {year} {2005})}\BibitemShut {NoStop}%
\bibitem [{\citenamefont {L'vov}\ \emph {et~al.}(2007)\citenamefont {L'vov},
  \citenamefont {Nazarenko},\ and\ \citenamefont {Rudenko}}]{Lvov_etal_2007}%
  \BibitemOpen
  \bibfield  {author} {\bibinfo {author} {\bibfnamefont {V.~S.}\ \bibnamefont
  {L'vov}}, \bibinfo {author} {\bibfnamefont {S.~V.}\ \bibnamefont
  {Nazarenko}},\ and\ \bibinfo {author} {\bibfnamefont {O.}~\bibnamefont
  {Rudenko}},\ }\bibfield  {title} {\bibinfo {title} {Bottleneck crossover
  between classical and quantum superfluid turbulence},\ }\href@noop {}
  {\bibfield  {journal} {\bibinfo  {journal} {Phys. Rev. B}\ }\textbf {\bibinfo
  {volume} {76}},\ \bibinfo {pages} {024520} (\bibinfo {year}
  {2007})}\BibitemShut {NoStop}%
\bibitem [{\citenamefont {L'vov}\ \emph {et~al.}(2008)\citenamefont {L'vov},
  \citenamefont {Nazarenko},\ and\ \citenamefont {Rudenko}}]{L'vov_etal_2008}%
  \BibitemOpen
  \bibfield  {author} {\bibinfo {author} {\bibfnamefont {V.~S.}\ \bibnamefont
  {L'vov}}, \bibinfo {author} {\bibfnamefont {S.~V.}\ \bibnamefont
  {Nazarenko}},\ and\ \bibinfo {author} {\bibfnamefont {O.}~\bibnamefont
  {Rudenko}},\ }\bibfield  {title} {\bibinfo {title} {Gradual {E}ddy-{W}ave
  {C}rossover in {S}uperfluid {T}urbulence},\ }\href@noop {} {\bibfield
  {journal} {\bibinfo  {journal} {J. Low Temp. Phys.}\ }\textbf {\bibinfo
  {volume} {153}},\ \bibinfo {pages} {140} (\bibinfo {year}
  {2008})}\BibitemShut {NoStop}%
\bibitem [{\citenamefont {Kozik}\ and\ \citenamefont
  {Svistunov}(2008)}]{Kozik_etal_2008}%
  \BibitemOpen
  \bibfield  {author} {\bibinfo {author} {\bibfnamefont {E.}~\bibnamefont
  {Kozik}}\ and\ \bibinfo {author} {\bibfnamefont {B.}~\bibnamefont
  {Svistunov}},\ }\bibfield  {title} {\bibinfo {title} {Kolmogorov and
  {K}elvin-wave cascades of superfluid turbulence at ${T}=0$: {W}hat lies
  between},\ }\href@noop {} {\bibfield  {journal} {\bibinfo  {journal} {Phys.
  Rev. B}\ }\textbf {\bibinfo {volume} {77}},\ \bibinfo {pages} {060502}
  (\bibinfo {year} {2008})}\BibitemShut {NoStop}%
\bibitem [{\citenamefont {Eyink}\ and\ \citenamefont
  {Aluie}(2009)}]{Eyink_Aluie_2009-1}%
  \BibitemOpen
  \bibfield  {author} {\bibinfo {author} {\bibfnamefont {G.~L.}\ \bibnamefont
  {Eyink}}\ and\ \bibinfo {author} {\bibfnamefont {H.}~\bibnamefont {Aluie}},\
  }\bibfield  {title} {\bibinfo {title} {Localness of energy cascade in
  hydrodynamic turbulence. {I}. {S}mooth coarse graining},\ }\href@noop {}
  {\bibfield  {journal} {\bibinfo  {journal} {Phys. Fluids}\ }\textbf {\bibinfo
  {volume} {21}},\ \bibinfo {pages} {115107} (\bibinfo {year}
  {2009})}\BibitemShut {NoStop}%
\bibitem [{\citenamefont {Aluie}\ and\ \citenamefont
  {Eyink}(2009)}]{Aluie_Eyink_2009-2}%
  \BibitemOpen
  \bibfield  {author} {\bibinfo {author} {\bibfnamefont {H.}~\bibnamefont
  {Aluie}}\ and\ \bibinfo {author} {\bibfnamefont {G.~L.}\ \bibnamefont
  {Eyink}},\ }\bibfield  {title} {\bibinfo {title} {Localness of energy cascade
  in hydrodynamic turbulence. {II}. {S}harp spectral filter},\ }\href@noop {}
  {\bibfield  {journal} {\bibinfo  {journal} {Phys. Fluids}\ }\textbf {\bibinfo
  {volume} {21}},\ \bibinfo {pages} {115108} (\bibinfo {year}
  {2009})}\BibitemShut {NoStop}%
\bibitem [{\citenamefont {Kraichnan}(1967)}]{Kraichnan_1967}%
  \BibitemOpen
  \bibfield  {author} {\bibinfo {author} {\bibfnamefont {R.~H.}\ \bibnamefont
  {Kraichnan}},\ }\bibfield  {title} {\bibinfo {title} {Inertial ranges in
  two-dimensional turbulence},\ }\href@noop {} {\bibfield  {journal} {\bibinfo
  {journal} {Phys. Fluids}\ }\textbf {\bibinfo {volume} {10}},\ \bibinfo
  {pages} {1417} (\bibinfo {year} {1967})}\BibitemShut {NoStop}%
\bibitem [{\citenamefont {Kraichnan}(1971)}]{Kraichnan_1971}%
  \BibitemOpen
  \bibfield  {author} {\bibinfo {author} {\bibfnamefont {R.~H.}\ \bibnamefont
  {Kraichnan}},\ }\bibfield  {title} {\bibinfo {title} {Inertial-range transfer
  in two-and three-dimensional turbulence},\ }\href@noop {} {\bibfield
  {journal} {\bibinfo  {journal} {J. Fluid Mech.}\ }\textbf {\bibinfo {volume}
  {47}},\ \bibinfo {pages} {525} (\bibinfo {year} {1971})}\BibitemShut
  {NoStop}%
\bibitem [{\citenamefont {Boldyrev}(2005)}]{Boldyrev_2005}%
  \BibitemOpen
  \bibfield  {author} {\bibinfo {author} {\bibfnamefont {S.}~\bibnamefont
  {Boldyrev}},\ }\bibfield  {title} {\bibinfo {title} {On the spectrum of
  magnetohydrodynamic turbulence},\ }\href@noop {} {\bibfield  {journal}
  {\bibinfo  {journal} {Astrophys. J. Lett.}\ }\textbf {\bibinfo {volume}
  {626}},\ \bibinfo {pages} {L37} (\bibinfo {year} {2005})}\BibitemShut
  {NoStop}%
\bibitem [{\citenamefont {Mason}\ \emph {et~al.}(2006)\citenamefont {Mason},
  \citenamefont {Cattaneo},\ and\ \citenamefont {Boldyrev}}]{Mason_etal_2006}%
  \BibitemOpen
  \bibfield  {author} {\bibinfo {author} {\bibfnamefont {J.}~\bibnamefont
  {Mason}}, \bibinfo {author} {\bibfnamefont {F.}~\bibnamefont {Cattaneo}},\
  and\ \bibinfo {author} {\bibfnamefont {S.}~\bibnamefont {Boldyrev}},\
  }\bibfield  {title} {\bibinfo {title} {Dynamic alignment in driven
  magnetohydrodynamic turbulence},\ }\href@noop {} {\bibfield  {journal}
  {\bibinfo  {journal} {Phys. Rev. Lett.}\ }\textbf {\bibinfo {volume} {97}},\
  \bibinfo {pages} {255002} (\bibinfo {year} {2006})}\BibitemShut {NoStop}%
\bibitem [{\citenamefont {Narita}(2017)}]{Narita_2017}%
  \BibitemOpen
  \bibfield  {author} {\bibinfo {author} {\bibfnamefont {Y.}~\bibnamefont
  {Narita}},\ }\bibfield  {title} {\bibinfo {title} {Scaling laws of
  wave-cascading superfluid turbulence},\ }\href@noop {} {\bibfield  {journal}
  {\bibinfo  {journal} {AIP Adv.}\ }\textbf {\bibinfo {volume} {7}},\ \bibinfo
  {pages} {065009} (\bibinfo {year} {2017})}\BibitemShut {NoStop}%
\bibitem [{\citenamefont {Yepez}\ \emph {et~al.}(2009)\citenamefont {Yepez},
  \citenamefont {Vahala}, \citenamefont {Vahala},\ and\ \citenamefont
  {Soe}}]{Yepez}%
  \BibitemOpen
  \bibfield  {author} {\bibinfo {author} {\bibfnamefont {J.}~\bibnamefont
  {Yepez}}, \bibinfo {author} {\bibfnamefont {G.}~\bibnamefont {Vahala}},
  \bibinfo {author} {\bibfnamefont {L.}~\bibnamefont {Vahala}},\ and\ \bibinfo
  {author} {\bibfnamefont {M.}~\bibnamefont {Soe}},\ }\bibfield  {title}
  {\bibinfo {title} {Superfluid {T}urbulence from {Q}uantum {K}elvin {W}ave to
  {C}lassical {K}olmogorov {C}ascades},\ }\href@noop {} {\bibfield  {journal}
  {\bibinfo  {journal} {Phys. Rev. Lett.}\ }\textbf {\bibinfo {volume} {103}},\
  \bibinfo {pages} {084501} (\bibinfo {year} {2009})}\BibitemShut {NoStop}%
\bibitem [{\citenamefont {Krstulovic}\ and\ \citenamefont
  {Brachet}(2010)}]{Comment_Krstulovic_2010}%
  \BibitemOpen
  \bibfield  {author} {\bibinfo {author} {\bibfnamefont {G.}~\bibnamefont
  {Krstulovic}}\ and\ \bibinfo {author} {\bibfnamefont {M.~E.}\ \bibnamefont
  {Brachet}},\ }\bibfield  {title} {\bibinfo {title} {Comment on ``{S}uperfluid
  {T}urbulence from {Q}uantum {K}elvin {W}ave to {C}lassical {K}olmogorov
  {C}ascades''},\ }\href@noop {} {\bibfield  {journal} {\bibinfo  {journal}
  {Phys. Rev. Lett.}\ }\textbf {\bibinfo {volume} {105}},\ \bibinfo {pages}
  {129401} (\bibinfo {year} {2010})}\BibitemShut {NoStop}%
\bibitem [{\citenamefont {L'vov}\ and\ \citenamefont
  {Nazarenko}(2010{\natexlab{b}})}]{Comment_L'vov_2010}%
  \BibitemOpen
  \bibfield  {author} {\bibinfo {author} {\bibfnamefont {V.~S.}\ \bibnamefont
  {L'vov}}\ and\ \bibinfo {author} {\bibfnamefont {S.}~\bibnamefont
  {Nazarenko}},\ }\bibfield  {title} {\bibinfo {title} {Comment on
  ``{S}uperfluid {T}urbulence from {Q}uantum {K}elvin {W}ave to {C}lassical
  {K}olmogorov {C}ascades''},\ }\href@noop {} {\bibfield  {journal} {\bibinfo
  {journal} {Phys. Rev. Lett.}\ }\textbf {\bibinfo {volume} {104}},\ \bibinfo
  {pages} {219401} (\bibinfo {year} {2010}{\natexlab{b}})}\BibitemShut
  {NoStop}%
\bibitem [{\citenamefont {L'vov}\ \emph {et~al.}(2004)\citenamefont {L'vov},
  \citenamefont {Nazarenko},\ and\ \citenamefont {Volovik}}]{Lvov_etal_2004}%
  \BibitemOpen
  \bibfield  {author} {\bibinfo {author} {\bibfnamefont {V.~S.}\ \bibnamefont
  {L'vov}}, \bibinfo {author} {\bibfnamefont {V.}~\bibnamefont {Nazarenko}},\
  and\ \bibinfo {author} {\bibfnamefont {G.~E.}\ \bibnamefont {Volovik}},\
  }\bibfield  {title} {\bibinfo {title} {Energy spectra of developed superfluid
  turbulence},\ }\href@noop {} {\bibfield  {journal} {\bibinfo  {journal} {JETP
  Lett.}\ }\textbf {\bibinfo {volume} {80}},\ \bibinfo {pages} {479} (\bibinfo
  {year} {2004})}\BibitemShut {NoStop}%
\bibitem [{\citenamefont {Volovik}(2004)}]{Volovik_2004}%
  \BibitemOpen
  \bibfield  {author} {\bibinfo {author} {\bibfnamefont {G.~E.}\ \bibnamefont
  {Volovik}},\ }\bibfield  {title} {\bibinfo {title} {On developed superfluid
  turbulence},\ }\href@noop {} {\bibfield  {journal} {\bibinfo  {journal} {J.
  Low Temp. Phys.}\ }\textbf {\bibinfo {volume} {136}},\ \bibinfo {pages} {309}
  (\bibinfo {year} {2004})}\BibitemShut {NoStop}%
\bibitem [{\citenamefont {Baggaley}\ \emph {et~al.}(2012)\citenamefont
  {Baggaley}, \citenamefont {Barenghi}, \citenamefont {Shukurov},\ and\
  \citenamefont {Sergeev}}]{Baggaley_2012}%
  \BibitemOpen
  \bibfield  {author} {\bibinfo {author} {\bibfnamefont {A.~W.}\ \bibnamefont
  {Baggaley}}, \bibinfo {author} {\bibfnamefont {C.~F.}\ \bibnamefont
  {Barenghi}}, \bibinfo {author} {\bibfnamefont {A.}~\bibnamefont {Shukurov}},\
  and\ \bibinfo {author} {\bibfnamefont {Y.~A.}\ \bibnamefont {Sergeev}},\
  }\bibfield  {title} {\bibinfo {title} {Coherent vortex structures in quantum
  turbulence},\ }\href@noop {} {\bibfield  {journal} {\bibinfo  {journal}
  {Europhys. Lett.}\ }\textbf {\bibinfo {volume} {98}},\ \bibinfo {pages}
  {26002} (\bibinfo {year} {2012})}\BibitemShut {NoStop}%
\bibitem [{\citenamefont {Nemirovskii}(2020)}]{Nemirovskii_2020}%
  \BibitemOpen
  \bibfield  {author} {\bibinfo {author} {\bibfnamefont {S.~K.}\ \bibnamefont
  {Nemirovskii}},\ }\bibfield  {title} {\bibinfo {title} {On the {C}losure
  {P}roblem of the {C}oarse-{G}rained {H}ydrodynamics of {T}urbulent
  {S}uperfluids},\ }\href@noop {} {\bibfield  {journal} {\bibinfo  {journal}
  {J. Low Temp. Phys.}\ }\textbf {\bibinfo {volume} {201}},\ \bibinfo {pages}
  {254} (\bibinfo {year} {2020})}\BibitemShut {NoStop}%
\bibitem [{\citenamefont {Eyink}(1995{\natexlab{b}})}]{Eyink_1995}%
  \BibitemOpen
  \bibfield  {author} {\bibinfo {author} {\bibfnamefont {G.~L.}\ \bibnamefont
  {Eyink}},\ }\bibfield  {title} {\bibinfo {title} {Besov {S}paces and the
  {M}ultifractal {H}ypothesis},\ }\href@noop {} {\bibfield  {journal} {\bibinfo
   {journal} {J. Stat. Phys.}\ }\textbf {\bibinfo {volume} {78}},\ \bibinfo
  {pages} {353} (\bibinfo {year} {1995}{\natexlab{b}})}\BibitemShut {NoStop}%
\bibitem [{\citenamefont {Perrier}\ and\ \citenamefont
  {Basdevant}(1996)}]{Perrier_1996}%
  \BibitemOpen
  \bibfield  {author} {\bibinfo {author} {\bibfnamefont {V.}~\bibnamefont
  {Perrier}}\ and\ \bibinfo {author} {\bibfnamefont {C.}~\bibnamefont
  {Basdevant}},\ }\bibfield  {title} {\bibinfo {title} {Besov norms in terms of
  the continuous wavelet transform. {A}pplication to structure functions},\
  }\href@noop {} {\bibfield  {journal} {\bibinfo  {journal} {Math. Mod. Meth.
  Appl. S.}\ }\textbf {\bibinfo {volume} {6}},\ \bibinfo {pages} {649}
  (\bibinfo {year} {1996})}\BibitemShut {NoStop}%
\bibitem [{\citenamefont {Rusaouen}\ \emph {et~al.}(2017)\citenamefont
  {Rusaouen}, \citenamefont {Chabaud}, \citenamefont {Salort},\ and\
  \citenamefont {Roche}}]{Rusaouen_etal_2017}%
  \BibitemOpen
  \bibfield  {author} {\bibinfo {author} {\bibfnamefont {E.}~\bibnamefont
  {Rusaouen}}, \bibinfo {author} {\bibfnamefont {B.}~\bibnamefont {Chabaud}},
  \bibinfo {author} {\bibfnamefont {J.}~\bibnamefont {Salort}},\ and\ \bibinfo
  {author} {\bibfnamefont {P.-E.}\ \bibnamefont {Roche}},\ }\bibfield  {title}
  {\bibinfo {title} {Intermittency of quantum turbulence with superfluid
  fractions from 0\% to 96\%},\ }\href@noop {} {\bibfield  {journal} {\bibinfo
  {journal} {Phys. Fluids}\ }\textbf {\bibinfo {volume} {29}},\ \bibinfo
  {pages} {105108} (\bibinfo {year} {2017})}\BibitemShut {NoStop}%
\bibitem [{\citenamefont {Kida}\ and\ \citenamefont
  {Orszag}(1990)}]{Kida_1990}%
  \BibitemOpen
  \bibfield  {author} {\bibinfo {author} {\bibfnamefont {S.}~\bibnamefont
  {Kida}}\ and\ \bibinfo {author} {\bibfnamefont {S.~A.}\ \bibnamefont
  {Orszag}},\ }\bibfield  {title} {\bibinfo {title} {Energy and {S}pectral
  {D}ynamics in {F}orced {C}ompressible {T}urbulence},\ }\href@noop {}
  {\bibfield  {journal} {\bibinfo  {journal} {J. Sci. Comput.}\ }\textbf
  {\bibinfo {volume} {5}},\ \bibinfo {pages} {85} (\bibinfo {year}
  {1990})}\BibitemShut {NoStop}%
\bibitem [{\citenamefont {Eyink}()}]{Eyink_lecture}%
  \BibitemOpen
  \bibfield  {author} {\bibinfo {author} {\bibfnamefont {G.~L.}\ \bibnamefont
  {Eyink}},\ }\href@noop {} {\bibinfo {title} {Turbulence {T}heory, {C}ourse
  {N}otes}},\ \bibinfo {note}
  {\url{http://www.ams.jhu.edu/~eyink/Turbulence/notes/}}\BibitemShut {NoStop}%
\end{thebibliography}%

\end{document}